\documentclass[preprint2]{emulateapj}
\usepackage{multirow}

\newcommand{\Reff}{$R_e$}

\newcommand{\ATLAS}{ATLAS$^{\rm 3D}$}
\newcommand{\kms}{km~s$^{-1}$}

\newcommand{\Vrot}{$V_{\rm rot}$}

\newcommand{\lambdae}{$\lambda_{\rm Re}$}
\newcommand{\lambdaec}{$\lambda^*_{\rm Re}$}

\newcommand{\sigmainst}{$\sigma_{\rm inst}$}

\newcommand{\reduceme}{\mbox{R\raisebox{-0.35ex}{E}D%
\hspace{-0.05em}\raisebox{0.85ex}{uc}\hspace{-0.90em}%
\raisebox{-.35ex}{{m}}\hspace{0.05em}E}}

\shorttitle{Structural and Kinematic Properties of Virgo cluster dEs. The survey.}
\shortauthors{Toloba et al.}

\begin{document}


\title{Stellar Kinematics and Structural Properties of Virgo Cluster Dwarf Early-Type Galaxies from the SMAKCED Project II. The Survey and a Systematic Analysis of Kinematic Anomalies and Asymmetries}


\author{E. Toloba\altaffilmark{1,2}\footnote{Fulbright Postdoctoral Fellow}}\email{toloba@ucolick.org}

\author{P. Guhathakurta\altaffilmark{1}}

\author{R. F. Peletier\altaffilmark{3}}

\author{A. Boselli\altaffilmark{4}}

\author{T. Lisker\altaffilmark{5}}

\author{J. Falc\'on-Barroso\altaffilmark{6,7}}

\author{J. D. Simon\altaffilmark{2}}

\author{G. van de Ven\altaffilmark{8}}

\author{S.~Paudel\altaffilmark{9}}

\author{E.~Emsellem\altaffilmark{10,11}}

\author{J. Janz\altaffilmark{12}}

\author{M. den Brok\altaffilmark{13}}

\author{J. Gorgas\altaffilmark{14}}

\author{G. Hensler\altaffilmark{15}}

\author{E. Laurikainen\altaffilmark{16,17}}

\author{S.-M Niemi\altaffilmark{18}}

\author{A. Ry\'s\altaffilmark{6,7}}

\author{H. Salo\altaffilmark{16}}

\affil{$^1$UCO/Lick Observatory, University of California, Santa Cruz, 1156 High Street, Santa Cruz, CA 95064, USA}
\affil{$^2$Observatories of the Carnegie Institution for Science, 813 Santa Barbara Street, Pasadena, CA 91101, USA}
\affil{$^{3}$ Kapteyn Astronomical Institute, Postbus 800, 9700 AV Groningen, The Netherlands}
\affil{$^{4}$ Laboratoire d'Astrophysique de Marseille-LAM, Universit\'e d'Aix-Marseille \& CNRS, UMR 7326, 38 rue F. Joliot-Curie, 13388 Marseille Cedex 13, France}
\affil{$^5$ Astronomisches Rechen-Institut, Zentrum f${\rm \ddot{u}}$r Astronomie der Universit${\rm \ddot{a}}$t Heidelberg, M${\rm \ddot{o}}$nchhofstra${\rm \ss}$e 12-14, D-69120 Heidelberg, Germany}
\affil{$^6$ Instituto de Astrof\'{i}sica de Canarias, V\'{i}a L\'{a}ctea s$/$n, La Laguna, Tenerife, Spain}
\affil{$^7$ Departamento de Astrof\'{i}sica, Universidad de La Laguna, E-38205, La Laguna, Tenerife, Spain}
\affil{$^8$ Max Planck Institute for Astronomy, K$\ddot{\rm o}$nigstuhl 17, 69117 Heidelberg, Germany}
\affil{$^{9}$ Korea Astronomy and Space Science Institute, Daejeon 305-348, Republic of Korea}
\affil{$^{10}$European Southern Observatory, Karl-Schwarzschild-Str. 2, 85748, Garching, Germany}
\affil{$^{11}$Universit\'e Lyon 1, Observatoire de Lyon, Centre de Recherche Astrophysique de Lyon and Ecole Normale Sup\'erieure de Lyon, 9 Avenue Charles Andr\'e, F-69230, Saint-Genis Laval, France}
\affil{$^{12}$ Centre for Astrophysics and Supercomputing, Swinburne University, Hawthorn, VIC 3122, Australia}
\affil{$^{13}$ Department of Physics and Astronomy, University of Utah, Salt Lake City, UT 84112, USA}
\affil{$^{14}$ Departamento de Astrof\'{i}sica y F\'isica de la Atm\'osfera, Universidad Complutense de Madrid, 28040, Madrid, Spain}
\affil{$^{15}$ University of Vienna, Department of Astrophysics, T${\rm \ddot{u}}$rkenschanzstra${\rm \ss}$e 17, 1180 Vienna, Austria}
\affil{$^{16}$ Division of Astronomy, Department of Physics, P.O. Box 3000, FI-90014 University of Oulu, Finland}
\affil{$^{17}$ Finnish Center for Astronomy with ESO (FINCA), University of Turku, Finland}
\affil{$^{18}$ Mullard Space Science Laboratory, University College London, Holmbury St. Mary, Dorking, Surrey RH5 6NT, United Kingdom}





\begin{abstract}

We present spatially resolved kinematics and global stellar populations and mass-to-light ratios for a sample of 39 dwarf early-type (dE) galaxies in the Virgo cluster studied as part of the SMAKCED stellar absorption-line spectroscopy and imaging survey. This sample is representative of the early-type population in the absolute magnitude range $-19.0 < M_r < -16.0$. For each dE, we measure the rotation curve and velocity dispersion profile and fit an analytic function to the rotation curve. We study the significance of the departure of the rotation curve from the best fit analytic function (poorly fit) and of the difference between the approaching and receding sides of the rotation curve (asymmetry). We find that $62\pm8~\%$ (23 out of the 39) of the dEs have a significant anomaly in their rotation curve. Analysis of the images reveals photometric anomalies for most galaxies. However, there is no  clear correlation between the significance of the photometric and kinematic anomalies. 
We measure age-sensitive and metallicity-sensitive Lick spectral indices and find a wide range of ages and metallicities. We also find that 4 dEs have emission partially filling in the Balmer absorption lines. Finally, we estimate the total masses and dark matter fractions of the dEs. They have a median total mass and dark matter fraction within the \Reff\ of $ {\rm \log M}_e = 9.1 \pm 0.2$ and $ f_{\rm DM} = 46 \pm 18~\%$. We plot several scaling relations and show that dEs seem to be the bridge between massive early-type and dwarf spheroidal galaxies.

\end{abstract}


\section{Introduction} \label{intro}

Galaxies in the local universe show a color bimodality where quiescent early-type galaxies (ETGs), including elliptical (E) and lenticular (S0) galaxies, populate a narrow red sequence and star-forming late-type galaxies populate a broader blue cloud \citep[e.g.][]{Strateva01,Baldry04}. This bimodality, already in place at redshifts beyond 1 \citep[e.g.][]{Bell04,Cooper07}, is related to morphology, mass, and environment. The fraction of luminous ETGs, at a fixed stellar mass, is higher in dense environments \citep[e.g.][]{Dress80,Sand85,Bing88,Kauff04}. At lower luminosities, this segregation is stronger. Dwarf early-type galaxies (dEs)\footnote{The term dE has traditionally been used to refer to dwarf elliptical galaxies, whereas we loosely use the term here to include dwarf ellipticals and dwarf lenticulars (dS0).}, the low luminosity  ($M_B \gtrsim -18$) and low surface brightness ($\mu_B \gtrsim 22$~mag~arcsec$^{-2}$) population of the ETG class, are found in high density environments and are very rare in isolation \citep[][]{Gavazzi10,Geha12}.

This color bimodality must be the result of some physical mechanism that suppresses the intense star formation and rapidly changes the color of galaxies, moving them from the blue cloud to the red sequence \citep{Faber07}. The mechanism responsible for that transformation is a long-standing problem. One way to approach this problem is to study the stellar kinematics of red sequence galaxies because the stellar kinematics have memory of the processes experienced in the course of their lives.  

Dwarf early-type galaxies are ideal objects with which to investigate this problem for several reasons: (1) dEs are the most numerous galaxy class in clusters and are very rare in isolation \citep{Sand85,Bing88,Gavazzi10,Geha12}, which suggests that the environment is playing a key role in quenching the progenitors of these galaxies; (2) they have low masses ($\sim 10^9$ M$_{\odot}$), and, thus, shallow potential wells, making them very sensitive to gravitational and/or hydrodynamical perturbations; and (3) they have little to no star formation and their dust content is negligible, simplifying the interpretation of the observations.

The dE galaxy class spans a wide range of internal properties. Structurally, dEs are very complex. Beneath their regular and smooth appearance some of them host disks, spiral arms, or irregular features \citep[e.g.][]{Jerjen00,Barazza02,Geha03,Graham03,DR03,Lisk06a,Ferrarese06,Janz12,Janz14}. This complexity is mirrored in their dynamics. Dwarf early-type galaxies with very similar photometric properties can have very different rotation speeds \citep{Ped02,SimPrugVI,Geha02,Geha03,VZ04,Chil09,etj09,etj11,etj14a,Koleva11,Rys13}. The stellar populations of dEs span a range of ages and sub-solar metallicities as well \citep[][]{Mich08,Paudel2010,Koleva09,Koleva11}.

With the goal of understanding the physical processes that form dEs, and therefore the low luminosity end of the red sequence, we have begun the SMAKCED \footnote{http://smakced.net} (Stellar content, MAss and Kinematics of Cluster Early-type Dwarf galaxies) project, a new spectroscopic and photometric survey of dEs in the Virgo cluster, the nearest dense galaxy cluster. This paper is part of a series in which we analyze the structural and kinematical properties of dEs in the Virgo cluster. In \citet[][hereafter Paper I]{etj14a} we focus on the analysis of two dEs in our sample that have kinematically decoupled cores. In (Toloba et al. 2014c, hereafter Paper III) we focus on the analysis of the stellar kinematics of dEs and investigate their relation with morphology and projected distance to the center of the Virgo cluster.
This paper, Paper II, is focused on the description of the spectroscopic part of the survey, the comparison with $H$ band photometry, and the analysis of possible anomalies and asymmetries present in the rotation curves. Because this paper is designed to be a survey paper, we present the spectroscopic and photometric measurements, along with some derived quantities based on these measurements, but we leave their discussion and interpretation for future papers (Toloba et al., in prep.).

This paper is organized as follows. In Sections \ref{sample_sec} and \ref{obs} we describe the sample, the observations, and the main steps of the data reduction. In Section \ref{kin_sec} we describe the kinematic measurements and test their accuracy and reliability. In Section \ref{indices_sec} we describe the measurements of age-sensitive and metallicity-sensitive Lick spectral indices. In Section \ref{phot_sec} we describe the photometric measurements. In Section \ref{anomalies} we analyze the shapes of the rotation curves. In Section \ref{stellarpops} we derive the ages and metallicities based on the Lick indices measured in Section \ref{indices_sec}. In Section \ref{scaling_rels} we infer the dynamical and stellar masses as well as the dark matter fractions, and plot some scaling relations to understand where the sample of dEs presented here lie with respect to other early-type galaxies. In Section \ref{summ} we summarize our findings and conclusions.

\section{Sample} \label{sample_sec}

\begin{figure}
\centering
\includegraphics[angle=-90,width=8cm]{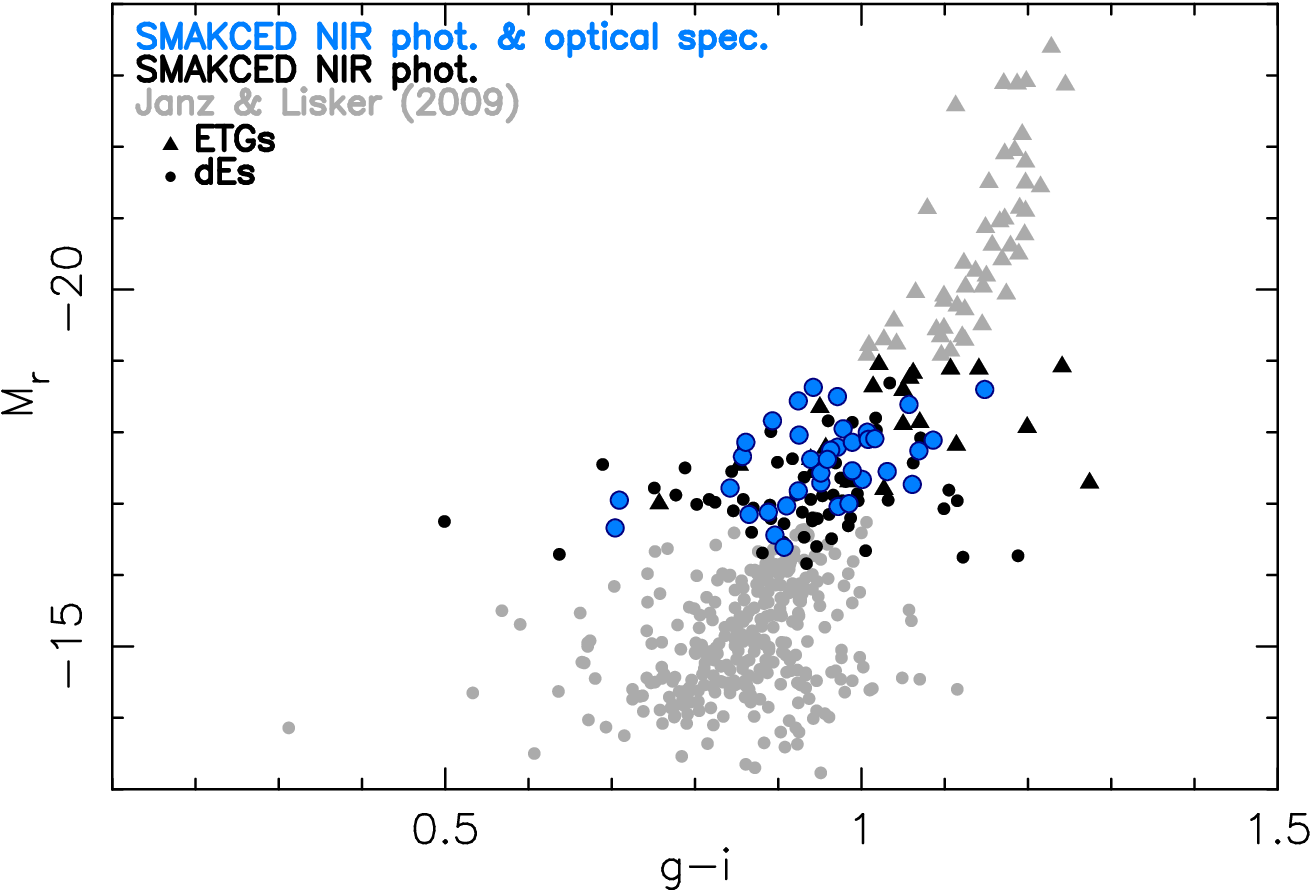}
\includegraphics[angle=-90,width=8cm]{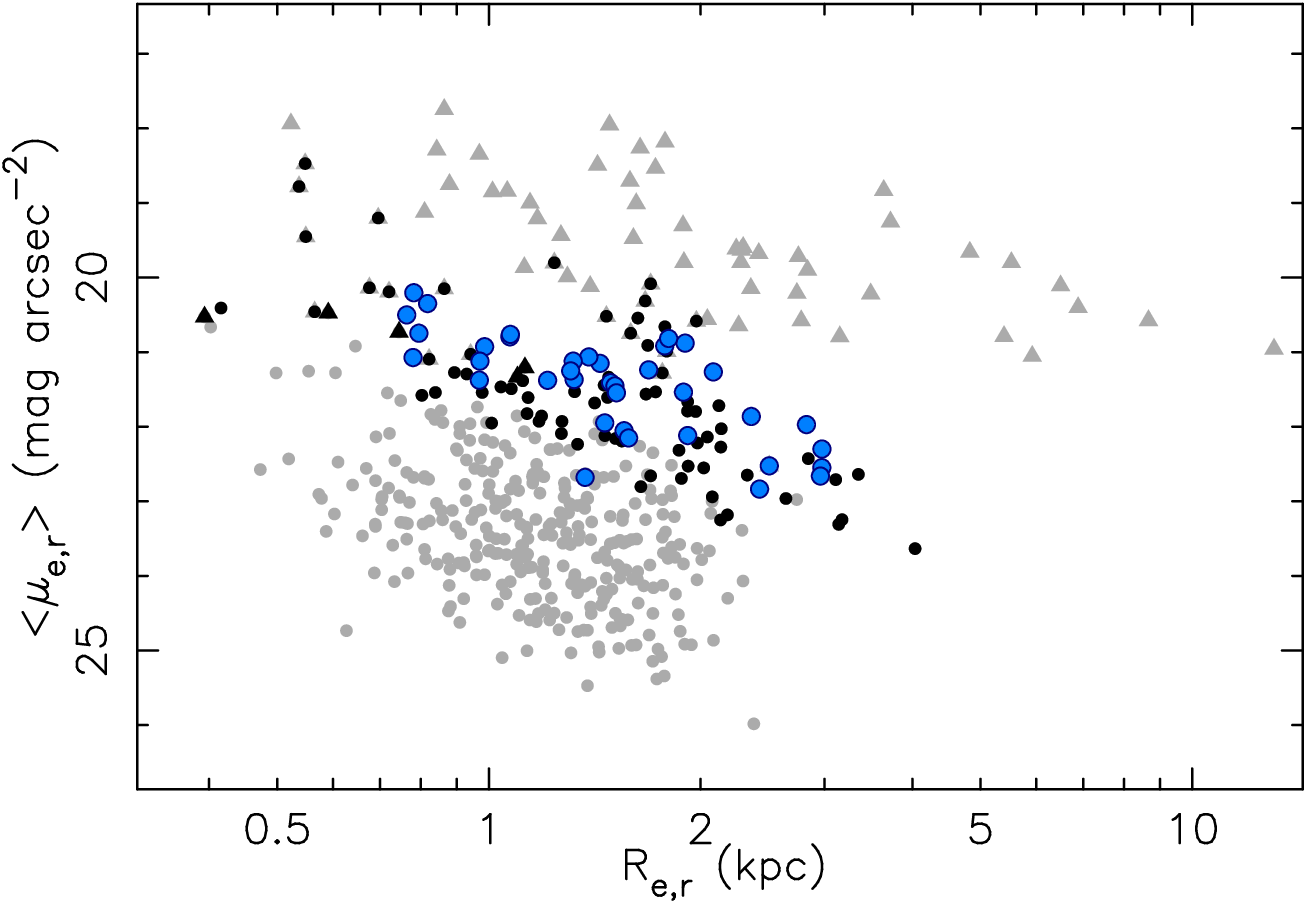}
\caption{Color-magnitude diagram (upper panel) and Kormendy relation (lower panel) of all early-type VCC  galaxy members of the Virgo cluster from \citet[][in grey]{Lisk07,Janz08,Janz09}. The morphological classification into massive and dwarf early-types is from the VCC catalog \citep[][triangles and dots, respectively]{Bing85}. The black and blue dots indicate the SMAKCED near-infrared ($H$ band) photometric survey presented by \citet{Janz14}.  The blue dots indicate the sample of 39 dEs targeted spectroscopically. The SMAKCED sample is representative of the early-type population of galaxies in the Virgo cluster in the absolute magnitude range $-19.0 < M_r < -16.0$.}
\label{sample}
\end{figure}

This paper uses optical spectroscopy and $H$ band photometry collected as part of the SMAKCED project. The sample of 39 dEs presented here are selected favoring high surface brightness dEs in the magnitude range $-19.0 < M_r < -16.0$ (see Figure \ref{sample}). The sample is selected  from the Virgo Cluster Catalog \citep[VCC,][]{Bing85} using updated memberships based on radial velocities from the literature \citep{Lisk06b}. Their selection is based on their early-type morphology and low luminosity; see \citet{Janz14} and Paper I for details.

This paper is focused on the spectroscopic observations of 39 dEs in the Virgo cluster. Although these 39 dEs are not a complete sample, Figure \ref{sample} shows that they are representative of the early-type population of galaxies in the Virgo cluster in the magnitude range $-19.0 < M_r < -16.0$. These 39 dEs are also representative of all the morphological sub-classes defined by \citet{Lisk06b,Lisk06a,Lisk07} using high-pass filtered Sloan Digital Sky Survey \citep[SDSS,][]{SDSS_DR4} images. Table \ref{sample_bins} shows the breakdown of the 39 SMAKCED dEs in bins of morphology, luminosity, and projected distance to the center of the cluster.

\begin{table}
\begin{center}
\caption{Sample of 39 dEs binned in different parameters\label{sample_bins}}
\begin{tabular}{c|c|c|c|c}
\hline \hline
                                  & dE(N)   & dE(nN)  & dE(di)  & dE(bc)\\
\hline
\multicolumn{5}{c}{$-19 < M_r < -17.5$}\\
\hline
$D \le 1.5^{\circ}$           &   2        &       2     &    0       &    0      \\
$1.5 < D \le 4^{\circ}$    &    6        &       5    &    1       &    1      \\
$D > 4^{\circ}$               &    2        &       2     &    0      &     2     \\
\hline
\multicolumn{5}{c}{$-17.5 \le M_r < -16$}\\
\hline
$D \le 1.5^{\circ}$           &   1        &       3     &    1       &    1      \\
$1.5 < D \le 4^{\circ}$    &    3        &       1    &    1       &    1      \\
$D > 4^{\circ}$               &    2        &       3     &    2      &     1     \\
\hline
\end{tabular}
\end{center}
\tablecomments{$D$ indicates the projected distance between each dE and M87, considered to be the center of the Virgo cluster. The different morphological sub-classes are defined by \citet{Lisk06b,Lisk06a,Lisk07}: dE(N) indicates nucleated dE, dE(nN) indicates non-nucleated dE, dE(di) indicates dE with disky underlying structures, dE(bc) indicates dE with a blue center. Four of the galaxies have a double morphological tag, thus, they are counted twice.}
\end{table}

\section{Observations and data reduction} \label{obs}

The observations were conducted at El Roque de los Muchachos Observatory (Spain) and the European Southern Observatory (ESO; Chile).
The data have been reduced following standard recipes for optical long-slit spectroscopy and near-infrared photometry. The details of the observations and the main steps of the data reduction are described below.

\subsection{Spectroscopy}

Eighteen out of the 39 dEs were observed as part of the MAGPOP-ITP collaboration (Multiwavelength Analysis of Galaxy POPulations-International Time Program). These 18 dEs have been previously presented in \citet{etj09,etj11,etj12}; the remaining 21 are analyzed for the first time in this series of papers.

The spectroscopic observations were conducted at three different telescopes. Ten out of the 39 dEs were observed at the INT 2.5~m telescope, 26 dEs were observed at the WHT 4.2~m telescope, and the remaining 3 dEs were observed at the VLT 8~m telescope.
The exposure times varied from 1 to 4 hours depending on the brightness of the dE and the weather conditions.

The observations at the INT were carried out using the IDS spectrograph with the 1200~l~mm$^{-1}$ grating covering the wavelength range 4600$-$5600~\AA. The spectral resolution obtained, using a slit width of $2''$, is 1.6~\AA\ (FWHM).

The observations at the WHT were carried out using the double-arm spectrograph ISIS which allowed us to get simultaneously two spectral ranges. The blue setup consisted of the 1200~l~mm$^{-1}$ grating and covered the wavelength range 4200$-$5000~\AA. The red setup consisted of the 600~l~mm$^{-1}$ grating and covered the wavelength range 5500$-$6700~\AA. The spectral resolution obtained, using a slit width of $2''$, is 1.4 and 3.2~\AA\ (FWHM) in the blue and red setups, respectively.

The observations at the VLT were carried out using the FORS2 spectrograph with the 1400V grism and covered the wavelength range 4500$-$5600~\AA. The spectral resolution obtained, using a slit width of $1.3''$, is 2.7~\AA\ (FWHM).

In Table \ref{instruments} we summarize the instrumental configurations used for the long-slit spectroscopic observations. In Table \ref{obs_table} we provide some details of the observations.

The reduction of the raw spectra is done following the standard procedure for long-slit spectroscopy using the package \reduceme\ \citep{Car99}. A full description of  the steps followed can be found in \citet{etj11}. The main steps are bias and dark current subtraction, flat fielding, and cosmic ray cleaning. The spectra are spatially aligned and wavelength calibrated, leading to typical wavelength residuals of 0.01~\AA. The spectra are then sky subtracted and flux calibrated using the response function derived from our observed flux standards.

\begin{table*}
\begin{center}
\caption{Instrumental configuration for the long-slit spectroscopic observations.\label{instruments}}
\begin{tabular}{|c|c|c|c|c|}
\hline \hline
                  &         INT (2.5m)          &\multicolumn{2}{c|}{WHT (4.2m)}&        VLT (8m)         \\
\hline
Spectrograph &     IDS           &\multicolumn{2}{c|}{ISIS} &       FORS2     \\
\hline
                  &                         &   Blue setup  & Red setup &                     \\
\hline
Grating (lines/mm) &  1200  &   1200          &    600        &      GRISM 1400v \\
Wavelength range (\AA) & 4600-5700 & 4100-4900 & 5400-6900 &  4500-5600 \\
Spectral resolution (FWHM, \AA) &   1.80     &   1.56 &     3.22      &     2.71 \\
Spectral resolution ($\sigma_{inst}$, km s$^{-1}$) &  45             &   44     &    67           &      69  \\
Spatial scale ($''$ pix$^{-1}$)  &  0.40           &     0.40      &    0.44        &      0.25      \\
Slit width ($''$) &   2              &              2    &       2         &                 1.3      \\
\hline
\end{tabular}
\end{center}
\tablecomments{The spectral or instrumental resolution \sigmainst\ is calculated at the central wavelength of each spectrograph setup.}
\end{table*}

\subsection{Photometry}

The observations and data reduction procedure for the $H$ band imaging are described in \citet{Janz14}. This Section is a summary of the most important steps.

The $H$ band images were  collected at three different telescopes. Sixteen out of the 39 dEs were observed at the NOT 2.6~m telescope, six at the TNG 3.6~m telescope, and the remaining 17 at the NTT 3.6~m telescope. One of the 16 dEs observed at the NOT and 12 of the 17 observed at the NTT were taken from archival images. The exposure times were estimated to achieve a S/N$\sim 1$ per pixel at 2 half-light radii (\Reff, see Table \ref{obs_table}). 

The observations at the NOT were carried out using the NOTCam camera, which has a pixel scale of 0.234~pix~arcsec$^{-1}$. The observations at the TNG were carried out using the NICS camera, which has a pixel scale of 0.25~pix~arcsec$^{-1}$. The observations at the NTT were carried out using the SOFI camera, which has a pixel scale of 0.288~pix~arcsec$^{-1}$.

The main steps in the data reduction,  performed with IRAF\footnote{IRAF is distributed by the National Optical Astronomy Observatory, which is operated by the Association of Universities for Research in Astronomy, Inc., under cooperative agreement with the National Science Foundation.}, included flat-fielding, sky subtraction, and correction for field distortions and illumination. 
The observations are done using standard dither patterns to get the sky level simultaneously with the target galaxy. The 22 dEs observed with SOFI and NICS are also corrected for {\it crosstalk}. This effect causes ghost images of bright sources that enhance the signal in the regions where they appear. The SOFI and NICS instrument teams provide scripts to correct for this effect.

The reduced images are flux calibrated using point sources in the field of view of the galaxy and comparing their fluxes to the magnitudes given in the 2MASS point source \citep{2MASS} and UKIDSS \citep{UKIDSS} catalogs. We convert the UKIDSS $H$ band filter into the 2MASS $H$ band filter following \citet{Hewett06}. We find a typical error of $2\%$ in the zeropoints obtained with this method.

\begin{table}
\begin{center}
\caption{Details of the observations\label{obs_table}}
\resizebox{9cm}{!} {
\begin{tabular}{|c|c|c|c|c|c|c|c|}
\hline \hline
               & \multicolumn{2}{c|}{Photometry} & \multicolumn{5}{c|}{Spectroscopy}\\
\hline
Galaxy    &  Telescope  &  $t_{exp}$  & Telescope & $t_{exp}$  & PA & S/N$_{0,B}$ & S/N$_{0,R}$\\
              &                     &       (s)     &                  &    (s)         & ($deg$)&(\AA$^{-1}$) &(\AA$^{-1}$)\\
 (1)         &       (2)          &   (3)         &     (4)         &   (5)         &   (6)     &     (7)    &    (8)   \\
\hline
VCC0009 &  TNG        &  13320  &  WHT &  11300  &  130  & 16.4 &  33.6 \\
VCC0021 &  NTT        &   720     &  INT   &   3600    &  $-81$  & 22.2 & ---\\
VCC0033 &  NTT        &   720    &  WHT  &  15950  &  $-156$ & 10.4 &  18.0 \\
VCC0170 &  NTT        &  10320 &  WHT  &  10800  &  $-8$ & 12.1 & 40.9 \\
VCC0308 &  NTT        &  720      &  WHT  &  2400   &  $-71$ & 24.6 & 44.1 \\
VCC0389 &  TNG        &  1800    &  WHT  &  9000  &  47  & 24.2 & 66.8 \\
VCC0397 &  NOT        &   1200   &  WHT  &  3600  &  $-47$ & 27.7 & ---\\
VCC0437 &  NTT        &   1080   &  WHT  &  10800 & $-103$ & 14.7 & 27.3 \\
VCC0523 &  NOT        &   1080   &  WHT  &  3400  &  $-36$ & 32.1 & 30.6 \\
VCC0543 &  TNG        & 3780     &  WHT  &  10800 &  117 & 19.2 & 53.3 \\
VCC0634 &  NOT        & 10800   &  WHT  &  8000  & $-81$ & 16.0 & 44.7 \\
VCC0750 &  NTT        &  1080    &  WHT  &  8000  &  65  & 21.8 & 52.7 \\
VCC0751 &  NTT        &  720     & WHT  &  10800  &  $-47$ & 17.3 & 31.4 \\
VCC0781 &  NTT        & 360     &  WHT  &  14400 & $-111$ & 18.2 & 26.8 \\
VCC0794 &  TNG        & 10980  &  WHT &  8000  & 168  & 13.0 & 33.9 \\
VCC0856 &  NTT        &  720     &  INT  &  2740   &  $-108$ & 19.1 & --- \\
VCC0917 &  TNG        &  1560   &  WHT &  3600  & $-123$ & 32.0 & 48.5\\
VCC0940 &  NTT        &  10440 &  VLT  &  6900  &  $-167$ & 45.6 & --- \\
VCC0990 &  NOT        & 480      &  INT  & 3000    & $-45$ & 35.4 & --- \\
VCC1010 &  NOT        & 480      &  WHT  &  9000  & $-4$ & 40.8 & 82.6 \\
VCC1087 &  NOT        & 4320    &  WHT  &  3600  & $-74$ & 26.2 & 39.2\\
VCC1122 &  TNG        &  3600  &  WHT  & 3600   &  132  & 28.5 & 19.3\\
VCC1183 &  NTT        &  720    &  INT  &  3600    & $-36$ & 25.8 & ---\\
VCC1261 &  NOT        & 3780   &  INT  &  6930    & 133  & 41.7 & --- \\
VCC1304 &  NOT        & 4320   &  WHT  &  10800 &  $-40$ & 25.6 & 58.2 \\
VCC1355 &  NOT        & 6480   &  WHT  & 10800  & 28  & 18.9 &  35.3 \\
VCC1407 &  NOT        & 3120   &  WHT  &  10800 &  $-28$ &  21.7 &  58.8 \\
VCC1431 &  NOT        & 4500   &  INT  & 3000  & $-45$ & 33.6 & --- \\
VCC1453 &  NOT        & 2820   &  WHT & 7000  & $-56$ & 37.2 & 64.9 \\
VCC1528 &  NTT        & 360     &  WHT &  10800 &  95 & 20.6 & 72.8 \\
VCC1549 &  NTT        & 720     &  INT  &  3300   &  13  & 23.5 & --- \\
VCC1684 &  NTT        & 10020  & VLT  &  13800 & $-318$ & 119.8 & --- \\
VCC1695 &  NTT        &  720    &  WHT  &  3600  & $-141$ & 28.1 & 36.4 \\
VCC1861 &  NOT        & 4560   &  WHT  &  3600  & 122 & 25.7 & 36.3 \\
VCC1895 &  NTT        &  720    &  WHT  &  3800   &  38 & 14.2 & 33.2 \\
VCC1910 &  NOT        &  480   &  INT  &  3800    &  135  &  34.7 & --- \\
VCC1912 &  NOT        &  480   &  INT  &  3600    &  $-14$ & 42.2 & --- \\
VCC1947 &  NTT        &  720   &  INT  &  3058   &  $-54$ & 37.2 & --- \\
VCC2083 &  NOT        &  10080 &  VLT  &  11500 &  $-86$ & 43.0 & --- \\
\hline
\end{tabular}}
\end{center}
\tablecomments{Column (1): Galaxy name. Column (2): Telescope used for the $H$ band images. Column (3): Exposure time for the $H$ band images. Column (4): Telescope used for the long-slit spectroscopy. Column (5): Exposure time for the long-slit spectroscopy. Column (6): Position angle, measured North-East, for the placement of the long-slit. The PA corresponds to the part of the slit with positive radial distances with respect to the center of the galaxy (receding side). Column (7): S/N in the central bin for the blue spectrograph setup. Column (8): S/N in the central bin for the red spectrograph setup. See Section \ref{SNsigma} for details on how the spatial bins are defined.}
\end{table}

\section{Stellar Kinematic Measurements}\label{kin_sec}

We measure the rotation curve and velocity dispersion profile of the SMAKCED dEs. We also evaluate the amplitude of the rotation curve at the \Reff\ and the velocity dispersion within the \Reff. In this Section we describe the steps followed to make these measurements and the tests performed to analyze their reliability.

\subsection{Software and stellar templates}\label{software_sec}

The line-of-sight radial velocities and velocity dispersions are measured using the penalized pixel-fitting method (pPXF) by \citet{PPXF}. This software finds the best fit composite stellar template for a target galaxy and provides the line-of-sight radial velocity and velocity dispersion.
The composite stellar template is created as a linear combination of the high S/N stellar templates that best reproduce the target galaxy spectrum by employing non-linear least-squares optimization. A different weight is given to each one of the templates. The optimal composite stellar template is created independently for each one of the spatial bins where the kinematics are measured.

The stellar templates used for the cross correlation are high S/N stars (S/N$>200$~\AA$^{-1}$) observed with the same instrumental configuration as the galaxies. For observations at the INT and WHT we defocused the stars to make them fill the slit homogeneously. We observed 13 stars in these conditions. Spectral types covered were B9, A0, A5V, G2III, G2V, G8III, G9III, K0I, K1V, K2III, K3III, K4III, M2III. This technique cannot be applied to the stars observed at the VLT given that the VLT telescope cannot be defocused. For the three galaxies observed at the VLT we use as stellar templates the ELODIE stellar library \citep{Prug07}. We convolved the ELODIE stellar library to the same instrumental resolution as the galaxies with a Gaussian function whose FWHM is the quadratic difference between the instrumental resolution of our observations at the VLT and the ELODIE resolution.

\begin{figure}
\centering
\includegraphics[angle=-90,width=8cm]{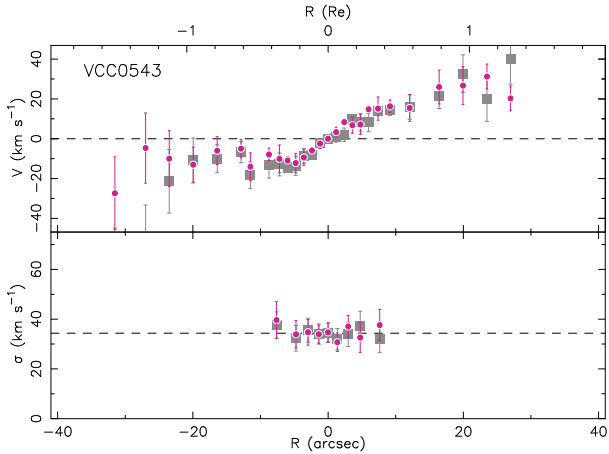}
\caption{Rotation curve (upper panel) and velocity dispersion profile (lower panel) for VCC~543. The red dots indicate the measurements performed using as templates the defocused stars observed with the same instrumental setup as the galaxy. The grey squares indicate the measurements performed using as templates the ELODIE stellar library. Both methods agree within the uncertainties.}
\label{ELODIE_MAGPOP}
\end{figure}

Figure \ref{ELODIE_MAGPOP} shows the rotation curve and velocity dispersion profile of a SMAKCED dE measured using two types of stellar templates: (1) the defocused stars observed with the same instrumental setup as the galaxy; and (2) the ELODIE stellar library convolved to the same instrumental resolution as the galaxy. Both methods lead to stellar kinematic profiles that agree within the error bars. The occasional difference between the radial velocity measured using ELODIE stars as stellar templates and the radial velocity measured using the observed stars only happens at large radii where the S/N is low and it is not a systematic effect. The amplitude of the rotation curve is not affected because it is measured by fitting an analytic function to all the data points. In Sections \ref{rotation} and \ref{rot_sig_atRe} we describe the fitting method and demonstrate that the amplitude of the rotation curve at the \Reff\ depends mainly on how rapidly the rotation increases in the central parts of the galaxy (0.4-0.6\Reff).

\subsection{Radial velocity and velocity dispersion estimations}\label{SNsigma}

The radial velocity ($V$) and velocity dispersion ($\sigma$) as a function of radius of each galaxy are measured by spatially co-adding the spectra. Each co-added spectrum must fulfill two conditions: (1) the minimum bin size for co-addition is 3 pixels, which represents the average seeing in the observations; and (2) the S/N must be above a minimum threshold. The radius for each co-added spectrum is calculated by weighting each pixel in the spatial direction by its luminosity.

The minimum S/N threshold is chosen based on our simulations described in \citet{etj11}. These simulations reproduce the stellar populations and velocity dispersions of Virgo cluster dEs at different S/N ratios. Measuring the radial velocities and velocity dispersions on these simulated dEs, we find that the reliability of the measured radial velocity is not guaranteed for spectra with S/N below 10~\AA$^{-1}$, and the same happens for velocity dispersion estimations with S/N$< 15$~\AA$^{-1}$.

We adopt a minimum S/N threshold of  10~\AA$^{-1}$ to measure radial velocities and  15~\AA$^{-1}$ to measure velocity dispersions. There are some exceptions for which we require a higher S/N ratio: (1)  when a large number of pixels are masked, i.e. when there are large skyline residuals or when the Galactic Na~I doublet is in the spectral range under analysis; (2) when the H${\beta}$ and/or H${\alpha}$ lines are found in emission, as is the case for four SMAKCED dEs (VCC~170, VCC~781, VCC~1304, and VCC~1684); and (3) the three dEs observed at the VLT because of their lower instrumental resolution (see Table \ref{instruments}). In those cases, the minimum S/N threshold is 15~\AA$^{-1}$ to measure $V$ and 25~\AA$^{-1}$ to measure $\sigma$.

The uncertainties in $V$ and $\sigma$ are calculated by running 100 Monte Carlo simulations. In each simulation, the flux of the spectrum is perturbed within a Gaussian function whose width is the uncertainty in the flux obtained in the reduction process. The parameters $V$ and $\sigma$ are measured in each simulation and their uncertainty is defined to be the biweight standard deviation of the Gaussian distribution ($1\sigma_G$; the distribution of all the individual $V$ and $\sigma$ measurements in the Monte Carlo simulations are visually inspected and only those with a Gaussian shape are included in the analysis, those without a Gaussian shape correspond to very poor fits and therefore are not trustworthy measurements).

\begin{figure*}
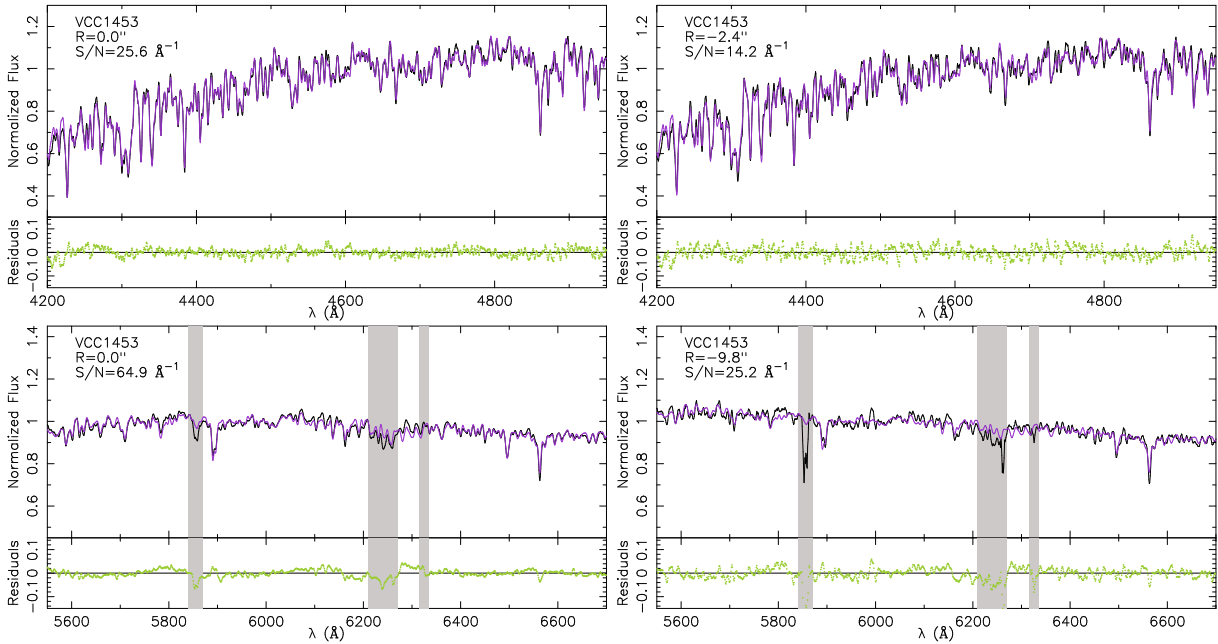

\centering
\includegraphics[angle=-90,width=8cm]{fig3a.ps}
\includegraphics[angle=-90,width=8cm]{fig3b.ps}
\includegraphics[angle=-90,width=8cm]{fig3c.ps}
\includegraphics[angle=-90,width=8cm]{fig3d.ps}
\caption{Examples of the best fit composite stellar template at different distances from the center of VCC~1453. The upper and lower panels show the best fit for the blue and red setups, respectively. The galaxy spectrum is shown in black, the best fit composite stellar template in purple, and the residuals in green. The grey areas indicate the masked regions not used in the fitting process. These regions correspond to the Galactic Na~I doublet absorption and areas severely affected by sky lines.}
\label{individual_fits}
\end{figure*}

Figure \ref{individual_fits} shows some examples of the best fit composite stellar template at different distances from the center of the galaxy, i.e. different S/N, for the blue and red instrumental setups for one of the SMAKCED dEs.

\subsection{Kinematic Profiles} \label{kinprofs_section}

\begin{figure}
\centering
\includegraphics[angle=-90,width=8cm]{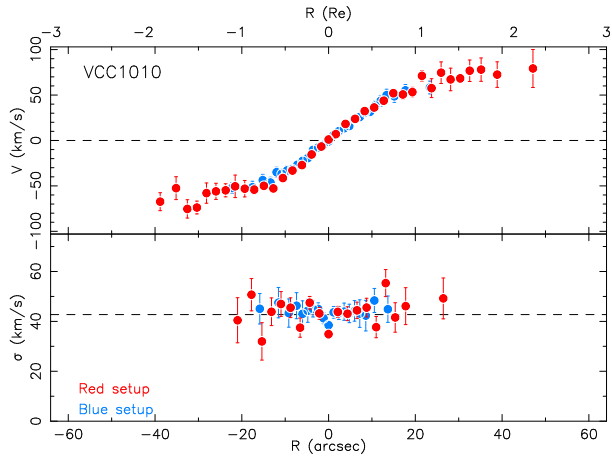}
\caption{Comparison of the kinematic profiles obtained independently for the blue and red instrumental setups for one of the SMAKCED dEs. Note that, besides the good agreement, the red setup is not used in the final $\sigma$ profile due to its lower resolution.}
\label{comparison}
\end{figure}

\begin{figure*}
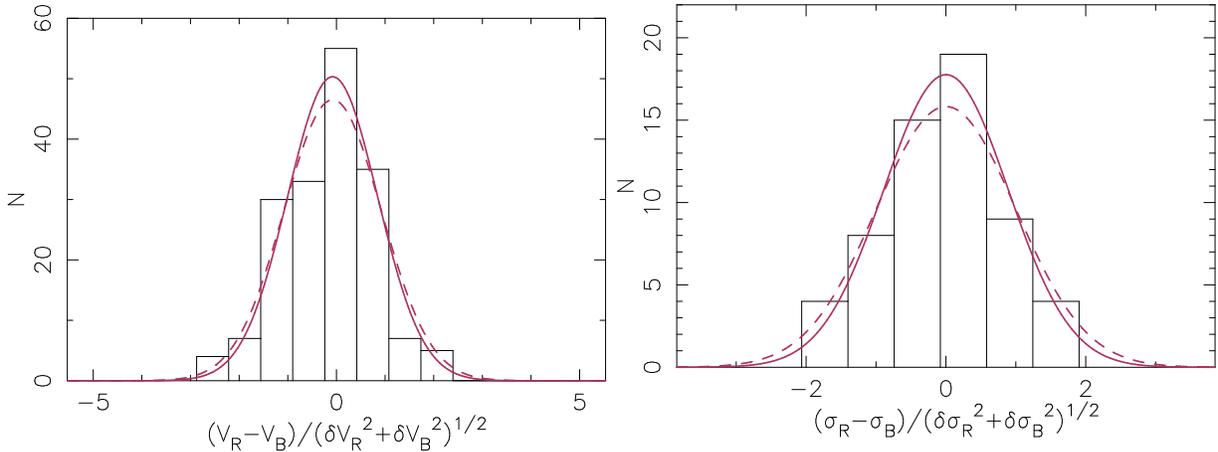

\centering
\includegraphics[angle=-90,width=8cm]{fig5a.ps}
\includegraphics[angle=-90,width=8cm]{fig5b.ps}
\caption{Distribution of differences between $V$ (left panel) or $\sigma$ (right panel) in the blue and red setups at similar radii divided by the estimated uncertainty of the difference. The red dashed line is a Gaussian with $\sigma_G=1$. The red solid line is the best fit Gaussian function  to the histograms, which have $\sigma_G=0.92\pm0.18$ and $\sigma_G=0.89\pm0.14$ for $V$ and $\sigma$, respectively. This indicates that the measurements and uncertainties are consistent between the blue and red setups.}
\label{hists}
\end{figure*}

The 26 SMAKCED dEs observed at the WHT have kinematic profiles measured independently in the blue and red instrumental setups. Figure \ref{comparison} shows the good agreement between the two setups for one galaxy.
We quantify this agreement by calculating the difference between $V$ or $\sigma$ measured in the blue and red setups at similar radii divided by the estimated uncertainty of the difference. We fit a Gaussian function to the distribution of these differences shown in Figure \ref{hists} and normalize the area of the Gaussian by the number of data points used in each histogram. The best fit Gaussian functions are centered on the origin and their widths are very close to unity ($\sigma_G=0.92\pm0.18$ and $\sigma_G=0.89\pm0.14$ for $V$ and $\sigma$, respectively). This indicates that there is good agreement between the kinematic measurements from the blue and red setups and that the uncertainties on those measurements are reliably estimated.

We combine the kinematic measurements obtained from blue and red setups with the following strategy. For the rotation curves, $V$ is the average of the velocities, weighted by their uncertainties, measured at a similar radius in the blue and red setups. At large radii only the red setup is available. For the velocity dispersion profiles, only the blue setup is used because of its higher instrumental resolution.

The stellar kinematic profiles are shown in panels {\it h} and {\it i} of Figures \ref{rotcurve_VCC0009}-\ref{rotcurve_VCC2083}. The rotation curves have the systemic velocity of the galaxy subtracted. This systemic velocity is the velocity of the nucleus, which is the value obtained in the central bin and has the highest signal-to-noise ratio (values in Table \ref{tabledEs}).

\begin{table*}
\begin{center}
\caption{Properties of the SMAKCED dEs. \label{tabledEs}}
{\renewcommand{\arraystretch}{1.}
\resizebox{18cm}{!} {
\begin{tabular}{|c|c|c|c|c|c|c|c|c|c|c|c|c|c|c|c|}
\hline \hline
Galaxy              &  RA (J2000)  &  Dec (J2000) &  M$_r$ & $R_{e,r}$ & M$_H$  & $R_{e,H}$ &  $R_{s,H}$  & Class  &  Comp. &  $n$     & $V_0$    &  $V_{sys}$ & $V_{rot}$ & $\sigma_e$ & $\langle \sigma \rangle$ \\
                        & hh:mm:ss    & dd:mm:ss     & mag     &   $''$     & mag   &       $''$     &   $''$     &            &               &                & \kms     & \kms     &   \kms   &   \kms          &           \kms                      \\
    (1)                &       (2)         &            (3)      &   (4)      &    (5)     &     (6)      &   (7)    &      (8)      &   (9)      &   (10)      &   (11)     &    (12)            &            (13)          &  (14)   &   (15)      &   (16)        \\
\hline \hline
VCC0009 & 12:09:22.25 & +13:59:32.74            &$  -18.2$  &  37.2  &$  -19.1$  &  33.9  &  18.2     $\pm$ 0.2  & dE (N)      &   multi  &  ---  &   23.4  $\pm$   5.0  &$  1674.0$  $\pm$   0.8  &    20.2$^{+  4.8}_{-  5.2 }$ &   26.0  $\pm$  3.9  &   26.0  $\pm$  4.6 \\       
VCC0021 & 12:10:23.15 & +10:11:19.04            &$  -17.1$  &  15.2  &$  -17.6$  &  14.2  &   6.6     $\pm$ 0.1  & dE (bc;N)   &  single  &  1.2  &   11.8  $\pm$   6.6  &$   483.0$  $\pm$   0.7  &    10.7$^{+  8.1}_{-  8.5 }$ &   28.9  $\pm$  2.9  &   27.7  $\pm$  3.0 \\       
VCC0033 & 12:11:07.79 & +14:16:29.19            &$  -16.9$  &   9.8  &$  -17.6$  &   9.7  &   5.3     $\pm$ 0.0  & dE (N)      &  single  &  1.0  &    1.0  $\pm$   3.2  &$  1179.3$  $\pm$   0.8  &     0.9$^{+  4.0}_{-  3.9 }$ &   20.8  $\pm$  4.9  &   21.8  $\pm$  4.6 \\       
VCC0170 & 12:15:56.34 & +14:26:00.33            &$  -17.6$  &  31.3  &$  -18.3$  &  27.3  &  11.2     $\pm$ 0.5  & dE (bc;nN)  &   multi  &  ---  &   16.2  $\pm$   7.0  &$  1398.3$  $\pm$   0.7  &    15.7$^{+  6.8}_{-  6.5 }$ &   26.6  $\pm$  4.6  &   25.0  $\pm$  4.4 \\       
VCC0308 & 12:18:50.90 & +07:51:43.38            &$  -18.0$  &  18.6  &$  -18.7$  &  17.5  &  11.2     $\pm$ 0.1  & dE (di;bc;  &   multi  &  1.9  &   14.0  $\pm$   2.4  &$  1530.2$  $\pm$   0.7  &    11.7$^{+  4.3}_{-  4.4 }$ &   24.1  $\pm$  2.4  &   24.8  $\pm$  1.9 \\       
VCC0389 & 12:20:03.29 & +14:57:41.70            &$  -18.1$  &  18.0  &$  -18.9$  &  16.3  &   6.1     $\pm$ 0.2  & dE (di;N)   &   multi  &  1.9  &   12.2  $\pm$   2.6  &$  1354.3$  $\pm$   0.7  &    12.1$^{+  2.8}_{-  2.6 }$ &   30.9  $\pm$  1.2  &   30.5  $\pm$  1.3 \\       
VCC0397 & 12:20:12.18 & +06:37:23.51            &$  -16.8$  &  13.6  &$  -17.8$  &  13.1  &   6.2     $\pm$ 0.1  & dE (di;N)   &     ---  &  ---  &   47.0  $\pm$   4.4  &$  2441.5$  $\pm$   0.7  &    41.9$^{+  3.0}_{-  2.8 }$ &   35.7  $\pm$  1.9  &   38.1  $\pm$  1.4 \\       
VCC0437 & 12:20:48.10 & +17:29:16.00            &$  -18.0$  &  29.5  &$  -18.8$  &  26.7  &  18.4     $\pm$ 1.1  & dE (N)      &   multi  &  1.7  &   61.8  $\pm$   5.2  &$  1412.3$  $\pm$   0.7  &    50.1$^{+  5.5}_{-  5.7 }$ &   40.9  $\pm$  4.0  &   36.0  $\pm$  4.1 \\       
VCC0523 & 12:22:04.14 & +12:47:14.60            &$  -18.7$  &  26.1  &$  -19.2$  &  18.9  &   8.2     $\pm$ 0.3  & dE (di;N)   &   multi  &  1.5  &   29.6  $\pm$   1.2  &$  1523.5$  $\pm$   0.7  &    27.5$^{+  2.7}_{-  2.8 }$ &   42.2  $\pm$  1.0  &   36.2  $\pm$  0.9 \\       
VCC0543 & 12:22:19.54 & +14:45:38.59            &$  -17.8$  &  23.6  &$  -18.5$  &  20.9  &   7.9     $\pm$ 0.2  & dE (nN)     &   multi  &  1.7  &   20.4  $\pm$   1.4  &$   977.1$  $\pm$   0.7  &    20.2$^{+  2.0}_{-  2.2 }$ &   35.1  $\pm$  1.4  &   34.3  $\pm$  1.6 \\       
VCC0634 & 12:23:20.01 & +15:49:13.25            &$  -18.5$  &  37.2  &$  -18.8$  &  20.5  &  13.9     $\pm$ 0.3  & dE (N)      &   multi  &  1.5  &   46.2  $\pm$   4.2  &$   484.5$  $\pm$   0.7  &    37.5$^{+  2.8}_{-  2.9 }$ &   31.3  $\pm$  1.6  &   29.2  $\pm$  1.7 \\       
VCC0750 & 12:24:49.58 & +06:45:34.49            &$  -17.0$  &  19.5  &$  -17.6$  &  15.0  &   9.2     $\pm$ 0.2  & dE (N)      &   multi  &  1.5  &   20.6  $\pm$   3.6  &$  1058.8$  $\pm$   0.8  &    16.7$^{+  2.9}_{-  2.9 }$ &   43.5  $\pm$  2.9  &   41.4  $\pm$  2.1 \\       
VCC0751 & 12:24:48.30 & +18:11:47.00            &$  -17.5$  &  12.3  &$  -18.3$  &  11.0  &   3.7     $\pm$ 0.0  & dE (di;N)   &   multi  &  2.0  &   13.8  $\pm$   2.6  &$   691.8$  $\pm$   0.8  &    13.7$^{+  4.4}_{-  4.3 }$ &   32.1  $\pm$  2.4  &   32.6  $\pm$  2.4 \\       
VCC0781 & 12:25:15.17 & +12:42:52.59            &$  -17.2$  &  13.4  &$  -18.0$  &  13.8  &   5.4     $\pm$ 0.1  & dE (bc;N)   &   multi  &  1.5  &   -0.0  $\pm$   4.0  &$  -342.1$  $\pm$   0.7  &     0.0$^{+  4.3}_{-  4.8 }$ &   38.0  $\pm$  2.8  &   36.4  $\pm$  2.6 \\       
VCC0794 & 12:25:22.10 & +16:25:47.00            &$  -17.3$  &  37.0  &$  -17.6$  &  28.1  &   9.2     $\pm$ 0.9  & dE (nN)     &     ---  &  ---  &   16.2  $\pm$   4.0  &$  1672.1$  $\pm$   0.8  &    16.1$^{+  5.3}_{-  5.7 }$ &   29.0  $\pm$  3.9  &   26.8  $\pm$  4.1 \\       
VCC0856 & 12:25:57.93 & +10:03:13.54            &$  -17.8$  &  16.5  &$  -18.5$  &  14.0  &   8.4     $\pm$ 0.0  & dE (di;N)   &  single  &  1.0  &   31.4  $\pm$   9.4  &$  1000.6$  $\pm$   0.8  &    25.8$^{+  7.5}_{-  7.2 }$ &   31.3  $\pm$  4.1  &   31.4  $\pm$  3.4 \\       
VCC0917 & 12:26:32.39 & +13:34:43.54            &$  -16.6$  &   9.9  &$  -17.3$  &  10.2  &   3.4     $\pm$ 0.1  & dE (nN)     &   multi  &  1.8  &   -0.6  $\pm$   1.4  &$  1244.8$  $\pm$   0.7  &     0.6$^{+  2.5}_{-  2.4 }$ &   28.4  $\pm$  1.4  &   27.4  $\pm$  1.3 \\       
VCC0940 & 12:26:47.07 & +12:27:14.17            &$  -17.4$  &  19.8  &$  -18.2$  &  18.1  &  10.6     $\pm$ 0.1  & dE (di;N)   &   multi  &  1.1  &   13.4  $\pm$   2.0  &$  1391.9$  $\pm$   0.8  &    11.6$^{+  1.9}_{-  2.0 }$ &   40.4  $\pm$  1.3  &   36.5  $\pm$  1.4 \\       
VCC0990 & 12:27:16.94 & +16:01:27.92            &$  -17.5$  &  10.2  &$  -18.2$  &  10.3  &   4.4     $\pm$ 0.1  & dE (di;N)   &   multi  &  1.7  &   29.0  $\pm$   4.0  &$  1704.1$  $\pm$   0.7  &    27.1$^{+  5.3}_{-  5.1 }$ &   38.7  $\pm$  1.3  &   37.6  $\pm$  1.2 \\       
VCC1010 & 12:27:27.39 & +12:17:25.09            &$  -18.4$  &  22.2  &$  -19.3$  &  20.4  &   8.8     $\pm$ 0.4  & dE (di;N)   &   multi  &  1.7  &   55.6  $\pm$   0.8  &$   930.0$  $\pm$   0.7  &    51.7$^{+  2.0}_{-  1.9 }$ &   44.6  $\pm$  0.9  &   42.7  $\pm$  0.8 \\       
VCC1087 & 12:28:14.90 & +11:47:23.58            &$  -18.6$  &  35.4  &$  -18.9$  &  18.6  &  12.3     $\pm$ 0.1  & dE (N)      &   multi  &  1.5  &    5.6  $\pm$   2.0  &$   658.6$  $\pm$   0.7  &     4.6$^{+  2.5}_{-  2.8 }$ &   42.0  $\pm$  1.5  &   38.6  $\pm$  1.2 \\       
VCC1122 & 12:28:41.71 & +12:54:57.08            &$  -17.2$  &  17.3  &$  -17.9$  &  16.9  &   7.8     $\pm$ 0.2  & dE (N)      &     ---  &  ---  &   16.4  $\pm$   2.0  &$   465.1$  $\pm$   0.7  &    14.9$^{+  2.8}_{-  2.8 }$ &   32.1  $\pm$  1.7  &   28.3  $\pm$  1.4 \\       
VCC1183 & 12:29:22.51 & +11:26:01.73            &$  -17.9$  &  21.1  &$  -18.6$  &  17.0  &  13.7     $\pm$ 0.1  & dE (di;N)   &   multi  &  1.7  &   25.2  $\pm$   9.0  &$  1326.6$  $\pm$   0.8  &    18.9$^{+  6.8}_{-  8.7 }$ &   44.3  $\pm$  2.4  &   40.4  $\pm$  1.4 \\       
VCC1261 & 12:30:10.32 & +10:46:46.51            &$  -18.5$  &  23.8  &$  -19.3$  &  21.7  &  14.0     $\pm$ 0.2  & dE (N)      &   multi  &  1.9  &   -2.2  $\pm$   3.6  &$  1825.3$  $\pm$   0.7  &     1.8$^{+  3.2}_{-  4.3 }$ &   44.8  $\pm$  1.4  &   44.6  $\pm$  0.9 \\       
VCC1304 & 12:30:39.90 & +15:07:46.68            &$  -16.9$  &  16.5  &$  -17.7$  &  16.4  &  10.0     $\pm$ 0.2  & dE (di;N)   &     ---  &  ---  &   47.8  $\pm$   1.4  &$   -37.0$  $\pm$   0.7  &    39.0$^{+  3.6}_{-  3.6 }$ &   25.9  $\pm$  2.7  &   23.8  $\pm$  2.1 \\       
VCC1355 & 12:31:20.21 & +14:06:54.93            &$  -17.6$  &  30.3  &$  -18.2$  &  22.3  &  11.6     $\pm$ 0.2  & dE (N)      &   multi  &  1.5  &    5.8  $\pm$   3.8  &$  1245.0$  $\pm$   0.8  &     5.2$^{+  4.8}_{-  5.0 }$ &   20.3  $\pm$  4.7  &   20.4  $\pm$  4.6 \\       
VCC1407 & 12:32:02.73 & +11:53:24.46            &$  -17.0$  &  12.1  &$  -17.9$  &  11.9  &   5.1     $\pm$ 0.1  & dE (N)      &  single  &  1.4  &    6.2  $\pm$   2.2  &$  1007.2$  $\pm$   0.7  &     5.9$^{+  2.4}_{-  2.5 }$ &   31.9  $\pm$  2.1  &   31.0  $\pm$  2.6 \\       
VCC1431 & 12:32:23.41 & +11:15:46.94            &$  -17.8$  &   9.8  &$  -18.7$  &   9.1  &   4.1     $\pm$ 0.1  & dE (N)      &  single  &  1.5  &   11.2  $\pm$   3.6  &$  1489.4$  $\pm$   0.7  &    10.6$^{+  4.6}_{-  4.8 }$ &   52.4  $\pm$  1.6  &   52.2  $\pm$  1.4 \\       
VCC1453 & 12:32:44.22 & +14:11:46.17            &$  -17.9$  &  18.9  &$  -18.7$  &  16.9  &  10.6     $\pm$ 0.2  & dE (N)      &   multi  &  2.2  &    7.0  $\pm$   4.6  &$  1880.0$  $\pm$   0.7  &     5.6$^{+  7.7}_{-  7.8 }$ &   35.6  $\pm$  1.4  &   32.9  $\pm$  1.2 \\       
VCC1528 & 12:33:51.61 & +13:19:21.03            &$  -17.5$  &   9.6  &$  -18.3$  &   8.4  &   2.6     $\pm$ 0.0  & dE (nN)     &   multi  &  2.1  &    0.8  $\pm$   1.2  &$  1615.4$  $\pm$   0.7  &     0.8$^{+  1.5}_{-  1.5 }$ &   47.0  $\pm$  1.4  &   48.0  $\pm$  1.6 \\       
VCC1549 & 12:34:14.83 & +11:04:17.51            &$  -17.3$  &  12.1  &$  -18.3$  &  11.4  &   4.8     $\pm$ 0.1  & dE (N)      &  single  &  1.7  &   27.0  $\pm$   3.4  &$  1389.3$  $\pm$   0.8  &    25.4$^{+  5.8}_{-  5.8 }$ &   36.7  $\pm$  2.3  &   36.6  $\pm$  1.6 \\       
VCC1684 & 12:36:39.40 & +11:06:06.97            &$  -16.7$  &  18.3  &$  -17.2$  &  18.5  &  10.3     $\pm$ 0.1  & dE (di;bc;  &     ---  &  ---  &   17.8  $\pm$   0.6  &$   660.9$  $\pm$   0.8  &    15.1$^{+  3.3}_{-  3.3 }$ &   28.0  $\pm$  0.9  &   32.2  $\pm$  0.9 \\       
VCC1695 & 12:36:54.85 & +12:31:11.93            &$  -17.7$  &  24.0  &$  -18.2$  &  16.2  &   4.2     $\pm$ 0.1  & dE (di;nN)  &   multi  &  ---  &   12.0  $\pm$   1.2  &$  1716.6$  $\pm$   0.7  &    12.6$^{+  3.2}_{-  3.2 }$ &   24.4  $\pm$  2.2  &   26.9  $\pm$  1.4 \\       
VCC1861 & 12:40:58.57 & +11:11:04.34            &$  -17.9$  &  19.0  &$  -18.6$  &  15.3  &   6.7     $\pm$ 0.1  & dE (N)      &   multi  &  1.5  &    5.6  $\pm$   1.6  &$   629.7$  $\pm$   0.7  &     5.3$^{+  2.5}_{-  2.5 }$ &   31.3  $\pm$  1.5  &   28.5  $\pm$  1.4 \\       
VCC1895 & 12:41:51.97 & +09:24:10.28            &$  -17.0$  &  16.3  &$  -17.7$  &  15.0  &   6.5     $\pm$ 0.1  & dE (nN)     &  single  &  1.3  &   15.4  $\pm$   2.0  &$   970.2$  $\pm$   0.7  &    14.7$^{+  2.7}_{-  2.6 }$ &   23.8  $\pm$  3.0  &   25.2  $\pm$  3.0 \\       
VCC1910 & 12:42:08.67 & +11:45:15.19            &$  -17.9$  &  13.4  &$  -18.9$  &  11.6  &   5.3     $\pm$ 0.1  & dE (di;N)   &   multi  &  1.6  &   11.0  $\pm$   1.8  &$   195.8$  $\pm$   0.7  &    10.0$^{+  3.4}_{-  3.3 }$ &   37.0  $\pm$  1.2  &   42.7  $\pm$  0.9 \\       
VCC1912 & 12:42:09.07 & +12:35:47.93            &$  -17.9$  &  22.5  &$  -18.7$  &  22.2  &  15.0     $\pm$ 0.9  & dE (di;bc;  &     ---  &  ---  &   32.6  $\pm$   3.8  &$   -97.6$  $\pm$   0.7  &    25.3$^{+  7.5}_{-  7.7 }$ &   36.0  $\pm$  1.5  &   35.1  $\pm$  1.1 \\       
VCC1947 & 12:42:56.34 & +03:40:35.78            &$  -17.6$  &   9.3  &$  -18.7$  &   9.1  &   4.1     $\pm$ 0.1  & dE (di;N)   &   multi  &  1.5  &   49.4  $\pm$   2.0  &$   973.5$  $\pm$   0.7  &    46.7$^{+  5.0}_{-  5.1 }$ &   48.3  $\pm$  1.3  &   44.2  $\pm$  1.0 \\       
VCC2083 & 12:50:14.48 & +10:32:24.07            &$  -16.4$  &  17.1  &$  -17.1$  &  14.1  &   8.9     $\pm$ 0.1  & dE (N)      &  single  &  0.9  &    4.8  $\pm$   2.2  &$   867.9$  $\pm$   0.7  &     3.9$^{+  4.8}_{-  4.7 }$ &   28.4  $\pm$  2.4  &   30.0  $\pm$  2.7 \\       
\hline
\end{tabular}
}}
\end{center}
\tablecomments{Column 1: galaxy name. Columns 2 and 3: right ascension and declination in $J2000$. Columns 4 and 5: $r$ band magnitude (in the AB system) and half-light radius by \citet{Janz08,Janz09}. Columns 6 and 7: $H$ band magnitude (in the AB system) and half-light radius by \citet{Janz14}. The transformation of the $H$ band magnitudes from the Vega system to the AB system is done following \citet{Blanton07}. Column 8: scale length of the $H$ band surface brightness profile. The nucleus, if present, is excluded from the fit. Column 9: morphological class based on the analysis of high-filtered optical images by \citet{Lisk06a,Lisk06b,Lisk07}. The classes are: {\it N} for nucleated, {\it nN} for non-nucleated, {\it di} for disky structures, and {\it bc} for blue center. Note that the {\it N/nN} classification is based on high-filtered SDSS optical images and checked also in the $H$ band images from \citet{Janz14}, however, when higher resolution and deeper images are analyzed, some {\it nN} dEs appear to be nucleated \citep{Cote06}. Column 10: whether the $H$ band surface brightness profile is best described by a single S\'ersic profile or by a multi-component profile based on the analysis by \citet{Janz14}. The nucleus, if present, is not included in the single or multi-component definition. Column 11: S\'ersic index $n$ that corresponds to the best fit single S\'ersic profile by \citet{Janz14}. Column 12: best fit $V_0$ parameter of the {\it Polyex} model. Column 13: heliocentric systemic velocity that corresponds to the central bin of the rotation curve. Column 14: rotation speed at the \Reff~ measured in the best fit {\it Polyex} model using $V_0$ as the only free parameter. Column 15: integrated line-of-sight velocity dispersion within an aperture with radius equal to the \Reff. Column 16: weighted average of the line-of-sight velocity dispersion profile.}
\end{table*}

\subsection{Comparison with the literature}\label{lit_comp}

\begin{figure*}
\centering
\resizebox{0.9\textwidth}{!}{\includegraphics[angle=-90]{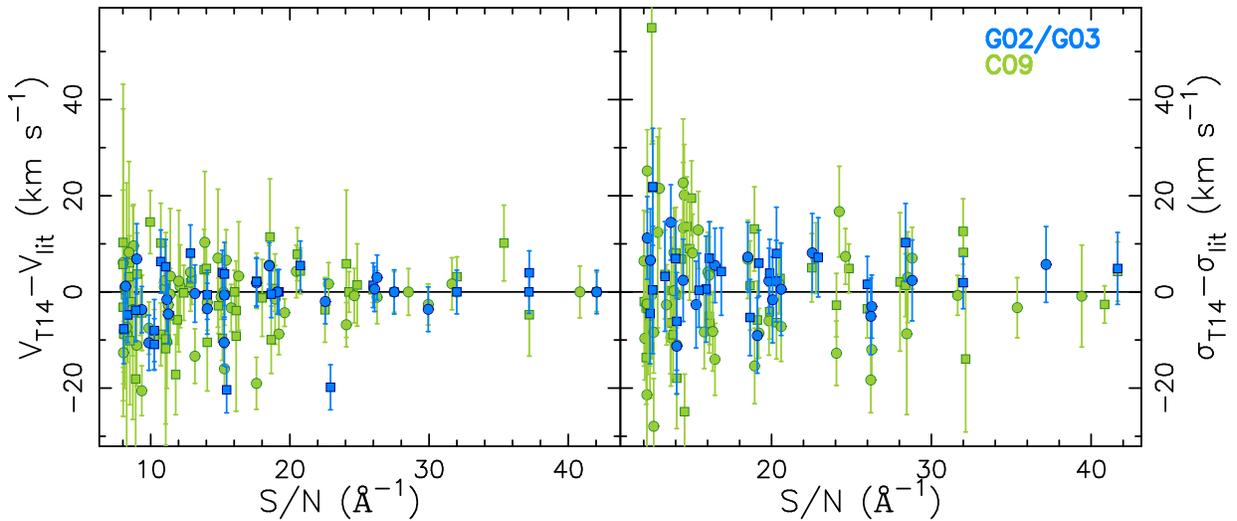}}
\caption{Comparison of the rotation curve (left panel) and velocity dispersion profile (right panel) for the 11 dEs in common with the literature as a function of S/N. For each radius, we calculate the difference between $V$ (left panel) or $\sigma$ (right panel) measured in this work (T14) and measured by other works in the literature (lit). The squares and dots indicate the positive and negative radial distances with respect to the center of the galaxy, measured along the long-slit. The uncertainties are the quadrature sum of the error bars measured in this work and the literature. The samples for comparison are \citet[][G02/G03]{Geha02,Geha03} in blue, and \citet[][C09]{Chil09} in green. The three samples compared cover the same range in S/N.}
\label{comp_lit}
\end{figure*}

\begin{figure*}
\centering
\resizebox{0.9\textwidth}{!}{\includegraphics[angle=-90]{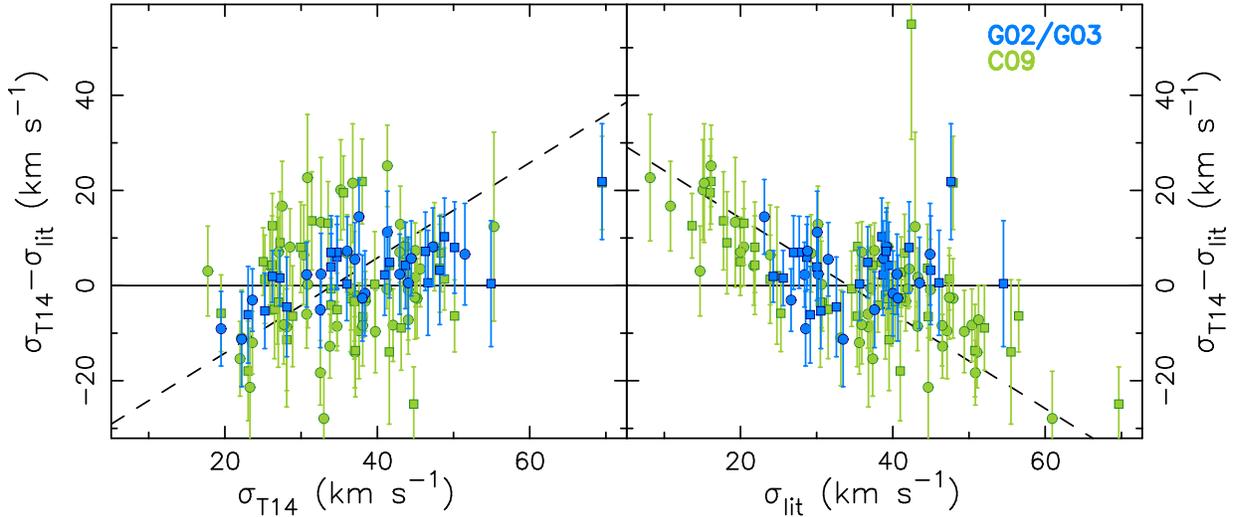}}
\caption{Comparison of the velocity dispersion $\sigma$ measured in this work (T14) and in the literature (G02/G03, C09) for the 11 dEs in common. Colors and symbols are as in Figure \ref{comp_lit}. The dashed line in the left panel has a slope of $+1$ and it is centered in the mean value covered by the data. The dashed line in the right panel has a slope of $-1$ and it is also centered in the mean value covered by the data. The elongation of the blue points along a line with slope $<+1$ suggests that the scatter  in T14's measurements is only slightly larger than in G02/G03's measurements. The elongation of the green points along the line with slope $-1$ suggests that the scatter in C09's measurements is larger than in T14, and consequently also larger than in G02/G03.}
\label{comp_sig}
\end{figure*}

\begin{figure*}
\centering
\resizebox{0.9\textwidth}{!}{\includegraphics[angle=-90]{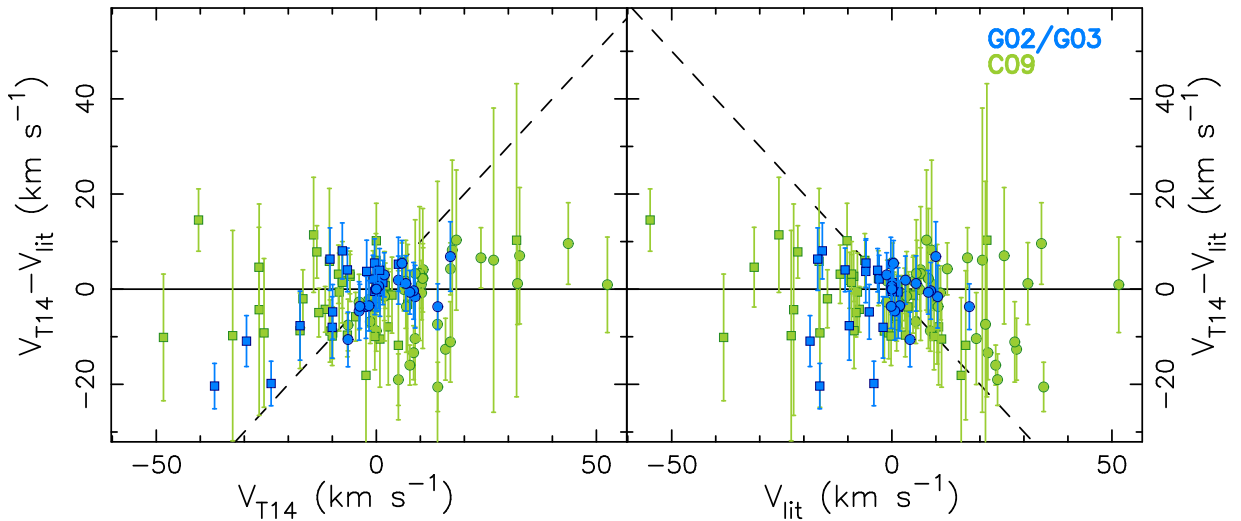}}
\caption{Same as Figure \ref{comp_sig} for radial velocities $V$. Our measurements are consistent with the published measurements.}
\label{comp_v}
\end{figure*}

\begin{figure*}
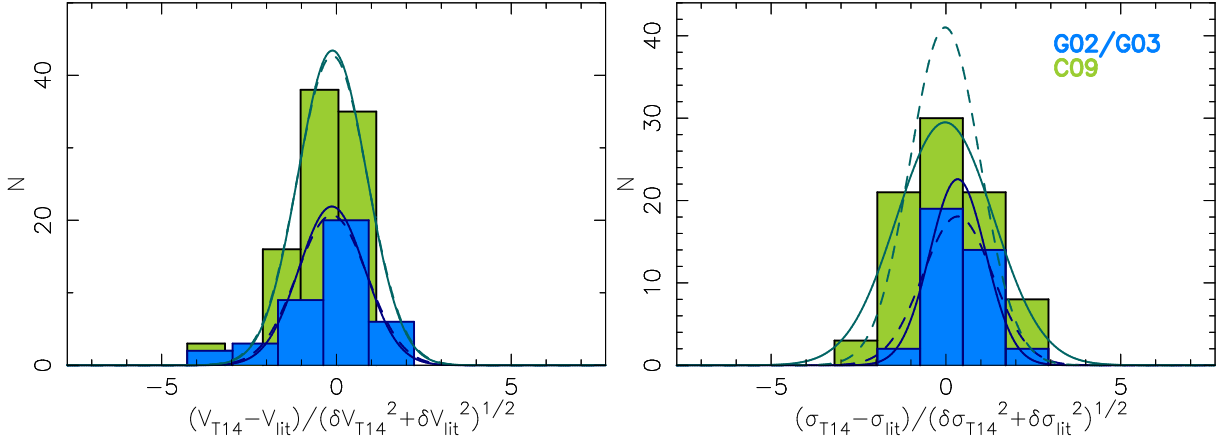

\centering
\includegraphics[angle=-90,width=8cm]{fig9a.ps}
\includegraphics[angle=-90,width=8cm]{fig9b.ps}
\caption{Distribution of the measured $V_{\rm T14}-V_{\rm lit}$ and $\sigma_{\rm T14}-\sigma_{\rm lit}$ of Figures \ref{comp_sig} and \ref{comp_v} normalized by their estimated uncertainty. The solid lines are the best fit Gaussian to the distributions. For $V$, the Gaussian's widths obtained are $\sigma_G=0.95\pm0.44$ and $\sigma_G=0.98\pm0.26$ for G02/G03 and C09, respectively. For $\sigma$, the Gaussian's widths are $\sigma_G=0.80\pm0.11$ and $\sigma_G=1.39\pm0.27$, respectively. The dashed line is a Gaussian whose width is $\sigma_G=1$. Colors are as in Figure \ref{comp_lit}. While the uncertainties in $V$ are enough to explain the scatter of values, the uncertainties in $\sigma$ of C09 are too small making the Gaussian broader than 1.}
\label{hists_comp_lit}
\end{figure*}

Fourteen of the SMAKCED dEs are in common with other works in the literature \citep{Ped02,SimPrugVI,Geha02,Geha03,VZ04,Chil09}. A direct comparison between the SMAKCED kinematic profiles and those in the literature is difficult to interpret because: (1) the position angle of the slit is not usually the same; (2) the set of templates used to approach the mismatch problem from stellar populations and the instrumental profile differs from work to work; and (3) the spatial co-addition scheme is also different in each work. 

To compare the kinematic profiles of the SMAKCED dEs and those in the literature we use only dEs observed with a long-slit whose PA is less than $20^{\circ}$ different from our PA. We do not use dEs for which the published kinematic profiles are folded, so that we do not know whether the positive distances with respect to the center of the dE have approaching or receding velocities. We make this comparison for 11 dEs in common with \citet[][G02/G03]{Geha02,Geha03} and \citet[][C09]{Chil09}.

For each radius, we calculate the difference between $V$ or $\sigma$ measured in this work (T14) and in the literature \citep{Geha02,Geha03,Chil09}. Figure \ref{comp_lit} shows these differences as a function of the signal-to-noise ratio. The range of S/N covered is the same as for G02/G03 and C09.

Figures \ref{comp_sig} and \ref{comp_v} show the same differences as Figure \ref{comp_lit} as a function of $\sigma$ and $V$, respectively. 
These Figures can be understood in a hypothetical situation where we have two sets of measurements of the same quantity ($x_a$, $x_b$) whose true value is $x_0$. If the set of measurements $x_a$ were perfect and the only scatter, here defined as the difference between the measured and true value, were that for the $x_b$ set of measurements, then the figure $x_a-x_b$ versus $x_a$ will show a vertical scatter about $x_0$ and the figure $x_a-x_b$ versus $x_b$ will show a scatter along the line with slope $-1$. In the same way, if the set of measurements $x_b$ were perfect and the only scatter were that of the $x_a$ set of measurements, then the figure $x_a-x_b$ versus $x_b$ will show a vertical scatter about $x_0$, and the figure $x_a-x_b$ versus $x_a$ will show a scatter along the line with slope $+1$.
If both sets of measurements have some scatter, the resulting figures would be a combination of vertical scatter and scatter along the lines with slope $+1/-1$. In addition, $x_a$ and $x_b$ may have a set of true values instead of only $x_0$ which adds some horizontal scatter to the figures. Due to the fact that the rotation speed varies with radius but the velocity dispersion is generally flat, the horizontal scatter will be larger for $V$ than for $\sigma$, which will make the interpretation of the scatter easier in $\sigma$ where the horizontal scatter will be nearly negligible.

Figures \ref{comp_sig} and \ref{comp_v} show that the uncertainties in the $\sigma$ measurements are slightly larger for T14 than for G02/G03 ($1.2$ times larger, the slope of the blue symbols on the left panel of Figure \ref{comp_sig} is lower than $+1$), and the uncertainties in the $\sigma$  measurements for C09 are larger than for T14 ($1.6$ times larger, the slope of the green symbols on the right panel of Figure \ref{comp_sig} is $\sim -1$), and consequently than G02/G03. On the other hand, the $V$ measurements of this work are consistent with the published $V$ measurements of G02/G03 and C09.

While the spread of the blue symbols in the x axis of the left and right panels of Figure \ref{comp_sig} is similar and nearly all the error bars are consistent with $\sigma_{\rm T14}-\sigma_{\rm lit}=0$, the spread of the green symbols is larger on the right panel and for $\sigma$ values of $\sigma_{\rm lit}<20$~\kms\ and $\sigma_{\rm lit}>50$~\kms\ the error bars are not consistent with $\sigma_{\rm T14}-\sigma_{\rm lit}=0$. This indicates that there are some systematic offsets in the data of C09 that are not accounted by their uncertainties (as it is also seen in Figure \ref{Dispersion_comparison}).

While in Figures \ref{comp_sig} and \ref{comp_v} we analyze the scatter of the measurements around the true value, in Figure \ref{hists_comp_lit} we analyze whether that scatter is consistent with the error bars reported. Figure \ref{hists_comp_lit} shows the distribution of the differences in $V$ and $\sigma$ measured in this work and in the literature divided by their estimated uncertainties. Each histogram is fitted by a Gaussian function whose area is normalized by the number of data points used. For $V$ measurements, the width of the best fit Gaussian function is $\sigma_G=0.95\pm0.44$ and $\sigma_G=0.98\pm0.26$ for G02/G03 and C09, respectively. For $\sigma$ measurements, the width of the best fit Gaussian function is $\sigma_G=0.80\pm0.11$ and $\sigma_G=1.39\pm0.27$ for G02/G03 and C09, respectively. The best fit Gaussian functions for $V$ and $\sigma$ are centered on the origin, indicating that there are no systematic offsets between our measurements and those in the literature. However, the width of the best fit Gaussian for $\sigma$ when compared to C09 is significantly larger than the unity. This large width is a consequence of the elongation of the green points along a line with slope $-1$ and the fact that the error bars do not reach the horizontal line $(\sigma_{\rm T14}-\sigma_{\rm lit}) = 0$ seen on the right panel of Figure \ref{comp_sig}. These suggest that the error bars for the $\sigma$ measurements by C09 are slightly underestimated (see also Figure \ref{Dispersion_comparison}).

\begin{figure*}
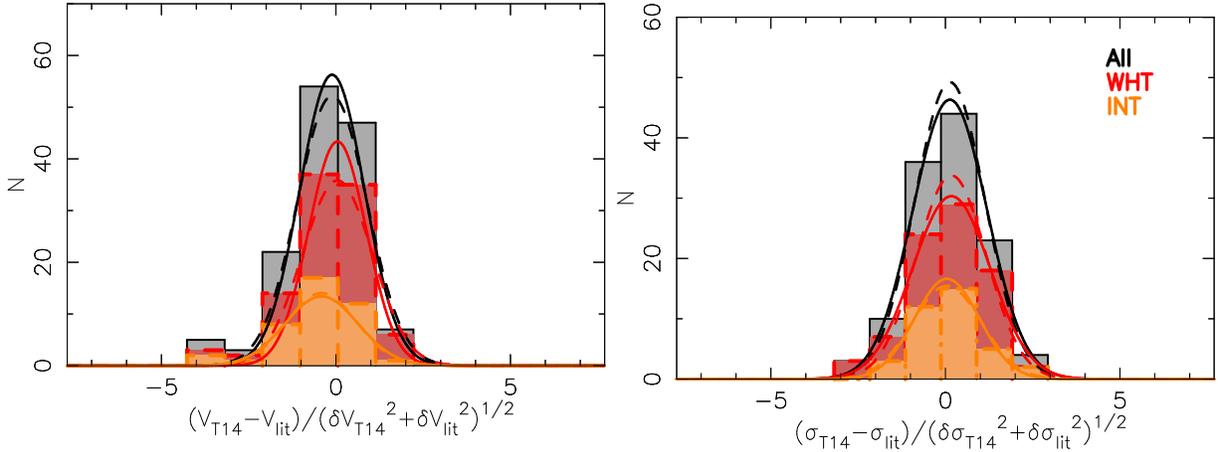

\centering
\includegraphics[angle=-90,width=8cm]{fig9c.ps}
\includegraphics[angle=-90,width=8cm]{fig9d.ps}
\caption{Same as Figure \ref{hists_comp_lit} with the dEs in common with G02/G03 and C09 combined together in grey, and split by the telescope used in our observations in red and orange. The solid lines are the best fit Gaussian to the distributions. For $V$, the Gaussian's widths obtained are  $\sigma_G=0.99\pm0.23$, $\sigma_G=0.98\pm0.21$, and $\sigma_G=0.97\pm0.27$ for all the measurements combined, only for those dEs observed at the WHT, and only for those dEs observed at the INT, respectively. For $\sigma$, the Gaussian's widths are $\sigma_G=1.06\pm0.10$, $\sigma_G=1.11\pm0.23$, and $\sigma_G=0.93\pm0.18$, respectively. The dashed line is a Gaussian whose width is $\sigma_G=1$. All of the best Gaussian fits are consistent with a Gaussian function whose width is 1, which indicates that the $V$ and $\sigma$ measurements done by this work and the literature agree within the error bars for the combined sample and for the subsamples split according to the telescope used in this work. }
\label{hists_telescopes}
\end{figure*}

Figure \ref{hists_telescopes} shows the distribution of differences between $V$ and $\sigma$ divided by their estimated uncertainty for the 11 dEs in common between this work and the literature. The different histograms indicate all the data combined together, and the data observed at the WHT and at the INT independently. For $V$ measurements, the width of the best fit Gaussian function is $\sigma_G=0.99\pm0.23$, $\sigma_G=0.98\pm0.21$, and $\sigma_G=0.97\pm0.27$ for all the measurements combined, only for those dEs observed at the WHT, and only for those dEs observed at the INT, respectively. For $\sigma$ measurements, the best fit Gaussian's widths are $\sigma_G=1.06\pm0.10$, $\sigma_G=1.11\pm0.23$, and $\sigma_G=0.93\pm0.18$, respectively. The shapes and widths of the histograms for dEs observed at the WHT and the INT are very similar to each other and also very similar to the histogram that combines the 11 dEs. All the best fit Gaussian functions are consistent within the $1\sigma_G$ uncertainty with a Gaussian function whose width is one. This indicates that the SMAKCED data is consistent with the different datasets obtained by other teams regardless of the telescope used to obtain the data.

We make three independent tests to check the accuracy and reliability of the $V$ and $\sigma$ measurements: (1) we check the internal agreement between the blue and red instrumental setups (Figure \ref{hists}); (2) we check the agreement between our measurements and those by G02/G03; and (3) we check the agreement between our measurements and those by C09 (Figures \ref{comp_lit},  \ref{comp_sig}, \ref{comp_v}, \ref{hists_comp_lit}, and \ref{hists_telescopes}). All these tests suggest that our measurements and uncertainties are accurate and reliable.

\subsection{Fitting a smooth function to the rotation curve}\label{rotation}

We measure the amplitude and shape of the rotation curve fitting the analytic function:

\begin{equation}\label{Polyexeq}
V(R)=V_0\left(1-e^{(-R/R_{PE})}\right) \left( 1+\frac{\alpha R}{R_{PE}} \right)
\end{equation}

This function, named {\it Polyex}, is described in \citet{GH02}, and used in \citet{Cat06} to fit the rotation curves of disk galaxies. It depends on three parameters: $V_0$, $R_{PE}$, and $\alpha$, which determine the amplitude, the exponential scale of the inner region, and the slope of the outer part of the rotation curve, respectively. 

We reduce the number of free parameters to only one, $V_0$.
The parameter $\alpha$, which determines the slope of the curve beyond the turnover radius $R_{PE}$, is not constrained in our rotation curves because of their limited radial coverage. The only dEs that have rotation curves that go well beyond the turnover radius are NGC~147 and NGC~185, which are two of the satellites of M31 \citep{Geha10}. Even though their radial coverage is larger than 8\Reff, $\alpha$ is still unconstrained due to the large uncertainties in their velocity measurements. \citet{Cat06} fitted the {\it Polyex} function to a master rotation curve that was the co-addition of several hundreds of individual rotation curves of disk galaxies. They obtained an exquisite curve with a very large radial coverage, several times the scale length of the co-added disk galaxies, and very small uncertainties (typically lower than $1\%$). \citet{Cat06} found that the best fit {\it Polyex} function has $\alpha \sim 0.02$ regardless the luminosity of the galaxy. We fix $\alpha$ to 0.02, and show in Section \ref{rot_sig_atRe} that this choice does not affect the measurement of the amplitude of the rotation curve at the \Reff.

We also fix $R_{PE}$ to $R_s$, where $R_s$ is the scale length of the $H$ band surface brightness profile. We estimate $R_s$ fitting an exponential profile. The nucleus, if present, is excluded from the fit. For low luminosity disk galaxies $R_{PE}/R_s \sim 1$ \citep{Cat06}. Fourteen of the SMAKCED dEs have enough data points in the slowly increasing or flat part of the rotation curve to constrain $R_{PE}$. For those 14 dEs we find $R_{PE}/R_s \sim 1$, within the uncertainties, when $V_0$ and $R_{PE}$ are left as free parameters. Panel  {\it h} of Figures \ref{rotcurve_VCC0009}-\ref{rotcurve_VCC2083} shows, in red, the best fit {\it Polyex} function.

The rotation curves of VCC~1183 and VCC~1453 show a different behaviour in the central region of the galaxy with respect to the outer region. These kinematically decoupled cores are not taken into account in the fit and are the main focus of Paper~I in this series.

The rotation curves of VCC~33, VCC~781, VCC~917, VCC~1261, and VCC~1528 are consistent with being non-rotators. The analysis of the dynamics of the SMAKCED galaxies (non-rotators, slow rotators, and fast rotators) is the main focus of Paper~III in this series.

\subsection{Amplitude of the rotation curve and velocity dispersion within the half-light radius}\label{rot_sig_atRe}

\begin{figure}
\centering
\includegraphics[angle=-90,width=8cm]{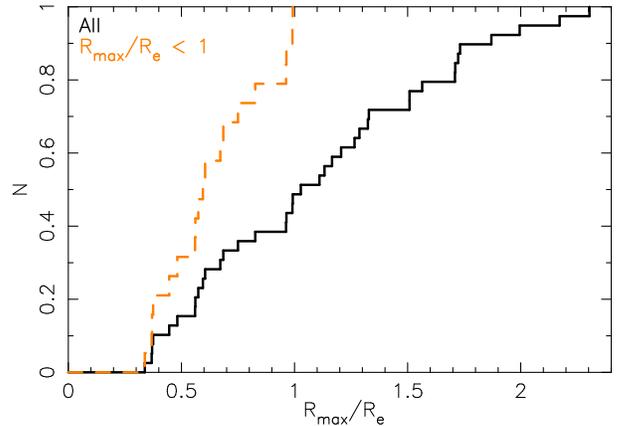}
\caption{Cumulative distribution of the maximum radial extent of the rotation curves. In black, all the galaxies are considered. In orange, only those dEs with $R_{\rm max}/$\Reff$<1$. Those galaxies with $R_{\rm max}/$\Reff$<1$ have a typical radial coverage of $R_{\rm max}/$\Reff$\sim 0.6$}
\label{RmaxRe}
\end{figure}

\begin{figure}
\centering
\includegraphics[angle=-90,width=8cm]{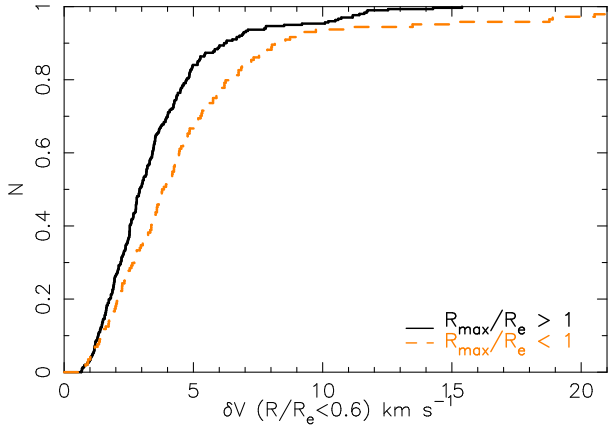}
\caption{Cumulative distribution of uncertainties in the region $R/$\Reff$<0.6$, which is the typical radial coverage for those galaxies with $R_{\rm max}/$\Reff$<1$. The solid line indicates the cumulative distribution for dEs with $R_{\rm max}/$\Reff$>1$. The dashed line indicate the cumulative distribution for dEs with $R_{\rm max}/$\Reff$<1$. To match both cumulative distributions, the uncertainties for dEs $R_{\rm max}/$\Reff$>1$ have to be boosted by $30\%$.}
\label{vel_uncertainties}
\end{figure}

\begin{figure}
\centering
\includegraphics[angle=-90,width=8cm]{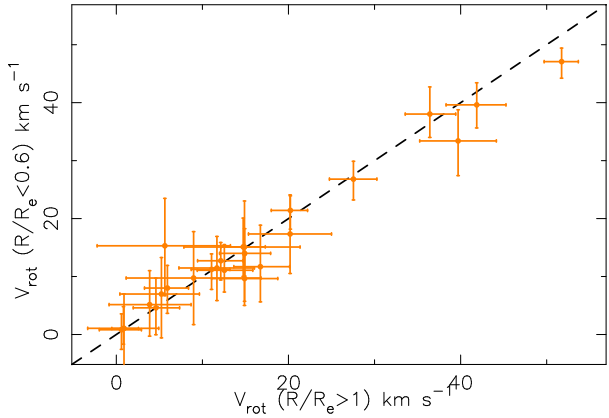}
\caption{Comparison of the \Vrot\ obtained when the {\it Polyex} function is fit to the full radial coverage of the rotation curve and when it is fit to the truncated rotation curve at 0.6\Reff. The dashed line indicates the $y=x$ relation. The radial coverage of the rotation curve does not have to be $\geq$\Reff\ to estimate \Vrot.}
\label{Vrotall}
\end{figure}

The maximum amplitude of the rotation curve is not generally reached in integrated light spectra, which are usually limited to $1-2$~\Reff\ \citep{Beasley09,Geha10}. To compare the rotation speed among different galaxies it has to be measured at a common radius. We define \Vrot\ as the value of the best fit {\it Polyex} function with $R_{PE}=R_s$ and $\alpha=0.02$ at the \Reff. 

We test the robustness of the best fit {\it Polyex} function and its effect on the rotation speed inferred from it (\Vrot). If we fit the {\it Polyex} function with two free parameters ($V_0$, $R_{PE}$), even though $R_{PE}$ is constrained for only 14 galaxies,  the \Vrot\ inferred for all dEs is in good agreement, within the $1\sigma_G$ uncertainty, with the \Vrot\ inferred when only $V_0$ is left as a free parameter ($R_{PE}=R_s$).

Not all of our galaxies reach the \Reff. We compare the \Vrot\ estimated from the best fit {\it Polyex} function leaving $V_0$ free when we use the full extent of the rotation curve and when we truncate the rotation curve and use only the inner regions. We can do this exercise for the 25 dEs that have a radial coverage of at least the \Reff. Figure \ref{RmaxRe} shows the cumulative distribution of the maximum radial coverage of the rotation curves of the SMAKCED dEs. For those galaxies that do not reach the \Reff, the typical radial coverage is $R_{\rm max}/$\Reff$=0.6$. This comparison is only valid if the velocity uncertainties in the inner regions of the truncated rotation curves are consistent with the uncertainties in the same regions for the dEs with $R_{\rm max}/$\Reff$<1$. Figure \ref{vel_uncertainties} shows the comparison of the velocity uncertainties in the region $R/$\Reff$<0.6$ for galaxies with $R_{\rm max}/$\Reff$\lessgtr1$. As expected, the uncertainties in the region $R/$\Reff$<0.6$ are smaller for the galaxies that have a larger radial coverage. We boost their velocity uncertainties by $30\%$ to match them to the distribution obtained for galaxies with $R_{\rm max}/$\Reff$<1$. Figure \ref{Vrotall} shows that the rotation speed \Vrot\ measured using the full and the truncated rotation curve agree well within the uncertainties. 

We repeat this test for an extreme situation in which the radial coverage is $R_{\rm max}/$\Reff$=0.4$. In that case, the uncertainties in the region $R/$\Reff$<0.4$ for the galaxies with $R_{\rm max}/$\Reff$>1$ are boosted by $39\%$ to match the velocity uncertainties measured in galaxies with $R/$\Reff$<0.4$. The uncertainties obtained in \Vrot\ using data in the region $R/$\Reff$<0.4$ are larger than the uncertainties obtained using data in the region $R/$\Reff$<0.6$, but, \Vrot\ is still in good agreement with the \Vrot\ obtained using the full radial coverage ($R/$\Reff$>1$).

\begin{figure}
\centering
\includegraphics[angle=-90,width=8cm]{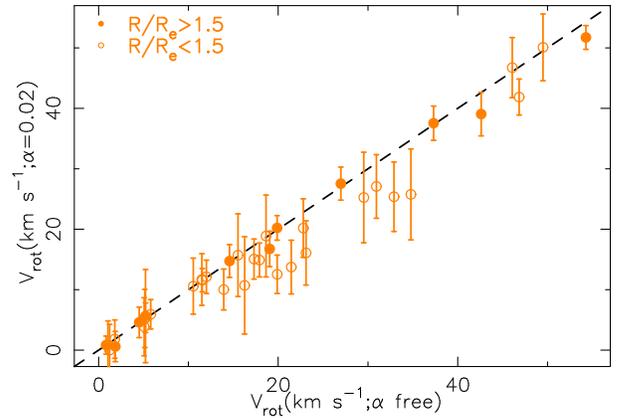}
\caption{Comparison of the \Vrot\ obtained when the {\it Polyex} function is fit using $R_{PE}=R_s$ and $\alpha=0.02$ and when it is fit using $R_{PE}=R_s$ and setting $\alpha$ free. The dashed line indicates the $y=x$ relation. The choice of $\alpha$ does not affect the measured \Vrot.}
\label{Vrotalphafree}
\end{figure}

Finally, we show that the choice of $\alpha$ does not affect the measured \Vrot. Figure \ref{Vrotalphafree} shows the measured \Vrot\ when $R_{PE}=R_s$ and $\alpha$ is fixed to 0.02 versus the measured \Vrot\ when $R_{PE}=R_s$ and alpha is left as a free parameter. Both values follow, within the $1\sigma_G$ uncertainties, the $y=x$ line. The parameter $\alpha$ measures the slope of the rotation curve beyond the turnover radius. For those galaxies that go beyond this radius, the parameter $S_{\rm PF}$, described in Section \ref{anomalies}, quantifies whether $\alpha=0.02$ is a good fit to that part of the curve or it is not.

These tests prove that the best fit {\it Polyex} function is robust even for those cases where the radial coverage of the rotation curve is $R/$\Reff$ \sim 0.5$. The main caveat of this method is the case in which the galaxy has a kinematically decoupled core (KDC) as the ones found in Paper~I. The fraction of dEs that contain KDCs is small ($5.9 \pm 2.4 \%$, Paper~I), and only one dE has been found to contain a KDC with a size larger than $0.4R/$\Reff. The fraction of dEs expected to have a KDC with a size larger than $R/$\Reff$ = 0.4$ in our subsample of 14 dEs with $R_{\rm max}/$\Reff$<1$ is $1\pm1 \%$. In that case, we would not be able to detect the KDC and the \Vrot\ measured would be that of the KDC and not of the main body of the galaxy.

The estimated uncertainty of \Vrot\ is the square root of the quadratic addition of two uncertainty components. The first uncertainty component is the width of the Gaussian distribution that results from measuring 100 times the rotation speed at the \Reff\ in the family of {\it Polyex} functions defined by the uncertainty in $V_0$ and $R_s$. The second uncertainty component is the RMS of all the data points in the rotation curve with respect to the best fit  {\it Polyex} function.
The measured values of \Vrot, along with the $H$ band scale length and the best fit $V_0$, can be found in Table \ref{tabledEs}.

The velocity dispersion $\sigma_e$ is measured by co-adding the spectra within the \Reff, so both rotation and dispersion are included in this parameter \citep[see][for a discussion of this measurement]{etj12}. The average velocity dispersion $\langle \sigma \rangle$, calculated using all the individual measurements in the $\sigma$ profile weighted by their uncertainties, is comparable to $\sigma_e$. This is expected because both measurements are luminosity weighted, which means that the error bars are smaller for the central regions of the kinematic profiles where the luminosity is also higher and $V\sim 0$~\kms.
In panels {\it h} and {\it i} of Figures \ref{rotcurve_VCC0009}-\ref{rotcurve_VCC2083}, $V_{rot}$  and $\sigma_e$ are indicated as yellow lines and the average $\langle \sigma \rangle$ is indicated with a dashed line. The values of $\sigma_e$ and $\langle \sigma \rangle$ can be found in Table \ref{tabledEs}.

To transform the long-slit velocity dispersion measurements into aperture measurements see Section \ref{scaling_rels}.

\section{Line-strength Measurements}\label{indices_sec}

\begin{table}
\begin{center}
\caption{Stellar Populations \label{stellarpopstable}}
\resizebox{9cm}{!} {
\begin{tabular}{|l|c|c|c|c|c|c|}
\hline \hline
Galaxy              &  H${\beta}$  &  H${\gamma A}$  & Fe4668 & Mgb & Age & [M/H] \\
                        &   \AA~          &   \AA~               &   \AA~  & \AA~ & Gyr   &  dex    \\
(1)                    & (2)                &  (3)                     &  (4)       &  (5)    &  (6)    &   (7)  \\ 
\hline
    VCC0009 & 2.59$\pm$0.08 & $-1.15\pm$ 0.15 & 2.91$\pm$0.20 &      ---       &  3.4$^{+ 0.5}_{- 0.5}$ & $-0.5^{+ 0.1}_{- 0.1}$\\
VCC0021$^*$ & 2.92$\pm$0.20 &       ---       & 1.67$\pm$0.50 & 1.24$\pm$0.21  &  5.8$^{+ 1.9}_{- 1.9}$ & $-1.2^{+ 0.4}_{- 0.4}$\\
    VCC0033 & 2.50$\pm$0.12 & $-0.30\pm$ 0.19 & 2.25$\pm$0.37 &      ---       &  4.9$^{+ 2.0}_{- 2.0}$ & $-0.9^{+ 0.2}_{- 0.2}$\\
    VCC0170 & 3.65$\pm$0.07 & $ 2.10\pm$ 0.13 & 1.86$\pm$0.21 &      ---       &  2.0$^{+ 0.4}_{- 0.4}$ & $-1.0^{+ 0.2}_{- 0.2}$\\
    VCC0308 &      ---      & $-1.11\pm$ 0.16 & 3.18$\pm$0.22 &      ---       &  2.8$^{+ 0.8}_{- 0.6}$ & $-0.4^{+ 0.1}_{- 0.2}$\\
    VCC0389 & 2.66$\pm$0.06 & $-2.98\pm$ 0.10 & 3.42$\pm$0.16 &      ---       &  3.0$^{+ 0.3}_{- 0.3}$ & $-0.4^{+ 0.1}_{- 0.1}$\\
    VCC0397 &      ---      & $-1.97\pm$ 0.13 & 4.21$\pm$0.19 &      ---       &  2.6$^{+ 0.3}_{- 0.3}$ & $-0.1^{+ 0.1}_{- 0.1}$\\
    VCC0437 & 2.77$\pm$0.17 & $-1.78\pm$ 0.19 & 1.39$\pm$0.37 &      ---       &  6.8$^{+ 4.5}_{- 2.1}$ & $-1.3^{+ 0.3}_{- 0.3}$\\
    VCC0523 &      ---      & $-2.10\pm$ 0.12 & 3.83$\pm$0.16 &      ---       &  2.7$^{+ 0.7}_{- 0.5}$ & $-0.2^{+ 0.1}_{- 0.1}$\\
    VCC0543 &      ---      & $-2.80\pm$ 0.10 & 2.90$\pm$0.16 &      ---       &  8.4$^{+ 2.9}_{- 2.8}$ & $-0.7^{+ 0.1}_{- 0.1}$\\
    VCC0634 & 2.40$\pm$0.04 & $-2.38\pm$ 0.07 & 3.48$\pm$0.10 &      ---       &  4.0$^{+ 0.4}_{- 0.4}$ & $-0.4^{+ 0.0}_{- 0.0}$\\
    VCC0750 & 2.35$\pm$0.06 & $-2.30\pm$ 0.11 & 3.66$\pm$0.17 &      ---       &  3.7$^{+ 0.6}_{- 0.6}$ & $-0.3^{+ 0.1}_{- 0.1}$\\
    VCC0751 & 2.25$\pm$0.11 & $-3.87\pm$ 0.21 & 3.89$\pm$0.28 &      ---       &  5.2$^{+ 1.7}_{- 1.7}$ & $-0.3^{+ 0.1}_{- 0.1}$\\
    VCC0781 & 3.32$\pm$0.12 & $ 1.95\pm$ 0.17 & 1.37$\pm$0.32 &      ---       &  3.9$^{+ 0.8}_{- 0.8}$ & $-1.3^{+ 0.2}_{- 0.2}$\\
    VCC0794 & 2.16$\pm$0.08 & $-1.93\pm$ 0.12 & 2.58$\pm$0.18 &      ---       &  7.8$^{+ 1.8}_{- 1.8}$ & $-0.8^{+ 0.1}_{- 0.1}$\\
VCC0856$^*$ & 1.61$\pm$0.24 &       ---       & 1.24$\pm$0.77 & 2.71$\pm$0.32  & 14.1                   & $-0.9^{+ 0.5}_{- 0.5}$\\
    VCC0917 &      ---      & $-2.07\pm$ 0.10 & 2.87$\pm$0.19 &      ---       &  5.6$^{+ 2.4}_{- 1.7}$ & $-0.6^{+ 0.1}_{- 0.1}$\\
VCC0940$^*$ & 2.24$\pm$0.13 &       ---       & 2.61$\pm$0.35 & 2.71$\pm$0.14  &  5.8$^{+ 2.2}_{- 2.2}$ & $-0.6^{+ 0.2}_{- 0.2}$\\
VCC0990$^*$ & 2.42$\pm$0.12 &       ---       & 3.52$\pm$0.31 & 2.30$\pm$0.14  &  4.0$^{+ 1.2}_{- 1.2}$ & $-0.4^{+ 0.2}_{- 0.2}$\\
    VCC1010 &      ---      & $-4.38\pm$ 0.07 & 3.98$\pm$0.12 &      ---       &  8.3$^{+ 2.1}_{- 1.6}$ & $-0.4^{+ 0.1}_{- 0.1}$\\
    VCC1087 &      ---      & $-3.44\pm$ 0.15 & 3.80$\pm$0.22 &      ---       &  5.6$^{+ 2.1}_{- 1.8}$ & $-0.4^{+ 0.1}_{- 0.1}$\\
    VCC1122 &      ---      & $-1.63\pm$ 0.10 & 3.33$\pm$0.14 &      ---       &  3.1$^{+ 0.5}_{- 0.4}$ & $-0.4^{+ 0.1}_{- 0.1}$\\
VCC1183$^*$ & 2.20$\pm$0.25 &       ---       & 3.12$\pm$0.57 & 2.52$\pm$0.26  &  6.2$^{+ 4.6}_{- 4.6}$ & $-0.6^{+ 0.3}_{- 0.3}$\\
VCC1261$^*$ & 2.29$\pm$0.14 &       ---       & 3.38$\pm$0.33 & 1.93$\pm$0.15  &  5.5$^{+ 2.3}_{- 2.3}$ & $-0.6^{+ 0.2}_{- 0.2}$\\
    VCC1304 &      ---      & $-0.40\pm$ 0.07 & 2.13$\pm$0.14 &      ---       &  5.2$^{+ 1.7}_{- 1.4}$ & $-0.9^{+ 0.1}_{- 0.2}$\\
    VCC1355 & 2.53$\pm$0.09 & $-1.98\pm$ 0.17 & 2.69$\pm$0.27 &      ---       &  4.5$^{+ 1.5}_{- 1.5}$ & $-0.7^{+ 0.1}_{- 0.1}$\\
    VCC1407 &      ---      & $-3.10\pm$ 0.08 & 1.79$\pm$0.14 &      ---       & 14.1                   & $-1.1^{+ 0.1}_{- 0.1}$\\
VCC1431$^*$ & 1.99$\pm$0.17 &       ---       & 1.99$\pm$0.40 & 3.13$\pm$0.16  &  7.7$^{+ 7.2}_{- 1.9}$ & $-0.3^{+ 0.2}_{- 0.4}$\\
    VCC1453 & 2.16$\pm$0.06 & $-3.58\pm$ 0.09 & 4.43$\pm$0.13 &      ---       &  4.5$^{+ 0.6}_{- 0.6}$ & $-0.2^{+ 0.0}_{- 0.0}$\\
    VCC1528 & 2.28$\pm$0.06 & $-3.92\pm$ 0.12 & 4.36$\pm$0.14 &      ---       &  4.6$^{+ 0.8}_{- 0.8}$ & $-0.2^{+ 0.0}_{- 0.0}$\\
VCC1549$^*$ & 1.88$\pm$0.22 &       ---       & 3.95$\pm$0.73 & 3.23$\pm$0.25  & 10.4$^{+ 7.1}_{- 7.1}$ & $-0.4^{+ 0.3}_{- 0.3}$\\
VCC1684$^*$ & 3.82$\pm$0.05 &       ---       & 1.13$\pm$0.13 & 1.11$\pm$0.05  &  2.2$^{+ 0.1}_{- 0.1}$ & $-1.2^{+ 0.1}_{- 0.1}$\\
    VCC1695 &      ---      & $-0.94\pm$ 0.14 & 3.01$\pm$0.21 &      ---       &  2.9$^{+ 0.8}_{- 0.6}$ & $-0.5^{+ 0.1}_{- 0.1}$\\
    VCC1861 &      ---      & $-3.82\pm$ 0.14 & 4.04$\pm$0.17 &      ---       &  5.9$^{+ 1.6}_{- 1.4}$ & $-0.3^{+ 0.1}_{- 0.1}$\\
    VCC1895 &      ---      & $-1.62\pm$ 0.09 & 2.19$\pm$0.15 &      ---       &  8.9$^{+ 2.5}_{- 2.2}$ & $-0.9^{+ 0.1}_{- 0.1}$\\
VCC1910$^*$ & 1.86$\pm$0.22 &       ---       & 5.73$\pm$0.53 & 2.45$\pm$0.18  &  9.0$^{+ 5.9}_{- 5.1}$ & $-0.0^{+ 0.2}_{- 0.2}$\\
VCC1912$^*$ & 2.80$\pm$0.16 &       ---       & 1.60$\pm$0.45 & 0.54$\pm$0.19  &  5.9$^{+ 4.1}_{- 2.0}$ & $-1.2^{+ 0.3}_{- 0.3}$\\
VCC1947$^*$ & 1.73$\pm$0.17 &       ---       & 3.51$\pm$0.35 & 3.13$\pm$0.19  & 14.1                   & $-0.6^{+ 0.2}_{- 0.2}$\\
VCC2083$^*$ & 1.68$\pm$0.19 &       ---       & 1.20$\pm$0.51 & 1.17$\pm$0.18  & 14.1                   & $-0.7^{+ 0.5}_{- 0.3}$\\
\hline
\end{tabular}}
\end{center}
\tablecomments{Column 1: Galaxy name. The asterisk indicates which galaxies do not have the H$_{\alpha}$ line covered in our observations. Columns 2$-$5: Lick spectral indices measured within the \Reff~ at LIS-5~\AA~ resolution. Columns 6 and 7: Ages and metallicities measured within the \Reff~ using the index-index diagrams of Figure \ref{indices} and the SSP models of \citet{Vazdekis10} with a Kroupa IMF \citep{KroupaIMF}.}
\end{table}

For consistency, the integrated line-strength indices are measured within the \Reff. We follow the same strategy as described in Paper~I.
The luminosity-weighted ages and metallicities are estimated using age-sensitive (H${\beta}$ and H${\gamma A}$) and metallicity-sensitive (Fe4668 and Mgb) Lick spectral indices \citep{Wor94} measured in the LIS-5~\AA\ system \citep{Vazdekis10}.

The uncertainties in the line-strength indices are estimated by running 100 Monte Carlo simulations. In each simulation, the flux of the science spectrum is randomly perturbed within a Gaussian function whose width is the difference between the science spectrum and the best fit composite stellar template used to obtain the kinematics. In addition, the perturbed science spectrum is convolved with a Gaussian function whose width is randomly chosen within the $1\sigma_G$ uncertainty of the velocity dispersion of the galaxy, and is shifted in wavelength by a randomly chosen value within the $1\sigma_G$ uncertainty of the radial velocity of the galaxy. The uncertainty in the line-strength indices is the standard deviation of the Gaussian distribution obtained from the measurements done in the 100 Monte Carlo simulations.

The spectral range covered for some galaxies allows the simultaneous measurement of H${\beta}$ and H${\gamma A}$, and for some others the simultaneous measurement of Fe4668 and Mgb (see Section \ref{obs}). In Table \ref{stellarpopstable} we provide the line-strength index measurements for each galaxy.

Four of the SMAKCED dEs have emission lines partially filling in the Balmer absorption lines. VCC~170, VCC~781, and VCC~1304 show emission in H$_{\alpha}$, H$_{\beta}$, and H$_{\gamma}$, and also show [NII] and [SII]. VCC~170 and VCC~1304 also show [OIII]. The spectral range of VCC~1684 does not cover H$_{\gamma}$ or H$_{\alpha}$, but some emission is seen in H$_{\beta}$ and [OIII].

In all four cases the emission lines are significantly narrower than the absorption lines, so both components can be decoupled. We use the software GANDALF \citep[Gas AND Absorption Line Fitting,][]{Sarzi06} to separate the absorption from the emission lines. This software simultaneously fits the stellar continuum and emission lines assuming that the emission lines are described by Gaussian functions. The stellar continuum is fitted by the same best combination of stellar templates used in the software pPXF to extract the stellar kinematics\footnote{The emission lines are masked to extract the stellar kinematics (see Section \ref{kin_sec})}. The Lick indices for VCC~170, VCC~781, VCC~1304, and VCC~1684 are measured in the spectrum that results from subtracting the emission line spectrum obtained with the GANDALF software.

Those galaxies observed at the INT and VLT telescopes do not cover the H$_{\alpha}$ region to check for emission. Three out of the 4 dEs for which we detect some emission are identified in SDSS optical images as dEs with a blue center \citep[dE(bc);][]{Lisk06b}. However, not all of the dEs classified as dE(bc) in our sample show emission lines. This can be an indication that the emission lines have the same width as the absorption lines or even broader.

In addition, if the emission is as broad as or broader than the absorption lines, the Balmer lines would appear shallower and the ages inferred would be older \citep[e.g.][]{Harker06}. However, we do not find any significant emission in any of the galaxies, but for the above mentioned VCC~170, VCC~781, VCC~1304, and VCC~1684. But, that does not rule out the possibility of them having some emission. Thus, the H${\beta}$ and H${\gamma A}$ spectral indices in Table \ref{stellarpopstable} should be taken as lower limits for galaxies observed at the INT and VLT telescopes, and their inferred ages should be taken as upper limits.

\section{Photometric Measurements}\label{phot_sec}

Dwarf early-type galaxies have a complex structure \citep[e.g.][]{Jerjen00,Barazza02,Geha03,Graham03,DR03,Lisk06b,Lisk06a,Lisk07,Ferrarese06,Janz12,Janz14}. In this Section, we measure the shapes and twists of the isophotes of the SMAKCED dEs in the $H$ band, and visually compare the features found with the kinematic profiles.

Panels {\it a} to {\it f} in Figures \ref{rotcurve_VCC0009}-\ref{rotcurve_VCC2083} show the $H$ band photometry of the SMAKCED dEs. The images used for this analysis are presented in \citet{Janz14}, with the exception of VCC~397 which is presented in \citet{etj12}.
Panels {\it a} and {\it b} show the $H$ band images in low and high contrast grey scales, respectively. The blue lines indicate the footprint of the long-slit used in the spectroscopic observations. 
Panel {\it c} shows high-pass filtered images created by subtracting a Gaussian-smoothed image with a $4''$ kernel from the original $H$ band image.
Panel {\it d} shows the departures from a single S\'ersic fit to the surface brightness profile with a nucleus when needed. Only 8 out of the 39 SMAKCED dEs are best fitted with a single S\'ersic profile; the remaining 31 needed one or more additional components \citep[see][]{Janz12,Janz14}.
Panels {\it e} and {\it f} are the result of fitting elliptical isophotes to the images using the IRAF task {\sc ellipse}. 

The isophotes are fitted with ellipses whose major axis are logarithmically increased to make every isophote $10\%$ larger than the previous one. The center, the position angle (P.A.), and the ellipticity ($\epsilon$) of the isophotes are left as free parameters. Panel {\it e} shows the residual image obtained by subtracting a smooth two-dimensional model based on the ellipse fitting (excluding higher order components) from the original image. A flat residual image indicates that the full structure of the galaxy is reproduced by elliptical isophotes whose parameters are shown in panel {\it f}. From top to bottom, panel {\it f} shows the surface brightness ($\mu$), the position angle P.A., the ellipticity ($\epsilon$), the $C_4$ parameter ($C_4 < 0$ indicates boxy isophotes and $C_4 > 0$ disky isophotes), and the drift of the center of the isophotes along the long-slit used in the spectroscopic observations. 

Figures \ref{rotcurve_VCC0009}-\ref{rotcurve_VCC2083} show a large diversity of structural and kinematic features, from concentric elliptical isophotes all oriented along the same position angle to twists in the position angle of the isophotes, from flat ellipticity profiles to ellipticity gradients, from non-rotating dEs to kinematically decoupled cores, and fast rotating dEs. 

\section{Shapes of the Rotation Curves: Anomalies and Asymmetries}\label{anomalies}

Galaxies affected by strong tidal interactions usually have distorted rotation curves, as is seen in the dE satellites of M31 \citep[][]{Geha06,Geha10}. Lopsidedness, i.e. the approaching and receding sides of the rotation curves have different shapes and speeds, is a common feature found in emission-line rotation curves of late-type star forming galaxies, especially in low mass star forming galaxies \citep[e.g.,][]{Swaters99,Swaters09}. Recently, it has been shown that the gas and stellar rotation of low mass star-forming galaxies closely follow each other \citep{Adams14}, thus the lopsidedness is also expected in the stellar rotation curves of low mass star-forming galaxies. In this Section, we analyze the shapes and amplitudes of the stellar rotation curves for the SMAKCED dEs and quantify the significance of any anomaly (poorly fit and/or asymmetry) found.
 
We define a {\it poorly fit (PF)} as a statistically significant difference between the measured velocities and the best fit  {\it Polyex} function. To quantify the significance of this velocity difference ($\Delta V = V - V_{polyex}$) we use the statistical parameter $S_{\rm PF}$, defined in Paper~I:

\begin{equation}\label{AnEq}
S_{\rm PF}(R) = \frac{\vert \langle \Delta V_{\rm inner}(<R) \rangle - \langle \Delta V_{\rm outer}(>R) \rangle \vert}{\sqrt{\delta \langle \Delta V_{\rm inner}(<R) \rangle^2+ \delta \langle \Delta V_{\rm outer}(<R) \rangle^2}}
\end{equation}

For each radius $R$, $\langle \Delta V_{\rm inner}(<R) \rangle$ is the mean difference between the mean $V$, weighted by the uncertainties in $V$, and the best fit {\it Polyex} function interior to that radius and $\langle \Delta V_{\rm outer}(>R) \rangle$ is the difference between the mean $V$, weighted by the uncertainties in $V$, and the best fit {\it Polyex} function exterior to that radius. The parameters $\delta \langle \Delta V_{\rm inner}(<R) \rangle$ and $\delta \langle \Delta V_{\rm outer}(>R) \rangle$ are the respective uncertainties.
We define the significance of the detection of a poorly fit rotation curve $\langle S_{{\rm PF,max}}\rangle$ as the average of the three maximum values of $S_{\rm PF}(R)$. Note that, in contrast to Paper~I, we do not require $\langle \Delta V_{\rm inner}(<R) \rangle$ and $\langle \Delta V_{\rm outer}(>R) \rangle$ to have opposite signs. That is a specific feature of kinematically decoupled cores which is the focus of that paper. The parameter $\langle R_{S_{\rm PF,max}}\rangle$  is the average radius of the three maximum values of $S_{\rm PF}(R)$ and indicates the radius where the PF anomaly is maximum. A PF anomaly is considered statistically significant when $\langle S_{{\rm PF,max}}\rangle \geq 3$, marginal when $2 \leq \langle S_{{\rm PF, max}}\rangle<3$, and not significant in the rest of cases. 

We define two kinds of {\it asymmetries: the amplitude asymmetry (AA)} and {\it the amplitude and shape asymmetry (AS)}. {\it AA} is sensitive to statistically significant differences between the amplitude of the approaching and receding sides of the rotation curve. {\it AS} is sensitive to statistically significant differences between both the amplitude and shape of the approaching and receding sides of the rotation curve, i.e. whether the two sides cross over each other.

The {\it AA} asymmetry is quantified by $S_{\rm AA}$:
 
\begin{equation}\label{AAEq}
S_{\rm AA} = \frac{|\langle \Delta V_{\rm app} \rangle - \langle \Delta V_{\rm rec} \rangle|}{\sqrt{\delta \langle V_{\rm app} \rangle^2+ \delta \langle V_{\rm rec} \rangle^2}}
\end{equation}

\noindent where $\langle \Delta V_{\rm app} \rangle$ and $\langle \Delta V_{\rm rec} \rangle$ are the mean, weighted by their uncertainty, of the distance in velocity of each data point to the best fit {\it Polyex} function for the approaching and receding sides of the rotation curve, respectively. The parameters $\delta \langle V_{\rm app} \rangle$ and $\delta \langle V_{\rm rec} \rangle$ are the uncertainties of the means assuming that the uncertainty in the {\it Polyex} function is negligible.

The {\it AS} asymmetry is quantified by $S_{\rm AS}$:

\begin{equation}\label{ASEq}
S_{\rm AS} = \frac{\langle |V_{\rm app}-V_{\rm rec,int}| \rangle+\langle |V_{\rm rec}-V_{\rm app,int}| \rangle}{2 \times \sqrt{\delta \langle V_{\rm app} \rangle^2+ \delta \langle V_{\rm rec} \rangle^2}}
\end{equation}

\noindent where $V_{\rm app}-V_{\rm rec,int}$ is the velocity difference between each data point in the approaching side of the rotation curve and the interpolated value at the same radius in the receding side of the rotation curve. The interpolation is done between the two nearest points in radius. Similarly, $V_{\rm rec}-V_{\rm app,int}$ is the velocity difference between each data point in the receding side of the rotation curve and the interpolated value at the same radius in the approaching side.

Figures \ref{Sstat_plots_VCC0009}-\ref{Sstat_plots_VCC2083} show the rotation curve, $\Delta V_{\rm app}$ and $\Delta V_{\rm rec}$, and the statistical parameter $S_{\rm PF}(R)$ for the SMAKCED dEs. 

The fraction of dEs with a significant PF anomaly in the rotation curve is $23\pm 7~\%$ ($9/39$). The fraction of dEs with a significant AA anomaly is $33 \pm 8~\%$ ($13/39$), and with a significant AS anomaly is $59 \pm 8~\%$ ($23/39$).
The anomalies do not seem to be related to the presence of disky subtle substructures as seen in high-pass filtered optical images or to the number of components that best fit the $H$ band surface brightness profile (see Table \ref{Sstat_fractions_table}).

In Sections \ref{kinprofs_section} and \ref{lit_comp} we make three independent tests to ensure the robustness of the velocity measurements and uncertainties. We make an internal check: we compare the measurements done using the blue and red instrumental setups. We make two external checks: we compare our measurements to those of G02/G03 and C09. We have a total of 11 dEs in common with the literature, some of them have significant anomalies in their rotation curves. The good agreement found with G02/G03 and C09 reassures that these anomalies are real (see Figures \ref{Rotation_comparison} and \ref{Dispersion_comparison}). The smaller AA asymmetry found in VCC~1947 by G02/G03 with respect to our measurement is likely due to the different position angle used to place the slit in the spectroscopy (while we used P.A.~$= -54^{\circ}$, G02/G03 used P.A.~$= -65^{\circ}$).

The calculation of the systemic velocity of the galaxy affects the kinematic anomalies. If instead of using the velocity of the center of galaxy (i.e. the brightest pixel which corresponds to the central bin of the rotation curve) we use the average of all the data points in the rotation curve, the significance of the anomalies of some dEs decreases but the anomaly does not disappear. However, in those cases the center of the rotation curve has a velocity different from zero, which is an anomaly on its own.

\begin{figure*}
\centering
\resizebox{0.87\textwidth}{!}{\includegraphics[bb= 20 226 563 609,angle=0]{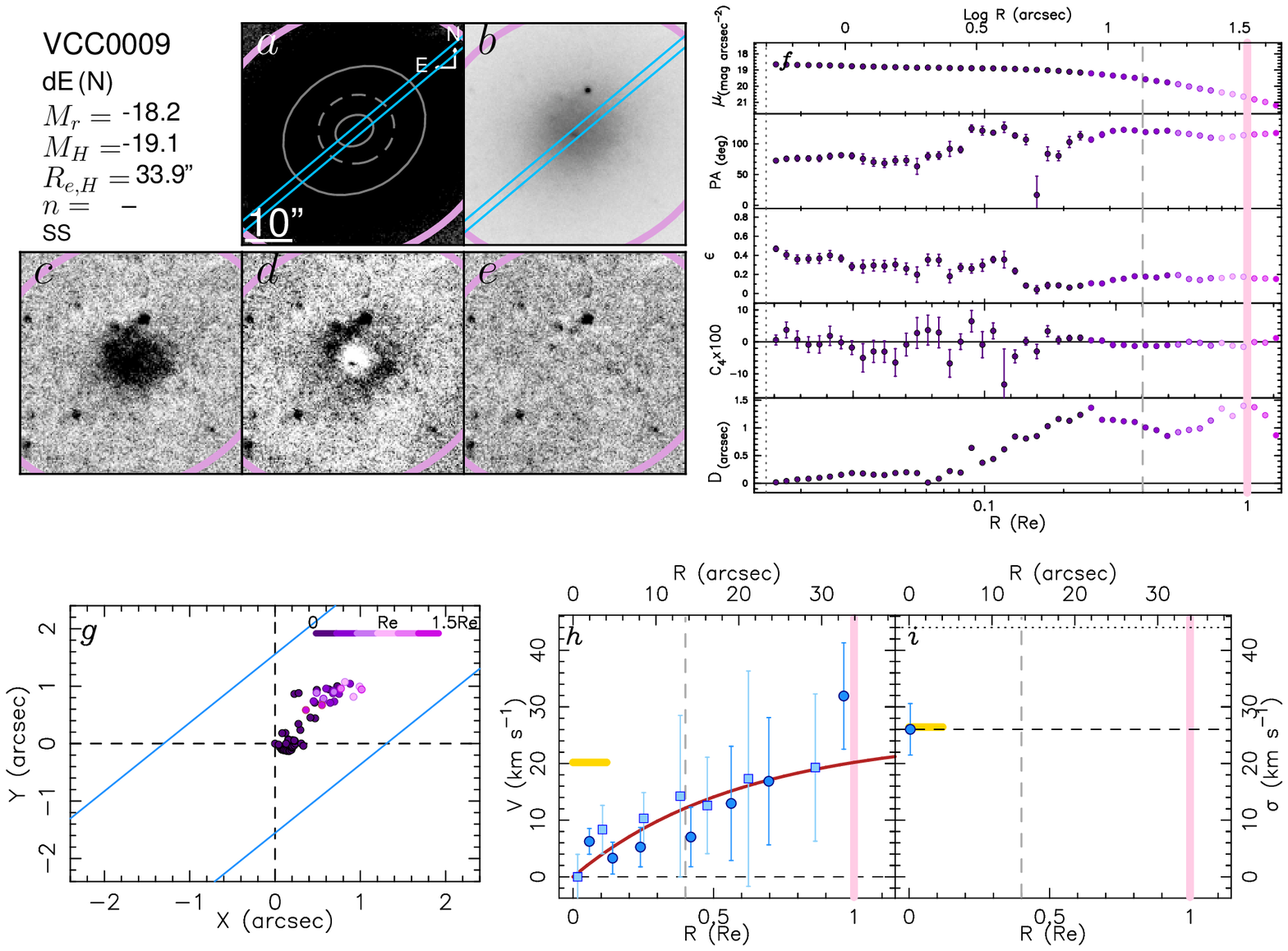}}
\caption{$H$ band images and kinematic profiles for VCC~9. Below the galaxy name there is some relevant information about that galaxy that is also provided in Tables \ref{tabledEs} and \ref{Sstat_table}. The first line indicates the galaxy class based on the analysis by \citet{Lisk06a,Lisk06b,Lisk07}. The second line is the absolute magnitude in the $r$ band by \citet[][in the AB system]{Janz08}. The third, fourth, and fifth lines are the absolute magnitude (in the AB system), half-light radius, and S\'ersic index in the $H$ band by \citet{Janz14}. The sixth line indicates whether the rotation curve is smooth and symmetric (SS), poorly fit (PF), and/or has amplitude or shape asymmetries (AA, AS) based on the analysis of Section \ref{anomalies}. Marginal anomalies are indicated within brackets. {\bf Panels a and b:} zoom in the central region in low and high contrast grey scales, respectively. The grey scales show bright regions in black. The pink ellipse shows the fitted isophote at the \Reff. The grey ellipses show the fitted isophotes at the \Reff/2, \Reff/4 (dashed line), and \Reff/8. The blue lines indicate the footprint of the long-slit used in the spectroscopic observations. The position angle, measured North-East, of the long-slit footprint corresponds with the receding side of the rotation curve (i.e. positive radial distances with respect to the center of the galaxy). {\bf Panel c:} high-filtered $H$ band image. {\bf Panel d:} departures from a single S\'ersic fit to the $H$ band surface brightness profile by \citet{Janz14}. {\bf Panel e:} residuals after subtracting a smooth model based on the ellipse-fitting. {\bf Panel f:} from top to bottom, surface brightness, position angle, ellipticity, $C_4$ parameter, and drift of the center of each isophote along the slit as a function of distance to the center of the galaxy. The pink and grey lines are as in panels {\it a}$-${\it e}. The dotted vertical line indicates a radius of 2 pixels or half the typical seeing of the observations. The color code is as indicated in panel {\it g}. {\bf Panel g:} spatial distribution of the centers of the best fit elliptical isophotes shown in panel {\it f}. The blue lines indicate the footprint of the long-slit used to get the kinematics. The colors indicate the distance from the center of the galaxy. Pink indicates the \Reff. {\bf Panel h:} stellar rotation curve. The light blue squares and dark blue dots indicate the approaching and receding sides, respectively. The red solid line is the best fit {\it Polyex} function leaving $V_0$ as the only free parameter. The yellow line indicates the rotation speed at the \Reff~ evaluated in the best fit {\it Polyex} function. The pink and grey lines are as in panels {\it a}$-${\it e}. {\bf Panel i:} velocity dispersion profile. Symbols are as in panel {\it h}. The dashed black line indicates the weighted average of the profile $\langle \sigma \rangle$. The yellow line indicates $\sigma_e$. The black dotted line indicates the instrumental resolution. The vertical pink and grey lines are as in panels {\it a}-{\it e}.}
             \label{rotcurve_VCC0009}
\end{figure*}

\begin{figure*}
\centering
\resizebox{0.87\textwidth}{!}{\includegraphics[bb= 20 226 563 609,angle=0]{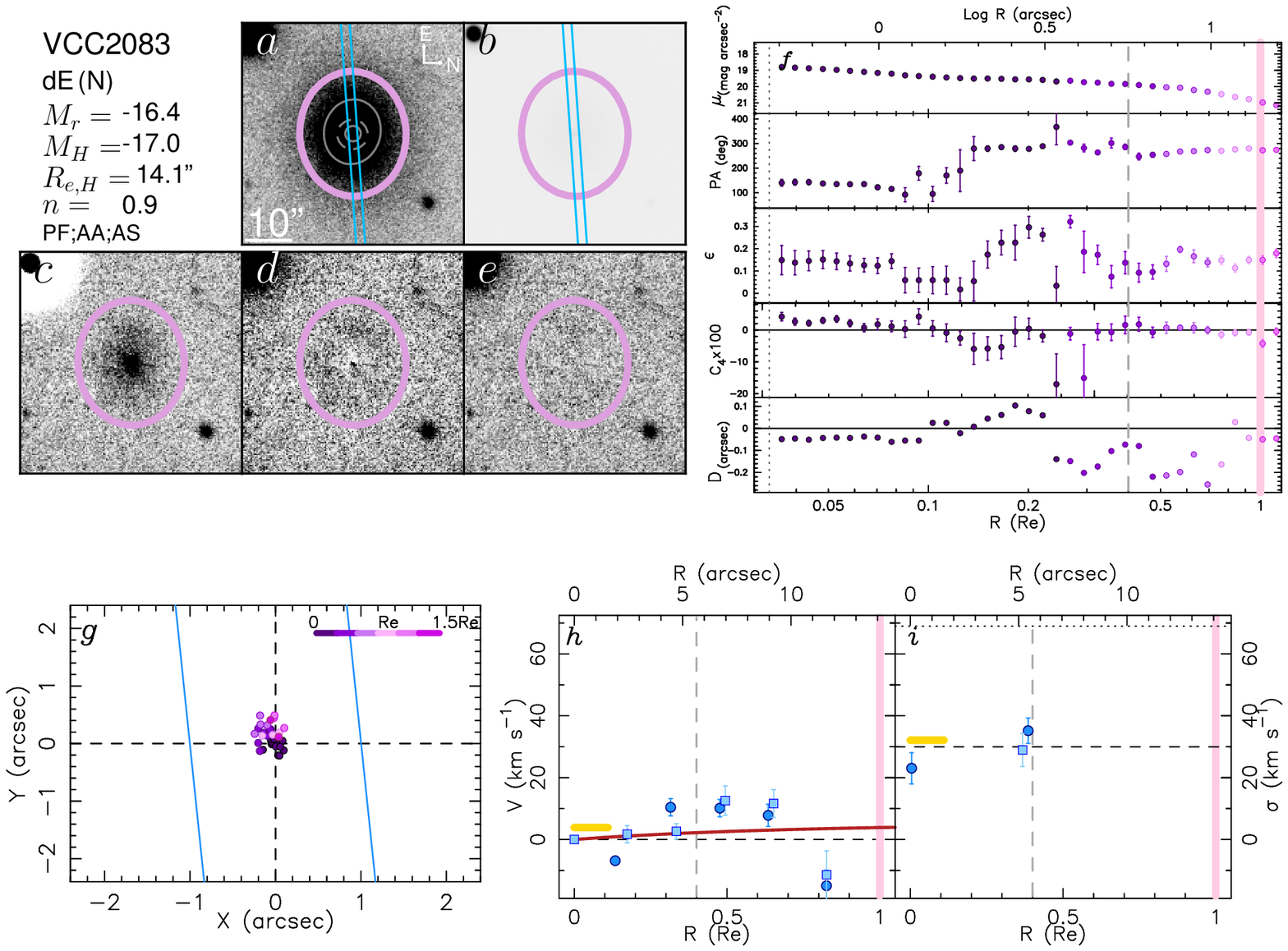}}
\caption{Same as Figure \ref{rotcurve_VCC0009} for VCC~2083.}
            \label{rotcurve_VCC2083}
\end{figure*}

The disky/no disky classification and the number of components that best fit the $H$ band surface brightness profile do not look for anomalies in the stellar light distribution. In the case of the disky structures, symmetric and asymmetric light distributions are mixed and, in the case of the multi-component analysis, the surface brightness profiles are assumed to be symmetric.
If the light distribution is asymmetric, the centers of the elliptical isophotes that best fit the light distribution will drift towards the brightest regions. We show this drift in panel {\it g} of Figures \ref{rotcurve_VCC0009}-\ref{rotcurve_VCC2083}. This drift will be important for the kinematics only if it is along the long-slit used in the spectroscopy. The last row of panel {\it f} in Figures \ref{rotcurve_VCC0009}-\ref{rotcurve_VCC2083} shows the drift of the centers of the isophotes along the long-slit used in the spectroscopy. The asymmetric light distribution would also appear in panel {\it d} of \ref{rotcurve_VCC0009}-\ref{rotcurve_VCC2083} as bright versus dark regions in opposite sides of the galaxy. 

Very few of the dEs analyzed here have a perfectly smooth and regular light distribution in the $H$ band. The majority of them show large drifts of the centers of the isophotes and P.A. and ellipticity gradients, which is in agreement with the large variety of kinematic anomalies found. Matching the photometric and kinematic irregularities is a difficult task. Even in the case of KDCs where the features are very prominent we do not find clear evidence of a one-to-one correspondence between the photometric and kinematic features (Paper~I). In addition, the optical spectroscopy and the infrared photometry are not probing the same stellar populations if young stars are present.

A more detailed study of the anomalies found in the rotation curves will be discussed in the paper Toloba et al. (in prep.)

\begin{table}
\begin{center}
\caption{Kinematic Anomalies\label{Sstat_table}}
\resizebox{9cm}{!} {
\begin{tabular}{c|c|c|c|c|c|c|c}
\hline
Galaxy      &     $\langle S_{{\rm PF,max}} \rangle$ & $\langle R_{S_{\rm PF,max}} \rangle$ & PF & $S_{\rm AA}$ & AA &  $S_{\rm AS}$ & AS \\
                &                    & (arcsec)  &     &    &   &   &  \\
  (1)          &    (2)            &  (3)          & (4) &  (5)  &  (6)  &  (7) & (8) \\
\hline \hline
VCC0009 &   1.6 $\pm$     0.4 &    12.6 $\pm$    15.9 &           No  &     0.4  &           No  &    1.8  &          No \\
VCC0021 &   2.0 $\pm$     0.6 &     2.4 $\pm$     0.6 &     Marginal  &     2.5  &     Marginal  &    3.2  & Significant \\
VCC0033 &   1.1 $\pm$     0.6 &     2.5 $\pm$     1.8 &           No  &     1.4  &           No  &    2.1  &    Marginal \\
VCC0170 &   1.2 $\pm$     0.1 &     7.6 $\pm$     1.2 &           No  &     0.2  &           No  &    1.2  &          No \\
VCC0308 &   1.3 $\pm$     0.5 &    10.9 $\pm$     9.6 &           No  &     6.1  &  Significant  &    6.7  & Significant \\
VCC0389 &   0.9 $\pm$     0.2 &     3.8 $\pm$     4.1 &           No  &     2.3  &     Marginal  &    2.5  &    Marginal \\
VCC0397 &   1.5 $\pm$     0.3 &     3.6 $\pm$     3.7 &           No  &     1.0  &           No  &    2.2  &    Marginal \\
VCC0437 &   2.4 $\pm$     0.4 &    11.6 $\pm$     4.3 &     Marginal  &     5.2  &  Significant  &    5.1  & Significant \\
VCC0523 &   3.4 $\pm$     0.3 &     4.9 $\pm$     1.5 &  Significant  &     2.6  &     Marginal  &    3.9  & Significant \\
VCC0543 &   2.2 $\pm$     0.1 &     6.9 $\pm$     1.1 &     Marginal  &     2.5  &     Marginal  &    2.5  &    Marginal \\
VCC0634 &   2.8 $\pm$     0.2 &    11.7 $\pm$     2.0 &     Marginal  &     1.4  &           No  &    3.5  & Significant \\
VCC0750 &   1.9 $\pm$     0.1 &     7.8 $\pm$     3.5 &           No  &     1.8  &           No  &    3.7  & Significant \\
VCC0751 &   1.8 $\pm$     0.3 &     3.2 $\pm$     0.8 &           No  &     3.8  &  Significant  &    4.6  & Significant \\
VCC0781 &   1.7 $\pm$     0.3 &     5.0 $\pm$     3.4 &           No  &     1.6  &           No  &    1.1  &          No \\
VCC0794 &   1.3 $\pm$     0.2 &     5.8 $\pm$     3.3 &           No  &     4.0  &  Significant  &    4.7  & Significant \\
VCC0856 &   1.1 $\pm$     0.3 &     2.3 $\pm$     1.4 &           No  &     1.6  &           No  &    3.2  & Significant \\
VCC0917 &   2.4 $\pm$     0.1 &     4.4 $\pm$     2.0 &     Marginal  &     1.7  &           No  &    2.7  &    Marginal \\
VCC0940 &   2.4 $\pm$     0.7 &     9.3 $\pm$     8.4 &     Marginal  &     2.8  &     Marginal  &    2.6  &    Marginal \\
VCC0990 &   2.2 $\pm$     0.5 &     2.6 $\pm$     2.0 &     Marginal  &     7.9  &  Significant  &    8.9  & Significant \\
VCC1010 &   6.5 $\pm$     0.1 &    12.7 $\pm$     0.7 &  Significant  &     0.2  &           No  &    3.6  & Significant \\
VCC1087 &   1.0 $\pm$     0.2 &    16.1 $\pm$    11.7 &           No  &     2.5  &     Marginal  &    3.3  & Significant \\
VCC1122 &   2.2 $\pm$     0.4 &     3.0 $\pm$     0.6 &     Marginal  &     2.8  &     Marginal  &    3.2  & Significant \\
VCC1183 &   2.8 $\pm$     0.7 &     2.2 $\pm$     0.4 &     Marginal  &     0.3  &           No  &    1.4  &          No \\
VCC1261 &   2.8 $\pm$     0.6 &     5.8 $\pm$     1.0 &     Marginal  &     2.5  &     Marginal  &    2.5  &    Marginal \\
VCC1304 &   3.4 $\pm$     0.2 &    19.9 $\pm$     7.7 &  Significant  &     2.3  &     Marginal  &    4.3  & Significant \\
VCC1355 &   1.6 $\pm$     0.3 &    13.6 $\pm$     6.2 &           No  &     1.7  &           No  &    1.0  &          No \\
VCC1407 &   1.5 $\pm$     0.2 &     9.6 $\pm$     3.7 &           No  &     0.8  &           No  &    1.8  &          No \\
VCC1431 &   2.5 $\pm$     0.4 &     4.4 $\pm$     0.9 &     Marginal  &     4.4  &  Significant  &    4.8  & Significant \\
VCC1453 &   2.2 $\pm$     0.1 &     9.6 $\pm$     6.3 &     Marginal  &     2.0  &     Marginal  &    2.5  &    Marginal \\
VCC1528 &   0.6 $\pm$     0.3 &    10.4 $\pm$     2.0 &           No  &     0.5  &           No  &    1.4  &          No \\
VCC1549 &   2.6 $\pm$     2.2 &     2.3 $\pm$     1.5 &     Marginal  &     8.4  &  Significant  &    9.4  & Significant \\
VCC1684 &   9.2 $\pm$     1.9 &    11.4 $\pm$     1.5 &  Significant  &     1.1  &           No  &    9.3  & Significant \\
VCC1695 &   3.7 $\pm$     1.0 &     7.7 $\pm$     4.5 &  Significant  &     5.3  &  Significant  &    5.8  & Significant \\
VCC1861 &   2.2 $\pm$     0.1 &     9.1 $\pm$     2.2 &     Marginal  &     5.7  &  Significant  &    6.6  & Significant \\
VCC1895 &   1.7 $\pm$     0.4 &    15.2 $\pm$     2.9 &           No  &     0.5  &           No  &    1.9  &          No \\
VCC1910 &   3.4 $\pm$     2.5 &     5.0 $\pm$     1.8 &  Significant  &     6.4  &  Significant  &    7.6  & Significant \\
VCC1912 &   5.2 $\pm$     3.4 &     7.8 $\pm$     6.1 &  Significant  &    14.7  &  Significant  &   19.5  & Significant \\
VCC1947 &   5.1 $\pm$     0.9 &     2.6 $\pm$     1.3 &  Significant  &    10.6  &  Significant  &   13.7  & Significant \\
VCC2083 &   5.7 $\pm$     1.4 &     3.4 $\pm$     1.2 &  Significant  &     3.2  &  Significant  &    8.1  & Significant \\
\hline
\end{tabular}}
\end{center}
\tablecomments{Column (1): galaxy name; Column (2): significance of the departure from the best fit {\it Polyex} function (poorly fit), i.e. average of the three maximum values of $S(V-V_{polyex})$; Column (3): average radius of the three maximum values of $S(V-V_{polyex})$. This parameter indicates the radius where the departure of the data points from the best fit {\it Polyex} fitting function is maximum. Column (4): is the poorly fit significant? It is significant if $\langle S_{{\rm PF,max}}(V-V_{polyex}) \rangle \geq 3$, marginal if $2 leq \langle S_{{\rm PF,max}}(V-V_{polyex}) \rangle < 3$, and it is not significant in the rest of the cases. Note that the marginal cases are a mixture of rotation curves with large uncertainties and rotation curves without enough data points to make a conclusive classification.; Column (5): significance of the amplitude asymmetry (AA). Column (6): is the AA asymmetry significant? It is significant if $\langle S_{\rm AA} \rangle \geq 3$, marginal if $2 \leq \langle S_{\rm AA} \rangle < 3$, and it is not significant in the rest of the cases. Columns (7) and (8): same as Columns (5) and (6) for the shape asymmetry (AS).}
\end{table}

\begin{table}
\begin{center}
\caption{Fraction of SMAKCED dEs with Anomalies and/or Asymmetries in the Rotation Curves\label{Sstat_fractions_table}}
{\renewcommand{\arraystretch}{1.}
\resizebox{9cm}{!} {
\begin{tabular}{c|c|c|c|c}
\hline
     & PF  &  AA &  AS &  Anomalous Rotation Curve \\
    & $\%$ & $\%$ & $\%$ & $\%$ \\
 (1)&  (2)    &  (3)    &  (4)   & (5)  \\
\hline
    SMAKCED dEs &  23 $\pm$   7 (9/39) &   33 $\pm$   8 (13/39) &   59 $\pm$   8 (23/39) &   62 $\pm$   8 (24/39) \\
       Disky    &  40 $\pm$  11 (8/20) &   35 $\pm$  11 (7/20) &  65 $\pm$  11 (13/20) &  65 $\pm$  11 (13/20) \\
       No disky &   5 $\pm$   5 (1/19) &   32 $\pm$  11 (6/19) &   53 $\pm$  11 (10/19) &  53 $\pm$  11 (10/19) \\
Single-component & 13 $\pm$  12 (1/8) &   38 $\pm$  17 (3/8) &   63 $\pm$  17 (5/8) &  63 $\pm$  17 (5/8) \\
Multi-component &  20 $\pm$   8 (5/25) &   32 $\pm$   9 (8/25) &   52 $\pm$  10 (13/25) &  52 $\pm$  10 (13/25) \\
 Not classified &  50 $\pm$  20 (3/6) &   33 $\pm$  19 (2/6) &   83 $\pm$  15 (5/6) & 83 $\pm$  15 (5/6) \\
\hline
\end{tabular}
}}
\end{center}
\tablecomments{Column (1): galaxy population studied. SMAKCED dEs refers to the full sample. Disky refers to those dEs with disky structures visible in high-pass filtered optical images. No disky refers to those dEs without visible structures in high-pass filtered optical images. The disky and no disky classification is based on the analysis of \citet{Lisk06a,Lisk07}. One component indicates that the $H$ band surface brightness profile is best fit by a single S\'ersic function. Multi-component indicates that the $H$ band surface brightness profile is best fit by more than one S\'ersic functions. Not-classified indicates galaxies not included in the decomposition analysis. The nucleus, if present, is not included in the number of components. This classification is based on the analysis by \citet{Janz14}. Column (2) fraction of galaxies with significant poorly fit rotation curves. The number within brackets indicates how many galaxies satisfy that condition. Column (3): same as Column (2) for amplitude asymmetries AA. Column (4): same as Column (2) for shape asymmetries AS. Column (5): same as Column (2) for any kind of anomaly found in the rotation curve. }
\end{table}

\section{Derived Ages and Metallicities}\label{stellarpops}

\begin{figure}
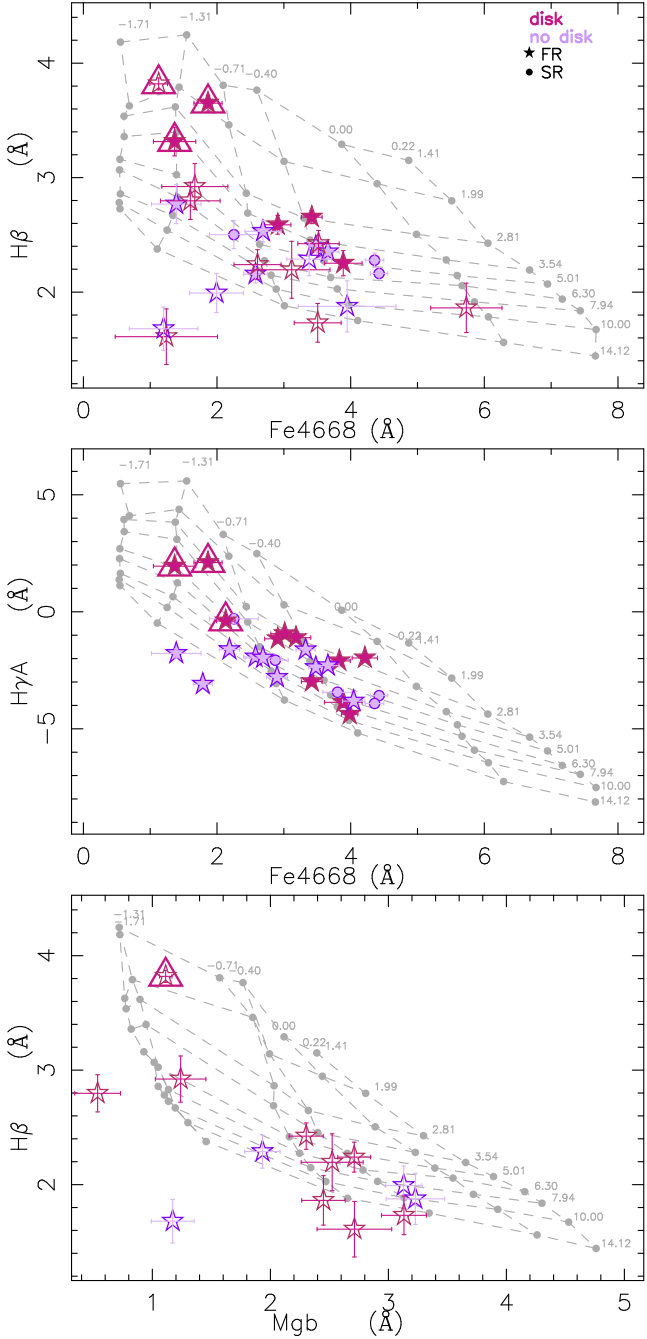

\centering
\includegraphics[angle=-90,width=8.5cm]{fig13a.ps}
\includegraphics[angle=-90,width=8.5cm]{fig13b.ps}
\includegraphics[angle=-90,width=8.5cm]{fig13c.ps}
\caption{Spectral index-index diagrams used to estimate the stellar populations of galaxies within the \Reff. The grey dashed lines represent the grid of SSP models by \citet{Vazdekis10} in the system LIS-5~\AA\ based on the MILES stellar library \citep{SB06lib} with a Kroupa initial mass function \citep{KroupaIMF}. Nearly horizontal lines indicate constant age, with values in Gyr printed at the right end of those lines, and nearly vertical lines indicate constant metallicity, with values printed in the upper part of the grid. Red and purple symbols indicate dEs with and without underlying disky structures seen in high-pass filtered optical images, respectively \citep{Lisk06a}. Dots and asterisks indicate slow and fast rotators, respectively, based on their specific angular momentum \lambdae\ and ellipticity described in Paper~III. Filled and open symbols indicate galaxies where the H$_{\alpha}$ region is or is not covered by our observations, respectively. Large open triangles indicate dEs with emission lines, which are cleaned using the GANDALF software \citep[][see the text for details]{Sarzi06}. The dEs have luminosity-weighted ages older than 1~Gyr and sub-solar metallicities.}
\label{indices}
\end{figure}

Using the single stellar population models of \citet[SSP;][]{Vazdekis10} based on the MILES stellar library \citep{SB06lib,MILEScen,FB11MILES} with a Kroupa initial mass function \citep{KroupaIMF} and also broadened to the LIS-5\AA\ system, we estimate the ages and metallicities ([M/H]) using the software {\sc rmodel\footnote{\url{http://www.ucm.es/info/Astrof/software/rmodel/rmodel.html}}} \citep[]{Cardiel03}. This software interpolates the age and metallicity inside an index-index grid. The errors in the age and [M/H] are calculated by running 1000 Monte Carlo simulations, varying the values of the spectral indices within a Gaussian function whose width is equal to their uncertainties. 

The index-index grids used to estimate the ages and metallicities are shown in Figure \ref{indices}. Some galaxies appear in more than one index-index diagram because their spectral range allows the simultaneous measurement of more than one of the age-sensitive or metallicity-sensitive indices shown in Figure \ref{indices}. For the dEs for which pairs of indices can be measured, the adopted ages and [M/H] are the uncertainty-weighted average of the estimations based on each independent index-index diagram.

The spread in ages and metallicities is remarkable in this galaxy class \citep{Mich08,Chil09,Paudel2010,Koleva09,Koleva11}. The SMAKCED dEs have luminosity-weighted ages that range from $\sim$2~Gyr to as old as the oldest models computed ($\sim$14~Gyr, see Section \ref{indices_sec} for a discussion of the effects that the presence of not detected emission lines would have in the derived ages). In addition, the SMAKCED dEs have metallicities from close to solar, [M/H]$\sim0.0$~dex, to as metal poor as some of the dwarf spheroidals in the Local Group, [M/H]$\sim -1.3$~dex, \citep{Kirby11,Kirby13}. 

The averaged age and [M/H] for dEs with underlying disky structures are $5.46\pm0.81$~Gyr and [M/H]$=-0.64\pm0.09$. The averaged age and [M/H] for dEs without underlying disky structures are $6.85\pm0.76$~Gyr and [M/H]$=-0.59\pm0.07$. 
The integrated ages and metallicities within the \Reff\ do not seem to be related to the presence or absence of subtle underlying disky structures. They also do not show a correlation with the position of the dEs within the Virgo cluster.

\section{Total Mass and Dark Matter Fraction}\label{scaling_rels}

\begin{table}
\begin{center}
\caption{Masses and Dark Matter Fractions. \label{masses_table}}
{\renewcommand{\arraystretch}{1.}
\resizebox{9cm}{!} {
\begin{tabular}{|c|c|c|c|c|c|}
\hline \hline
Galaxy              &  $\log M_e$  &  $\log M_e^*$  &  $f_{\rm DM}$  & $(M/L)_{dyn,r}$ & $(M/L)_{dyn,H}$\\
                        &   M$_{\odot}$  &   M$_{\odot}$    &                       &   M$_{\odot}/$L$_{\odot,r}$  & M$_{\odot}/$L$_{\odot,H}$\\
    (1)                &       (2)           &            (3)       &   (4)                &          (5)          &         (6)            \\
\hline \hline
VCC0009  &9.24$\pm$0.14 & 8.27$\pm$0.12 & $ 0.28\pm$0.30 & 2.62$\pm$0.83  &0.67$\pm$0.21\\
VCC0021  &8.88$\pm$0.11 & 8.34$\pm$0.12 & $ 0.61\pm$0.15 & 3.19$\pm$0.81  &1.22$\pm$0.31\\
VCC0033  &8.56$\pm$0.19 & 8.37$\pm$0.12 & $ 0.19\pm$0.41 & 1.82$\pm$0.78  &0.59$\pm$0.25\\
VCC0170  &9.11$\pm$0.15 & 8.47$\pm$0.12 & $ 0.56\pm$0.20 & 3.16$\pm$1.08  &1.08$\pm$0.37\\
VCC0308  &8.93$\pm$0.12 & 8.47$\pm$0.12 & $-0.03\pm$0.40 & 1.53$\pm$0.42  &0.47$\pm$0.13\\
VCC0389  &9.04$\pm$0.09 & 8.48$\pm$0.12 & $ 0.08\pm$0.31 & 1.82$\pm$0.36  &0.52$\pm$0.10\\
VCC0397  &9.02$\pm$0.08 & 8.50$\pm$0.12 & $ 0.64\pm$0.12 & 5.73$\pm$1.08  &1.35$\pm$0.25\\
VCC0437  &9.42$\pm$0.10 & 8.52$\pm$0.12 & $ 0.65\pm$0.13 & 4.54$\pm$1.02  &1.37$\pm$0.31\\
VCC0523  &9.31$\pm$0.07 & 8.52$\pm$0.12 & $ 0.37\pm$0.20 & 1.98$\pm$0.30  &0.76$\pm$0.11\\
VCC0543  &9.16$\pm$0.08 & 8.57$\pm$0.12 & $ 0.52\pm$0.16 & 3.05$\pm$0.56  &1.00$\pm$0.19\\
VCC0634  &9.15$\pm$0.09 & 8.59$\pm$0.12 & $ 0.36\pm$0.22 & 1.66$\pm$0.34  &0.75$\pm$0.15\\
VCC0750  &9.26$\pm$0.08 & 8.62$\pm$0.12 & $ 0.83\pm$0.06 & 8.35$\pm$1.52  &2.78$\pm$0.51\\
VCC0751  &8.83$\pm$0.10 & 8.66$\pm$0.12 & $ 0.17\pm$0.29 & 1.97$\pm$0.43  &0.58$\pm$0.13\\
VCC0781  &9.09$\pm$0.09 & 8.71$\pm$0.12 & $ 0.63\pm$0.13 & 4.36$\pm$0.87  &1.29$\pm$0.26\\
VCC0794  &9.15$\pm$0.13 & 8.71$\pm$0.12 & $ 0.79\pm$0.08 & 4.73$\pm$1.37  &2.30$\pm$0.66\\
VCC0856  &9.01$\pm$0.12 & 8.72$\pm$0.12 & $ 0.33\pm$0.26 & 2.27$\pm$0.64  &0.72$\pm$0.20\\
VCC0917  &8.75$\pm$0.09 & 8.74$\pm$0.12 & $ 0.58\pm$0.15 & 3.70$\pm$0.80  &1.15$\pm$0.25\\
VCC0940  &9.30$\pm$0.08 & 8.75$\pm$0.12 & $ 0.74\pm$0.09 & 6.32$\pm$1.16  &1.85$\pm$0.34\\
VCC0990  &8.99$\pm$0.07 & 8.75$\pm$0.12 & $ 0.43\pm$0.18 & 2.87$\pm$0.47  &0.85$\pm$0.14\\
VCC1010  &9.33$\pm$0.07 & 8.78$\pm$0.12 & $ 0.29\pm$0.23 & 2.57$\pm$0.41  &0.68$\pm$0.11\\
VCC1087  &9.26$\pm$0.07 & 8.79$\pm$0.12 & $ 0.46\pm$0.17 & 1.82$\pm$0.29  &0.88$\pm$0.14\\
VCC1122  &9.01$\pm$0.09 & 8.84$\pm$0.12 & $ 0.59\pm$0.14 & 3.81$\pm$0.76  &1.18$\pm$0.24\\
VCC1183  &9.33$\pm$0.07 & 8.84$\pm$0.12 & $ 0.64\pm$0.12 & 4.27$\pm$0.71  &1.33$\pm$0.22\\
VCC1261  &9.41$\pm$0.06 & 8.89$\pm$0.12 & $ 0.42\pm$0.18 & 2.85$\pm$0.43  &0.82$\pm$0.12\\
VCC1304  &8.81$\pm$0.12 & 8.89$\pm$0.12 & $ 0.48\pm$0.20 & 3.13$\pm$0.84  &0.92$\pm$0.25\\
VCC1355  &8.89$\pm$0.18 & 8.91$\pm$0.12 & $ 0.32\pm$0.34 & 1.90$\pm$0.80  &0.71$\pm$0.30\\
VCC1407  &8.93$\pm$0.09 & 8.91$\pm$0.12 & $ 0.54\pm$0.16 & 3.74$\pm$0.80  &1.04$\pm$0.22\\
VCC1431  &9.20$\pm$0.06 & 8.93$\pm$0.12 & $ 0.46\pm$0.17 & 3.49$\pm$0.47  &0.89$\pm$0.12\\
VCC1453  &9.15$\pm$0.08 & 8.94$\pm$0.12 & $ 0.40\pm$0.20 & 2.72$\pm$0.49  &0.79$\pm$0.14\\
VCC1528  &9.05$\pm$0.06 & 8.94$\pm$0.12 & $ 0.46\pm$0.17 & 3.22$\pm$0.47  &0.88$\pm$0.13\\
VCC1549  &9.01$\pm$0.08 & 8.96$\pm$0.12 & $ 0.42\pm$0.20 & 3.53$\pm$0.68  &0.83$\pm$0.16\\
VCC1684  &8.85$\pm$0.09 & 8.96$\pm$0.12 & $ 0.69\pm$0.11 & 4.24$\pm$0.88  &1.54$\pm$0.32\\
VCC1695  &8.84$\pm$0.11 & 8.99$\pm$0.12 & $ 0.27\pm$0.28 & 1.67$\pm$0.44  &0.66$\pm$0.17\\
VCC1861  &9.06$\pm$0.09 & 9.00$\pm$0.12 & $ 0.33\pm$0.23 & 2.17$\pm$0.44  &0.72$\pm$0.14\\
VCC1895  &8.73$\pm$0.13 & 9.03$\pm$0.12 & $ 0.38\pm$0.25 & 2.41$\pm$0.72  &0.78$\pm$0.23\\
VCC1910  &9.03$\pm$0.07 & 9.10$\pm$0.12 & $ 0.00\pm$0.33 & 2.05$\pm$0.35  &0.48$\pm$0.08\\
VCC1912  &9.23$\pm$0.08 & 9.11$\pm$0.12 & $ 0.52\pm$0.16 & 3.36$\pm$0.60  &0.99$\pm$0.18\\
VCC1947  &9.11$\pm$0.06 & 9.17$\pm$0.12 & $ 0.36\pm$0.20 & 3.19$\pm$0.44  &0.75$\pm$0.10\\
VCC2083  &8.93$\pm$0.10 & 9.18$\pm$0.12 & $ 0.78\pm$0.08 & 6.51$\pm$1.55  &2.17$\pm$0.52\\
\hline
\end{tabular}
}}
\end{center}
\tablecomments{Column 1: galaxy name. Column 2: dynamical mass within
  the \Reff~ estimated as described in Equation \ref{eqn_mdyn}. Column
  3: stellar mass within the \Reff~ estimated assuming a stellar
  mass-to-light ratio of $(M/L)^*_H = 0.73 \pm 0.19$ for all dEs. The
  average mass does not change if we assume a different $(M/L)^*_H$ or
  $(M/L)^*_V$ for each dE (see Section \ref{scaling_rels}). The
    total dynamical masses and the total stellar masses can be
    calculated by multiplying by 2 the masses in columns 2
    and 3. Column 4: dark matter fraction within the \Reff~ estimated as described in Equation \ref{eqn_fDM}. Note that negative values of $f_{\rm DM}$ are consistent with no dark matter within the uncertainties. Columns 5 and 6: dynamical mass-to-light ratio calculated dividing the dynamical masses in Column 1 by half the luminosities obtained from the $r$ and $H$ band absolute magnitudes in Table \ref{tabledEs}, respectively.}
\end{table}

We measure the dynamical mass and dark matter fraction of the Virgo cluster dEs.
The dynamical mass is calculated following the Equation:

\begin{equation}\label{mass}
M \simeq c~ G^{-1} \sigma_R^2 R
\end{equation}

\noindent where $\sigma_{R}$ is the velocity dispersion within an aperture of radius $R$ and $G$ is the gravitational constant. This Equation, based on the virial theorem, assumes that the galaxies are in equilibrium. For an aperture with $R= $~\Reff, Equation \ref{mass} is a reliable estimator of the enclosed mass \citep{Cappellari06,Cappellari13}. The constant $c$ depends, within other parameters, on the light distribution of the galaxy, i.e. the S\'ersic index ($n$) that best fits the surface brightness profile. For galaxies with $n\sim2$, $c=3.63$ \citep{Cappellari06,Courteau14}. The SMAKCED sample of Virgo cluster dEs has, on average, $n=1.52^{+0.33}_{-0.30}$. Then, the dynamical mass is estimated as:

\begin{equation}\label{eqn_mdyn}
M_{e} = 3.63~ G^{-1} (\sigma^{\rm 2D}_{e})^2 R_e
\end{equation}

To transform the measured long-slit $\sigma_{e}$ into the integrated velocity dispersion within an aperture with radius \Reff, $\sigma^{\rm 2D}_{e}$, we simulate the two dimensional distribution of the flux, $V$, and $\sigma$ based on the long-slit spectroscopic measurements. We define elliptical isophotes using the $H$ band ellipticity gradients where $\sigma$ is constant and $V$ follows a cosine function that makes the rotation maximum along the major axis and zero along the minor axis. These simulations are generated only in the regions where we have spectroscopic data. We calculate $\sigma^{\rm 2D}_{e}$ in the following way

\begin{equation}
\sigma^{\rm 2D}_{e}=\frac{\sum_{i=0}^{Re}F_i\sqrt{V_i^2+\sigma_i^2}}{\sum_{i=0}^{Re}F_i}
\end{equation}

\noindent where $F_i$, $V_i$, and $\sigma_i$ are the flux, velocity, and velocity dispersion of the $i$th elliptical isophote.
We calculate $\sigma^{\rm 2D}_{e}$, the integrated velocity dispersion within and ellipse with semi-major axis equal to the \Reff, for those dEs with spectroscopic information within $R/R_e \gtrsim 1$. The best fit between $\sigma^{\rm 2D}_{e}$ and $\sigma_{e}$ is

\begin{equation}\label{sigma2D}
\sigma^{\rm 2D}_{e}=(6.7 \pm 3.1)+(0.9 \pm 0.0) \sigma_{e}
\end{equation}

We use this relation to convert the long-slit velocity dispersion measurements into aperture values and estimate the dynamical masses within the \Reff\ using the Equation \ref{eqn_mdyn}. These dynamical masses are in agreement, within the $1\sigma_G$ uncertainties, with those estimated using dynamical models by \citet{Geha02} and \citet{Rys14}.

The inclination affects the rotation amplitude measured in a galaxy. However, both parameters are compensated in a way that the resulting dynamical mass-to-light ratio is independent from the assumed inclination \citep{vandermarel91,Cappellari06}. \citet{Rys14} applies Jeans axisymmetric models to a sample of dEs assuming different inclination values and the resulting dynamical masses are always consistent within the $1\sigma_G$ uncertainties.

The dark matter fraction within the \Reff\ for galaxies with negligible amounts of gas is defined as:

\begin{equation}\label{eqn_fDM}
f_{\rm DM}=\frac{M_e^{\rm DM}}{M_e^*+M_e^{\rm DM}}=\frac{M_e-M_e^*}{M_e}
\end{equation}

\noindent where $M_e^{\rm DM}$ is the mass of the dark matter, $M_e^*$ is the stellar mass, and $M_e$ is the dynamical mass all of them within the \Reff.

We use three different methods to estimate the $M_e^*$: (1) we assume a common stellar mass-to-light ratio in the $H$ band $(M/L)^*_H$ for all the SMAKCED dEs. We estimate the $(M/L)^*_H$ using the SSP models of \citet{Vazdekis10} and the median age and metallicity of the dEs, and get $(M/L)_H^*=0.73 \pm 0.19$; (2) we estimate the $(M/L)^*_H$ independently for each dE using the SSP models of \citet{Vazdekis10} and the inferred ages and metallicities. This method gets the same median stellar mass as in method (1) but with a scatter $1.15$ times larger; (3) we follow the technique described in \citet{etj12}, where the stellar mass-to-light ratio in the $V$ band $(M/L)^*_V$ is estimated from the best linear fit $(M/L)_V^*-{\rm H{\beta}}$ and $(M/L)_V^*-{\rm H{\gamma A}}$. The $(M/L)_V^*$, ${\rm H{\beta}}$, and ${\rm H{\gamma A}}$ are obtained using exponentially declining star formation histories with values of $\tau$, the declining time scale, from 0.1 to 10.0~Gyr using the models of \citet{Vazdekis10}. Method (3) also gets the same median stellar mass as method (1) but with a scatter that is $1.27$ times larger.

In conclusion, the stellar masses obtained are independent of the method used.
However, the large uncertainties in the stellar populations inferred from the measured optical spectral indices (see Section \ref{stellarpops}) include a scatter in the stellar masses estimated using method (2). This reflects the limitations of the procedure used to derive the ages and metallicities. Method (3) also suffers from large uncertainties because, although the uncertainties in the measured Lick spectral indices are smaller than in the inferred ages and metallicities, there are some dEs outside the index-index grid of models (Figure \ref{indices}), which suggests that for these dEs, and probably others, their ${\rm H{\beta}}$ and ${\rm H{\gamma A}}$ values are overestimated. This could be an effect of not detected emission as discussed in Section \ref{indices_sec}. 
In summary, the larger scatter in the stellar masses obtained using these two methods is mainly  due to the fact that the Lick spectral indices and the quantities derived from them are noisy, thus, it is dominated by noise rather than by a real scatter in the stellar masses. Given that the $(M/L)^*_H$ is fairly constant for different stellar populations \citep[e.g.][]{Vazdekis10}, the $M_e^*$ reported in this work are based on the common value of $(M/L)_H^*=0.73 \pm 0.19$ for all the SMAKCED dEs. The values obtained are in good agreement, within the $1\sigma_G$ uncertainty, with the values obtained by \citet{Rys14} for the galaxies in common.

In Table \ref{masses_table} we provide the dynamical and stellar masses as well as the dark matter fractions for the SMAKCED dEs. While the dynamical masses are estimated within a sphere with radius the \Reff\ following the Equation \ref{eqn_mdyn}, the stellar masses are estimated within a projected cylinder with radius the \Reff. Although this integration effect can affect the derived dark matter fractions \citep[see e.g.][]{Dutton11}, the obtained dark matter content is consistent, within the $1\sigma_G$ uncertainties, with the values obtained, based on dynamical models, for the dEs in common with \citet{Geha02,Rys14}. Given that the SMAKCED dEs have very similar luminosity distribution, i.e. similar Sersic indices (see Table \ref{tabledEs}), all dEs will be affected by this effect in the same way. Full dynamical models are needed to better address this issue, however, these are beyond the scope of this paper.

The median dynamical mass for our sample of dEs is ${\rm \log M}_e = 9.1 \pm 0.2$, the median stellar mass is ${\rm \log M}_e^* = 8.8 \pm 0.2$, and the median dark matter fraction is $f_{\rm DM} = 46 \pm 18~\%$. This dark matter fraction is consistent with the previous estimations by \citet{etj11,etj12}, and it is significantly higher than the dark matter found for the ETGs in the \ATLAS\ sample \citep[$13~\%$;][]{Cappellari13}.

The luminosity in the $H$ band is calculated using the apparent magnitudes from \citet{Janz14} with the exception of VCC~397 whose $K$ band magnitude comes from \citet{etj12} and it is transformed into the $H$ band using a color of $H-K=0.21$ \citep{Pel99}. The luminosity in the $V$ band, used to estimate $(M/L)^*_V$, is calculated using the SDSS $r$ band and $g-r$ color by \citet{Janz08} and applying the transformation by \citet{Blanton07}\footnote{All the magnitudes used in this work are referred to the AB system.}

\begin{equation}\label{Vband_eqn}
V= g-0.3516-0.7585 \times (g-r-0.6102)
\end{equation}

\begin{figure}
\centering
\includegraphics[angle=-90,width=8.5cm]{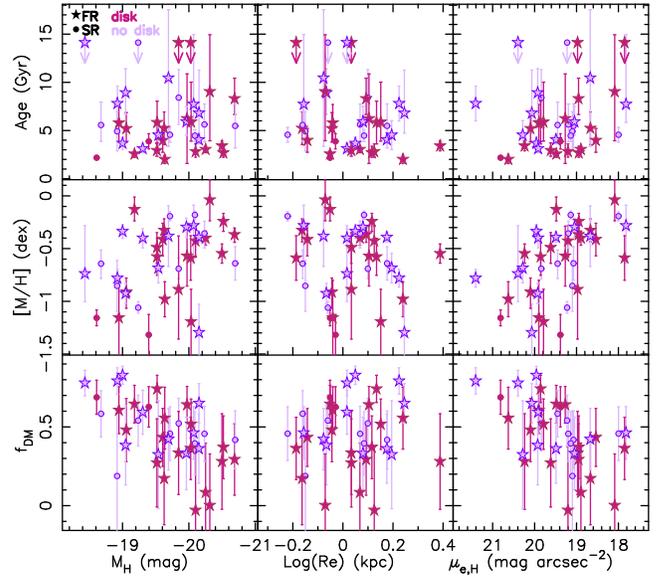}
\caption{Derived optical ages, metallicities, and dark matter fractions as a function of $H$ band luminosity, \Reff, and mean surface brightness within the \Reff\ for the SMAKCED dEs. Red and purple symbols indicate dEs with and without underlying disky structures seen in high-pass filtered optical images, respectively \citep{Lisk06a}. Dots and asterisks indicate slow and fast rotators, respectively, based on their specific angular momentum \lambdae\ and ellipticity described in Paper~III. The trend of more metal rich dEs being brighter and having a higher surface brightness is also seen for dwarf spheroidal galaxies \citep{Kirby13}.}
\label{stellarpops_fDM}
\end{figure}

Figure \ref{stellarpops_fDM} shows the derived ages, metallicities, and dark matter fractions as a function of the luminosity, size, and surface brightness of the SMAKCED galaxies. While the ages do not seem to have a strong dependence on any of these photometric properties, the metallicities and the dark matter fraction of the dEs show some trends. These trends, similar to those found for dwarf spheroidal galaxies \citep{SimonGeha07,Kirby13}, will be discussed in a future paper. In the case of the dark matter fraction, full dynamical models are needed to interpret the nature of this possible trend.

\begin{figure}
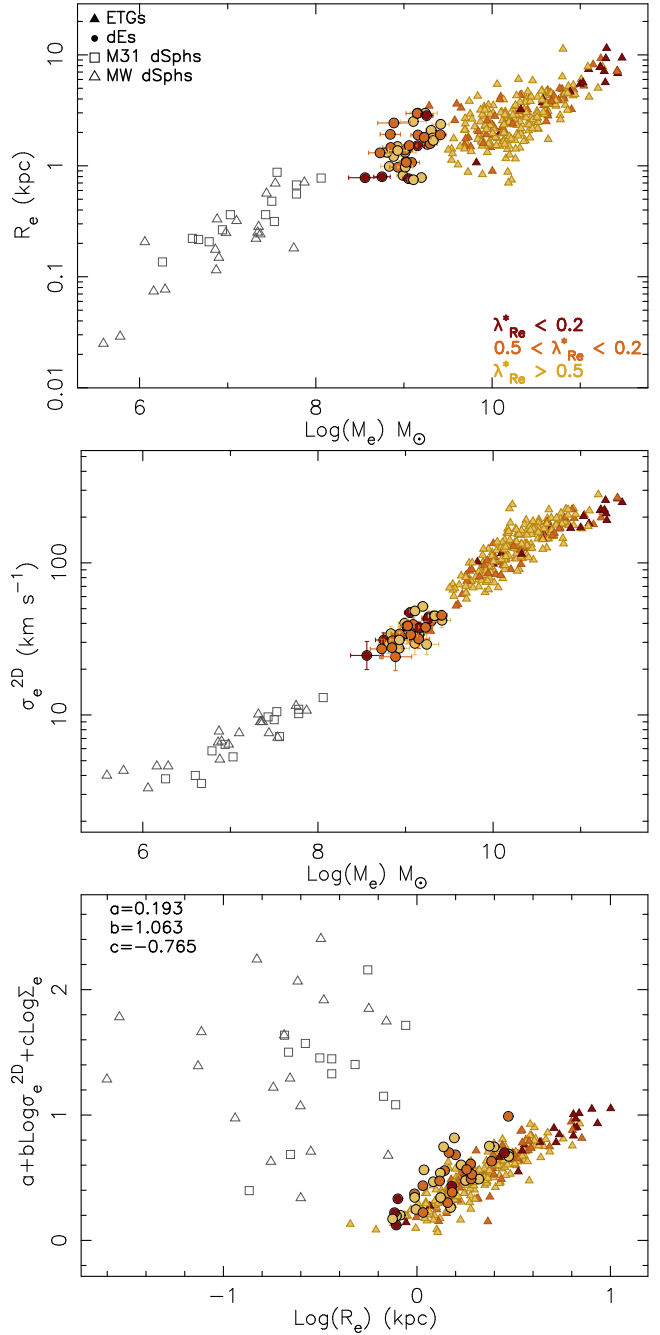

\centering
\includegraphics[angle=-90,width=8.5cm]{fig14a.ps}
\includegraphics[angle=-90,width=8.5cm]{fig14b.ps}
\includegraphics[angle=-90,width=8.5cm]{fig14c.ps}
\caption{{\bf Upper panel:} dynamical mass-size relation for the SMAKCED dEs (filled dots) in comparison with the \ATLAS\ ETGs (filled triangles) and the Milky Way and M31 dSphs (open triangles and squares, respectively). The colors for the ETGs and the dEs indicate whether the galaxies are slow or fast rotators based on their specific stellar angular momentum normalized by the square root of the ellipticity (\lambdaec, see Paper~III). {\bf Middle panel:} dynamical mass-velocity dispersion relation. {\bf Lower panel:} fundamental plane relation. The coefficients of the fundamental plane are those calculated for the \ATLAS\ sample by \citep{Cappellari13}. The dEs are the extension of ETGs in the direction of dSphs in the upper and middle panels. In the fundamental plane some dEs begin to separate from the plane defined by the ETGs in the direction where the dSphs lie. The fraction of slow rotating dEs is higher than expected given that the slow rotating ETGs tend to be within the most massive galaxies.}
\label{mass_size}
\end{figure}

Figure \ref{mass_size} shows the dynamical mass-size, dynamical mass-velocity dispersion, and fundamental plane scaling relations in the SDSS $r$ band for the SMAKCED dEs in comparison with the \ATLAS\ ETGs by \citet{Cappellari13} and the  Milky Way and M31 dwarf spheroidals by \citet{Wolf10,Tollerud12,McConnachie12}. 

The luminosities in the $V$ band for the dSphs are transformed into the $r$ band considering a color of $V-r=0.16$ \citep{Girardi04}. The surface brightness $\Sigma_e$ is calculated as $L_r/(2\pi R_{e,r})$, where $L_r$ and $R_{e,r}$ are the luminosity and the half-light radius in the $r$ band. 

The color-code of Figure \ref{mass_size} is based on the specific stellar angular momentum \lambdae\ normalized by the square root of the ellipticity \lambdaec. This parameter, which is the main focus of Paper~III of this series, indicates that galaxies with a small value of \lambdaec\ rotate slower than galaxies with a larger value of \lambdaec. 

In the mass-size and mass-$\sigma_e$ scaling relations, the dEs are the extension of ETGs in the direction of dSphs. However, in the fundamental plane, the dEs and the ETGs follow a sequence from which dSphs deviate. Some of the dEs show some deviation in the same direction as the dSphs but with a significantly smaller offset \citep[see e.g.][for a detailed discussion]{Zaritsky06,Zaritsky11,Tollerud11,etj12}. 

The fraction of slow rotating dEs (\lambdaec $<0.2$) is surprisingly high given that the majority of ETGs with \lambdaec $<0.2$ have masses of $\sim 10^{11}$~M$_{\odot}$. Looking at the distribution of ETGs in these three diagrams, it was expected that the majority, if not all, of the dEs have \lambdaec$>0.5$ (see Paper~III).

\section{Summary}\label{summ}

In this work, we present the analysis of the kinematic properties of a sample of 39 dEs in the Virgo cluster observed as part of the SMAKCED project. This sample is representative of the early-type population of galaxies in the Virgo cluster in the absolute magnitude range $-19.0 < M_r < -16.0$ and it is also representative of all the morphological sub-classes found for dEs by \citet{Lisk06b,Lisk06a,Lisk07}. In this paper, the second one on this series, we present the survey and analyze the shapes and amplitudes of the kinematic curves, the stellar populations, and the mass-to-light ratios.

We use optical spectroscopy to measure the rotation curves and velocity dispersion profiles of the SMAKCED dEs. We fit the rotation curves with an analytic function, called {\it Polyex}, and evaluate the amplitude at the \Reff\ (\Vrot).
We complement the spectroscopy with the $H$ band images and measure the surface brightness, position angle, ellipticity, and $C_4$ profiles. 
We find that dEs have a wide range of kinematic properties, from non-rotating to high rotation speeds. Two of the dEs in our sample have kinematically decoupled cores (which were the focus of Paper~I). These properties confirm previous results indicating that dEs are structurally very complex. 

For each galaxy, we quantify the significance of the departure of the rotation curve with respect to the {\it Polyex} function (poorly fit, PF), and also the significance of the different shape and amplitude between the approaching and receding sides of the rotation curve (AA and AS asymmetries). We find that more than half of the dEs have a significant kinematic anomaly (PF, AA, and/or AS, $62\pm8~\%$, 24/39). We also find a hint that dEs with smooth and symmetric rotation curves have smaller rotation speeds than those with kinematic anomalies. 

These kinematic anomalies do not seem to be related to the presence or lack of subtle disky structures visible in high-pass filtered optical images or to the number of components in which the $H$ band surface brightness profiles are best fitted.
However, the disk/no disk or single/multi-component classifications are not specifically designed to seek for asymmetries in the light distribution. In the case of the multi-component analysis, the light distribution is assumed symmetric, so, not finding a correlation with the kinematic anomalies is not surprising.

We find that the centers of the isophotes of the majority of the dEs ($64\pm8~\%$, 25/39) drift. This drift indicates that one side of the galaxy is brighter than the other, i.e. the light distribution is asymmetric. Sometimes the drift is found along the slit used for the spectroscopic observations and sometimes with an angle with respect to it. Even though we do not find a clear correlation between the degree of photometric and kinematic asymmetry, our analysis reveals that these asymmetries are frequent within the dE galaxy class.

Low luminosity star forming galaxies also show anomalous gas rotation curves. Stellar kinematic profiles of star forming galaxies are beginning to emerge in the literature and they seem to closely follow the gas kinematics \citep[see e.g.][]{Adams14,Koleva14}. We compare the stellar rotation curves of dwarf star forming galaxies with the stellar rotation curves of dEs in Paper~III. 

The large variety of kinematic features found in this work (non-rotators, slow rotators, fast rotators, kinematically decoupled cores, anomalous rotation curves) must be accounted in models to explain the physical mechanisms involved in the formation of this galaxy class. We discuss the origin and evolution of the dEs in the Virgo cluster based on the kinematic data presented here in Papers~I and III of this series.

\acknowledgments

E.T. thanks Laura Ferrarese for very useful suggestions on ellipse-fitting and Tomer Tal, Guillermo Barro, and Michele Cappellari for very helpful discussions. E.T. and P.G. thank Alice Wu and Ajinkya Nene, students of the 2014 Science Internship Program (SIP) for high school students that takes place every summer at UCSC, for their contribution to this manuscript.
The authors thank the anonymous referee for useful suggestions that helped improve the manuscript.
E.T. acknowledges the financial support of the Fulbright Program jointly with the Spanish Ministry of Education. PG acknowledges the NSF grant AST-1010039. T.L. was supported within the framework of the Excellence Initiative by the German Research Foundation (DFG) through the Heidelberg Graduate School of Fundamental Physics (grant number GSC 129/1). RFP, JFB, GvdV, E.L., and H.S. acknowledge the DAGAL network from the People Programme (Marie Curie Actions) of the European Union’s Seventh Framework Programme FP7/2007-2013/ under REA grant agreement number PITN-GA-2011-289313. J.J. thanks the ARC for financial support (DP130100388). G.H. acknowledges support by the FWF project P21097-N16. This work has made use of the GOLDMine database \citep{GOLDMine} and the NASA/IPAC Extragalactic Database (NED.\footnote{http://ned.ipac.caltech.edu})

\bibliographystyle{aa}
\bibliography{references}{}

\begin{figure*}
\centering
\resizebox{0.87\textwidth}{!}{\includegraphics[bb= 20 226 563 609,angle=0]{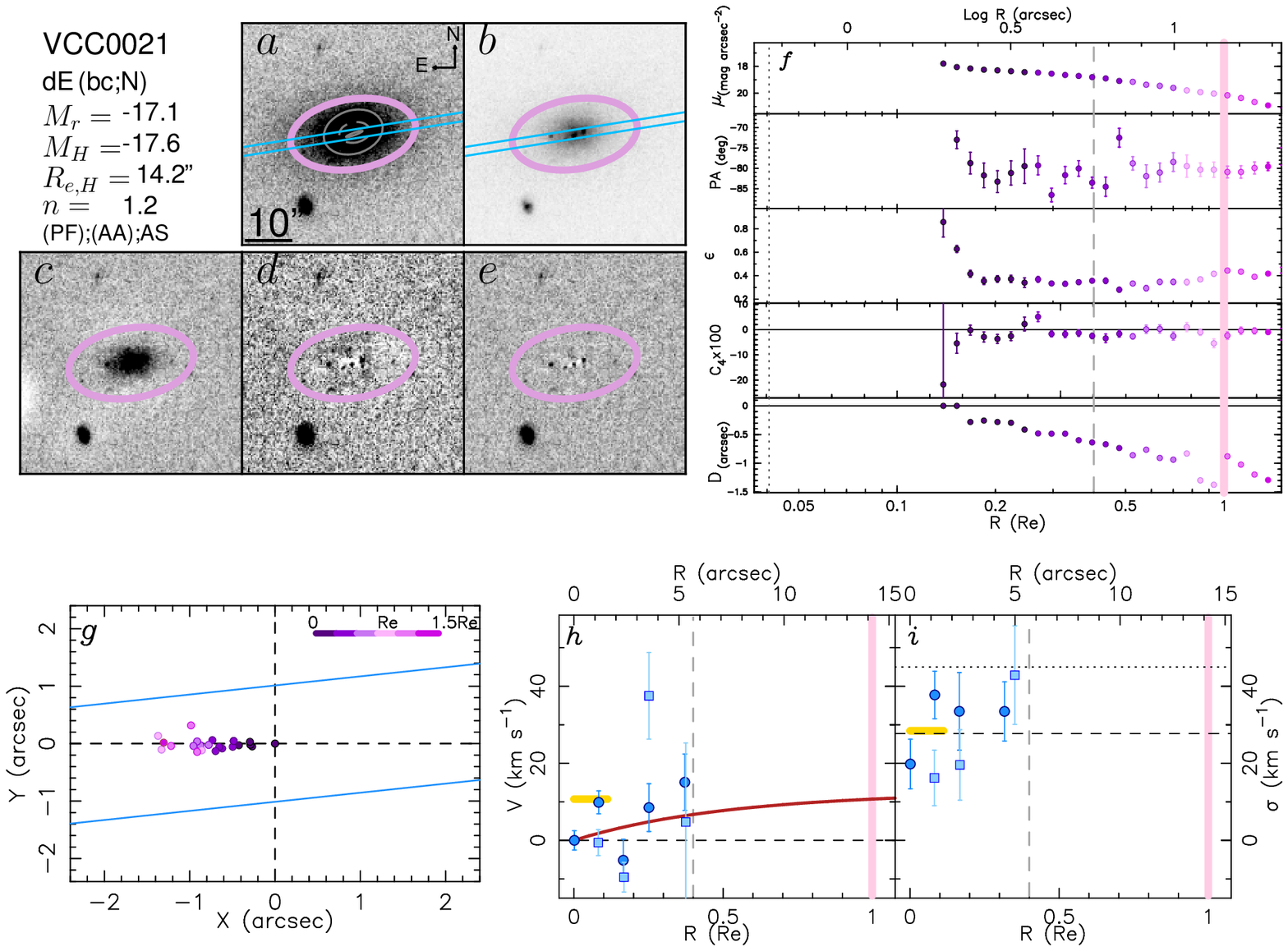}}
\caption{Same as Figure \ref{rotcurve_VCC0009} for VCC~21.}
\end{figure*}

\begin{figure*}
\centering
\resizebox{0.87\textwidth}{!}{\includegraphics[bb= 20 226 563 609,angle=0]{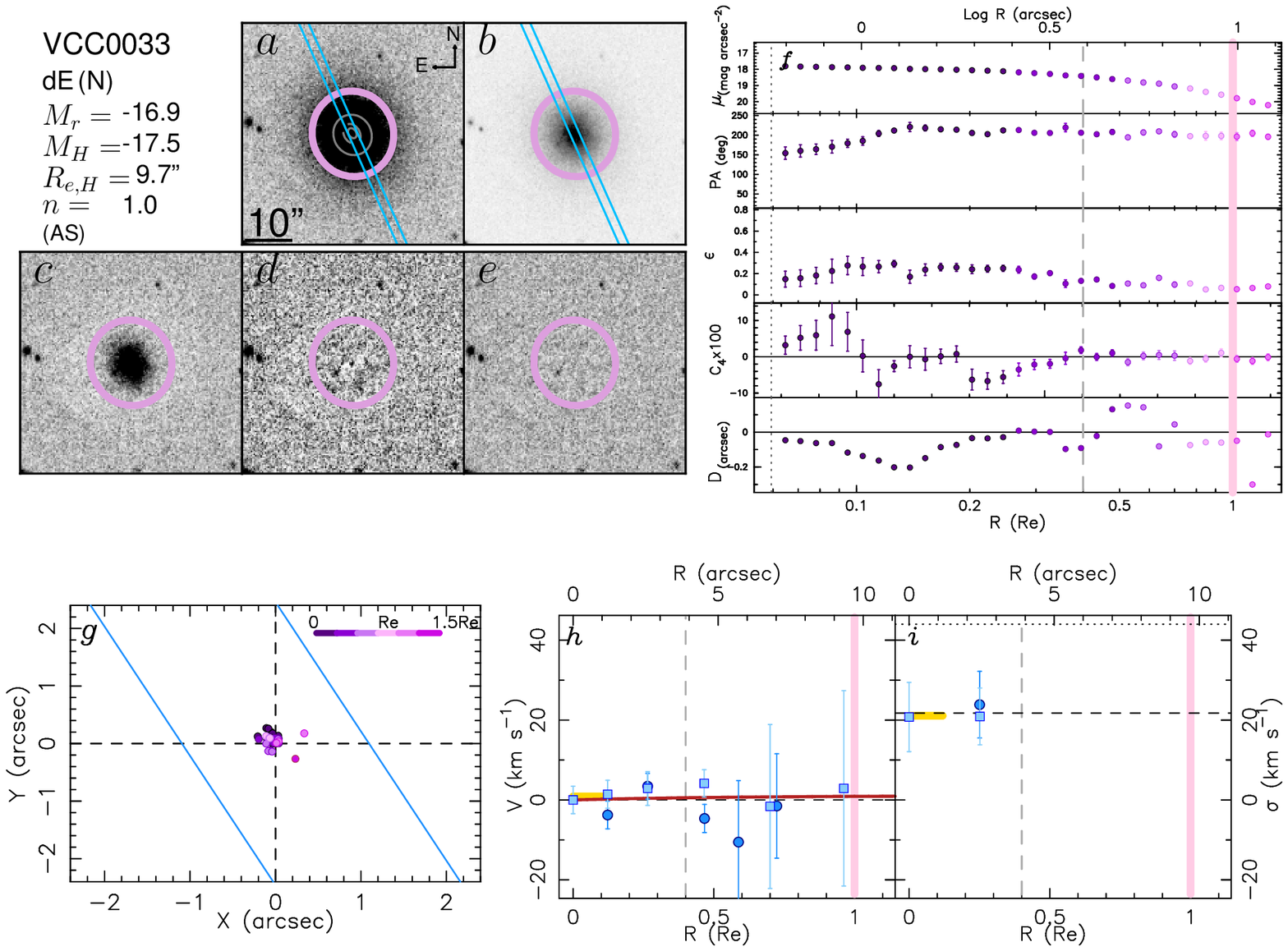}}
\caption{Same as Figure \ref{rotcurve_VCC0009} for VCC~33.}
\end{figure*}

\begin{figure*}
\centering
\resizebox{0.87\textwidth}{!}{\includegraphics[bb= 20 226 563 609,angle=0]{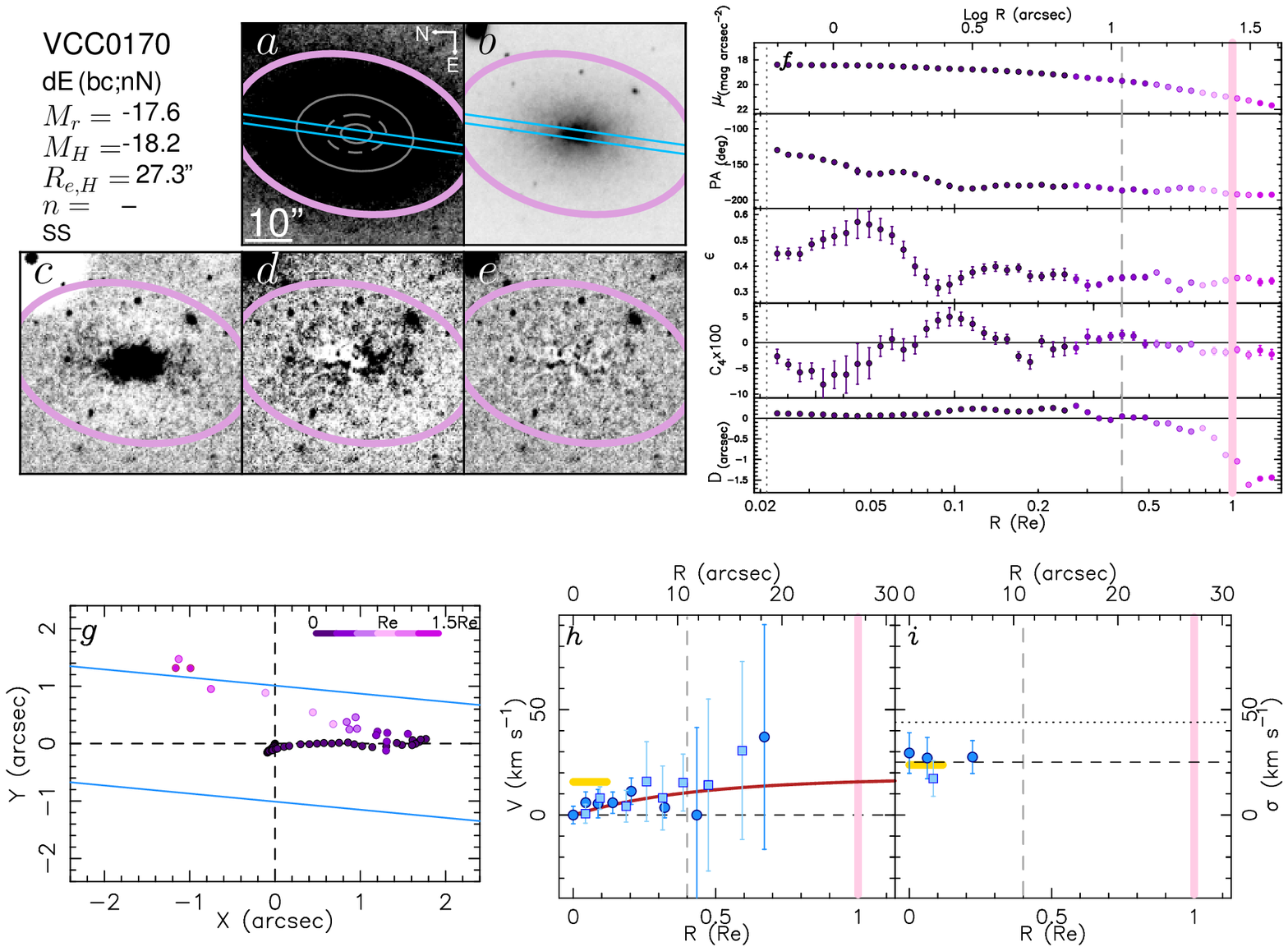}}
\caption{Same as Figure \ref{rotcurve_VCC0009} for VCC~170.}
\end{figure*}

\begin{figure*}
\centering
\resizebox{0.87\textwidth}{!}{\includegraphics[bb= 20 226 563 609,angle=0]{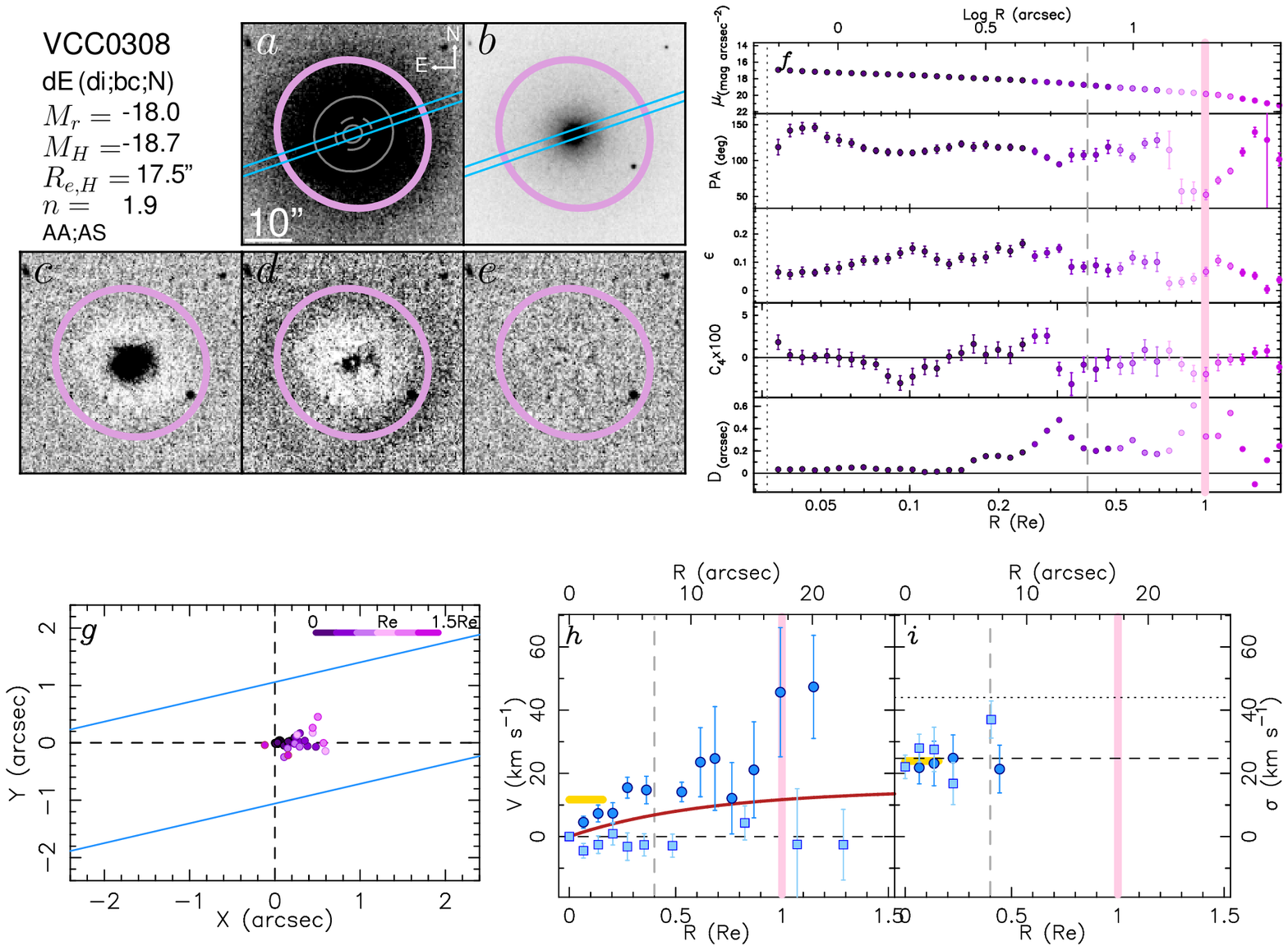}}
\caption{Same as Figure \ref{rotcurve_VCC0009} for VCC~308.}
\end{figure*}

\begin{figure*}
\centering
\resizebox{0.87\textwidth}{!}{\includegraphics[bb= 20 226 563 609,angle=0]{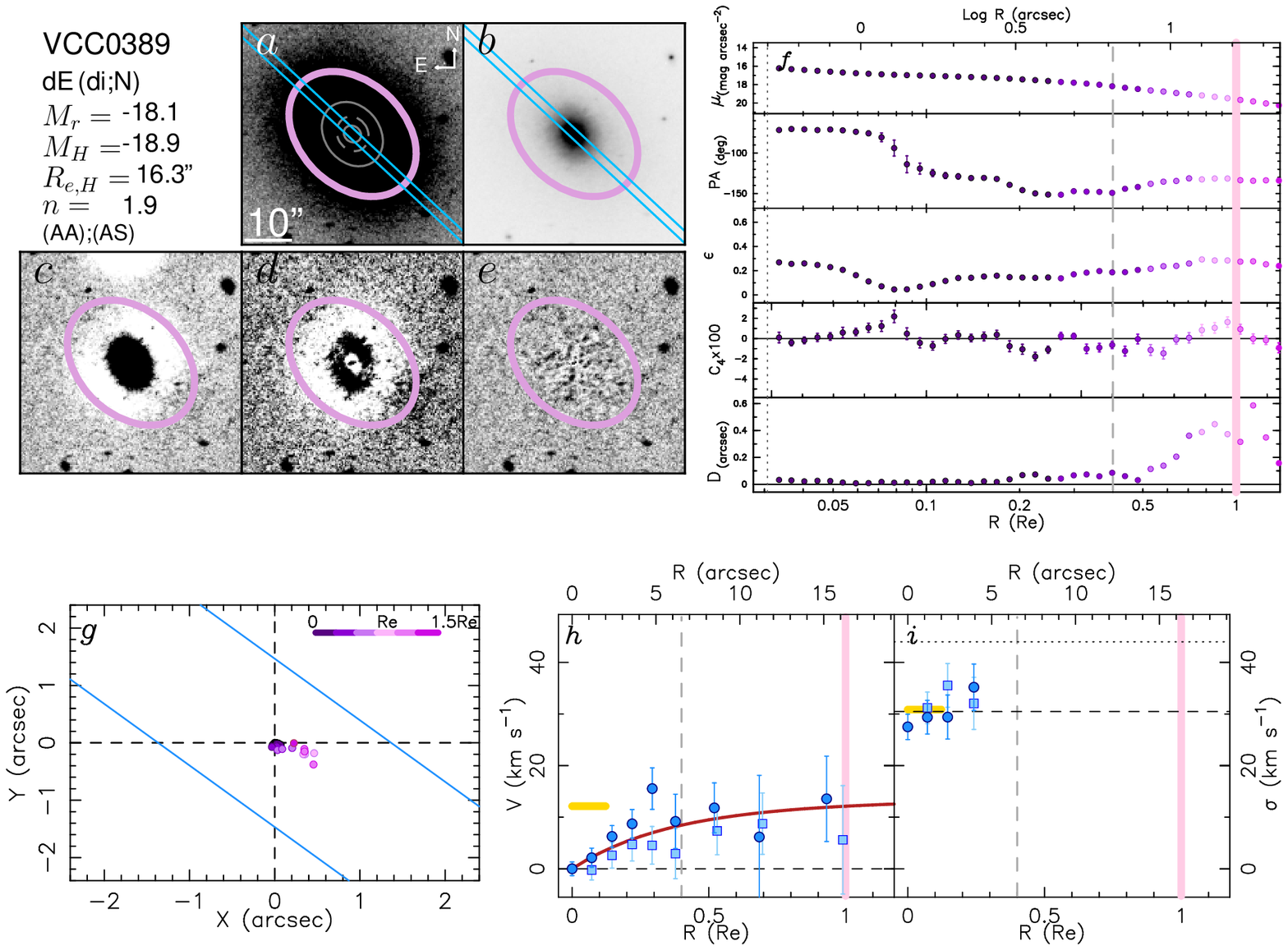}}
\caption{Same as Figure \ref{rotcurve_VCC0009} for VCC~389.}
\end{figure*}

\begin{figure*}
\centering
\resizebox{0.87\textwidth}{!}{\includegraphics[bb= 20 226 563 609,angle=0]{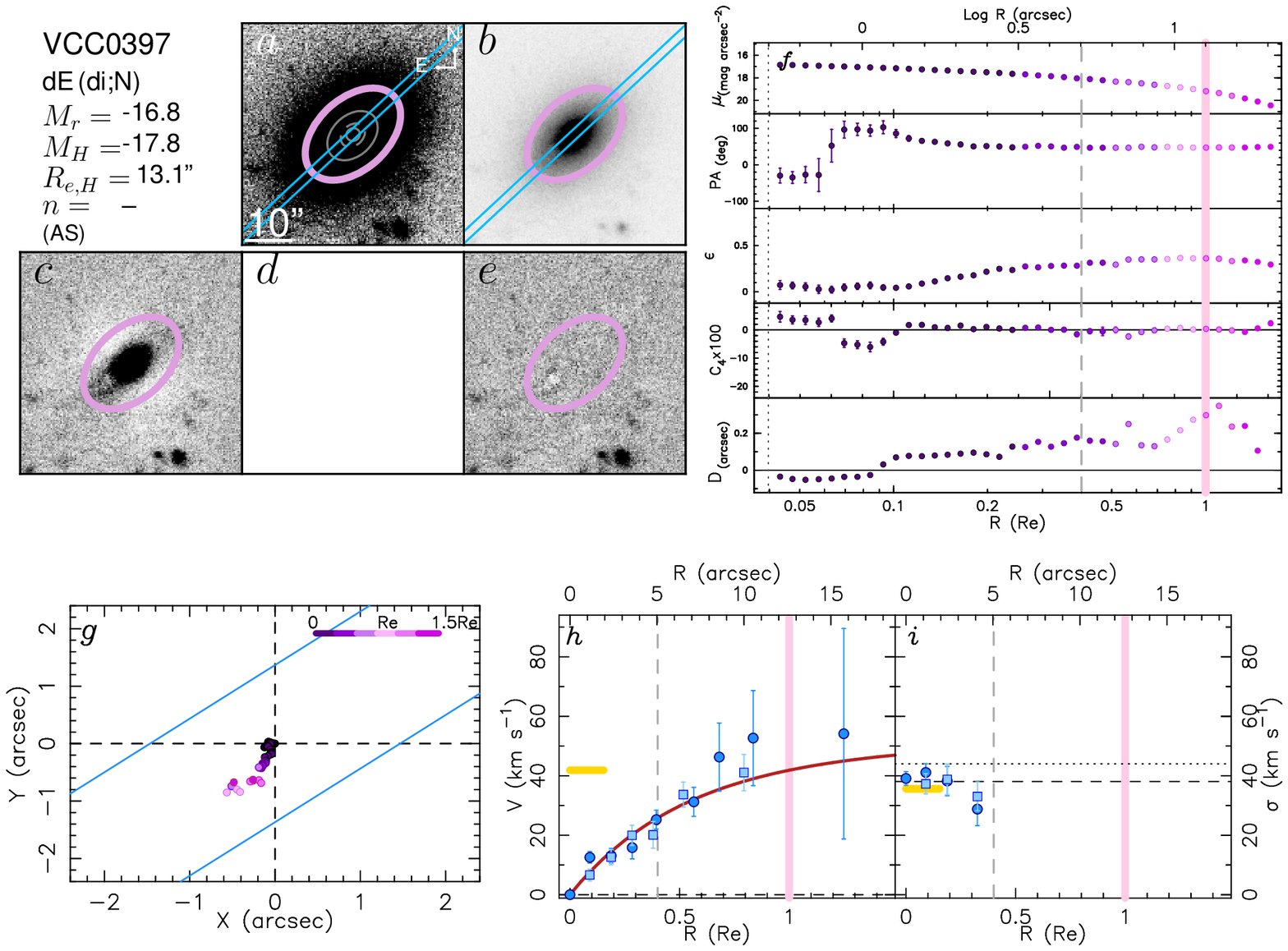}}
\caption{Same as Figure \ref{rotcurve_VCC0009} for VCC~397.}
\end{figure*}

\begin{figure*}
\centering
\resizebox{0.87\textwidth}{!}{\includegraphics[bb= 20 226 563 609,angle=0]{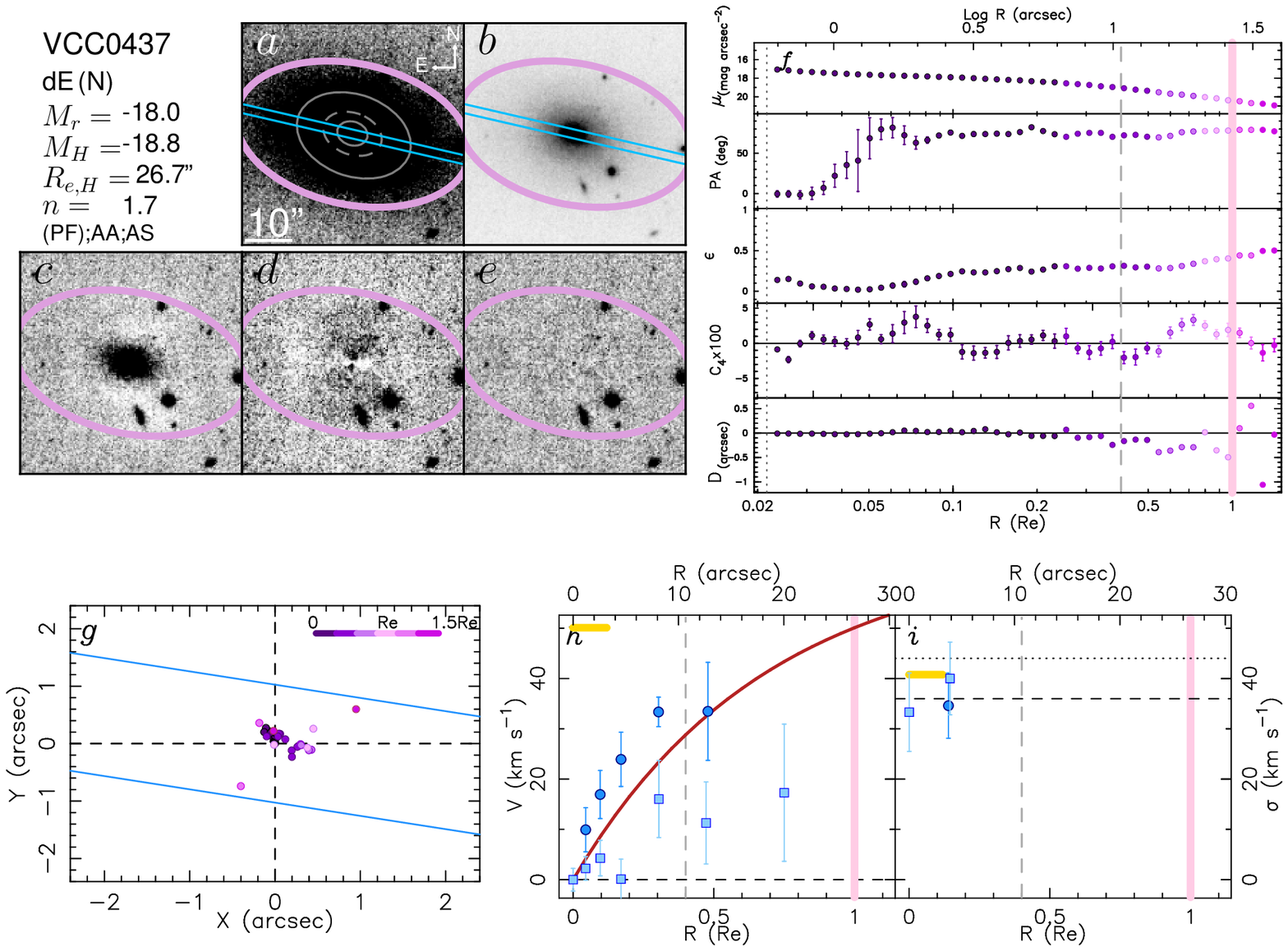}}
\caption{Same as Figure \ref{rotcurve_VCC0009} for VCC~437.}
\end{figure*}

\begin{figure*}
\centering
\resizebox{0.87\textwidth}{!}{\includegraphics[bb= 20 226 563 609,angle=0]{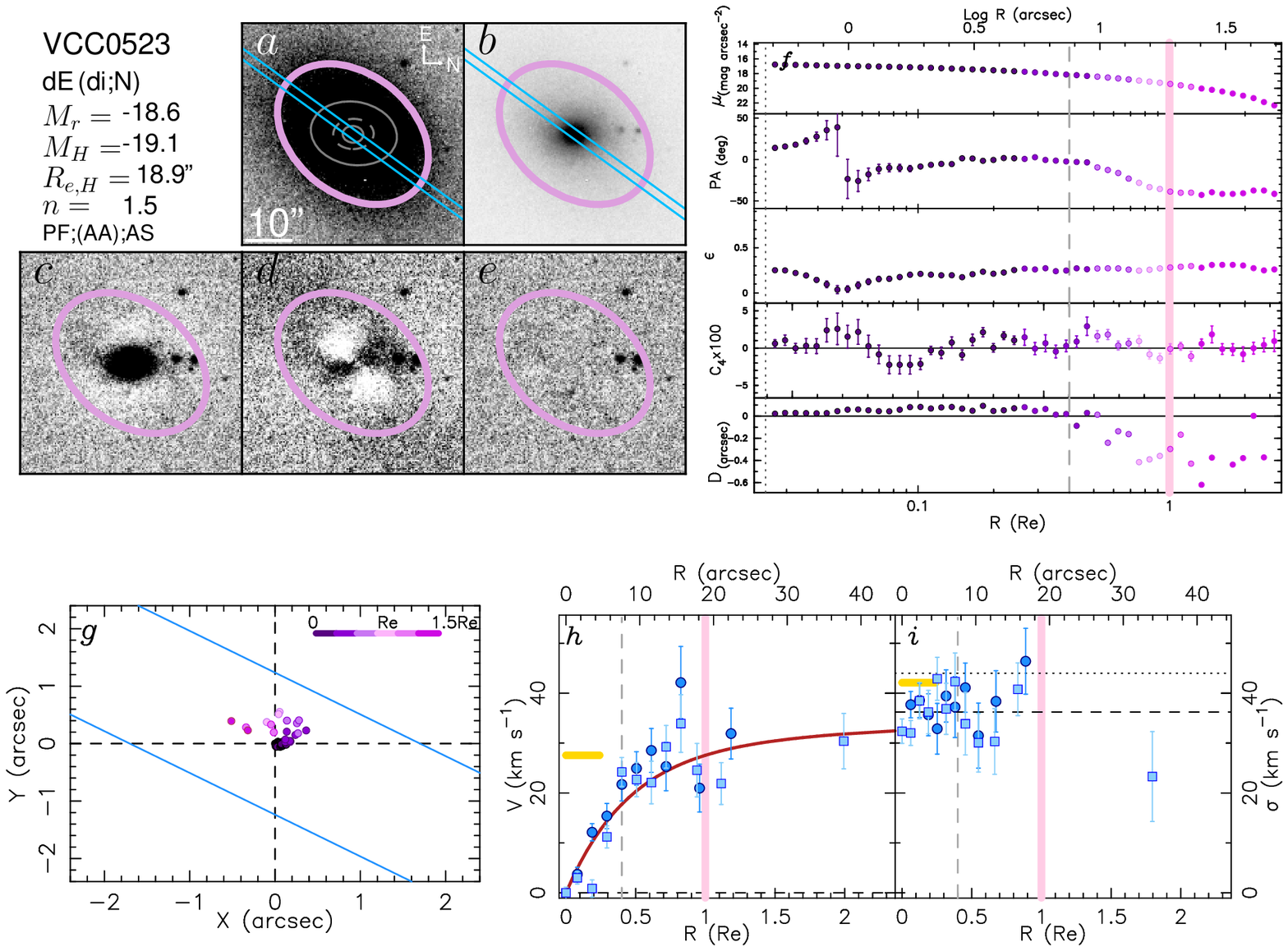}}
\caption{Same as Figure \ref{rotcurve_VCC0009} for VCC~523.}
\end{figure*}

\begin{figure*}
\centering
\resizebox{0.87\textwidth}{!}{\includegraphics[bb= 20 226 563 609,angle=0]{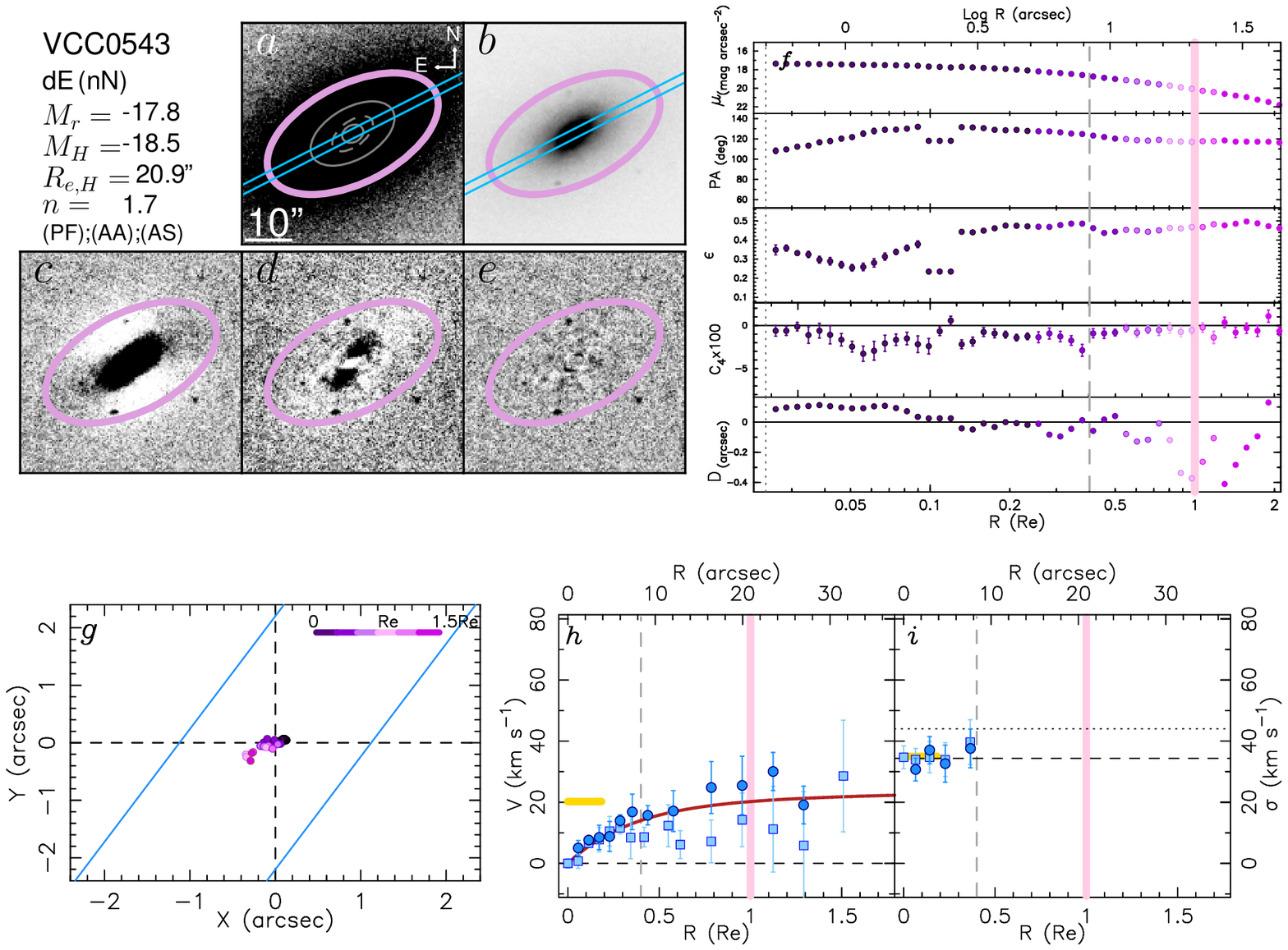}}
\caption{Same as Figure \ref{rotcurve_VCC0009} for VCC~543.}
\end{figure*}

\begin{figure*}
\centering
\resizebox{0.87\textwidth}{!}{\includegraphics[bb= 20 226 563 609,angle=0]{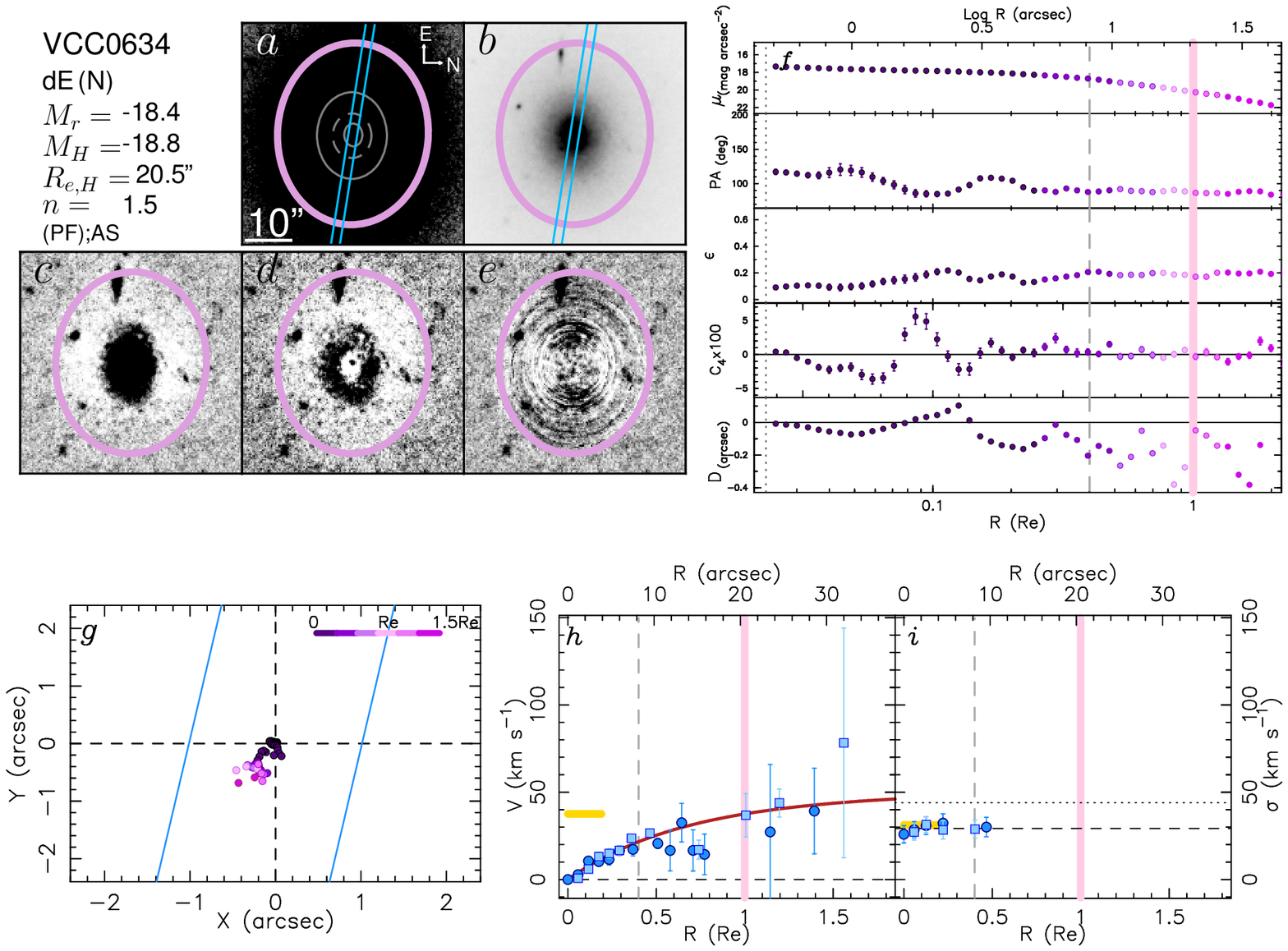}}
\caption{Same as Figure \ref{rotcurve_VCC0009} for VCC~634.}
\end{figure*}

\begin{figure*}
\centering
\resizebox{0.87\textwidth}{!}{\includegraphics[bb= 20 226 563 609,angle=0]{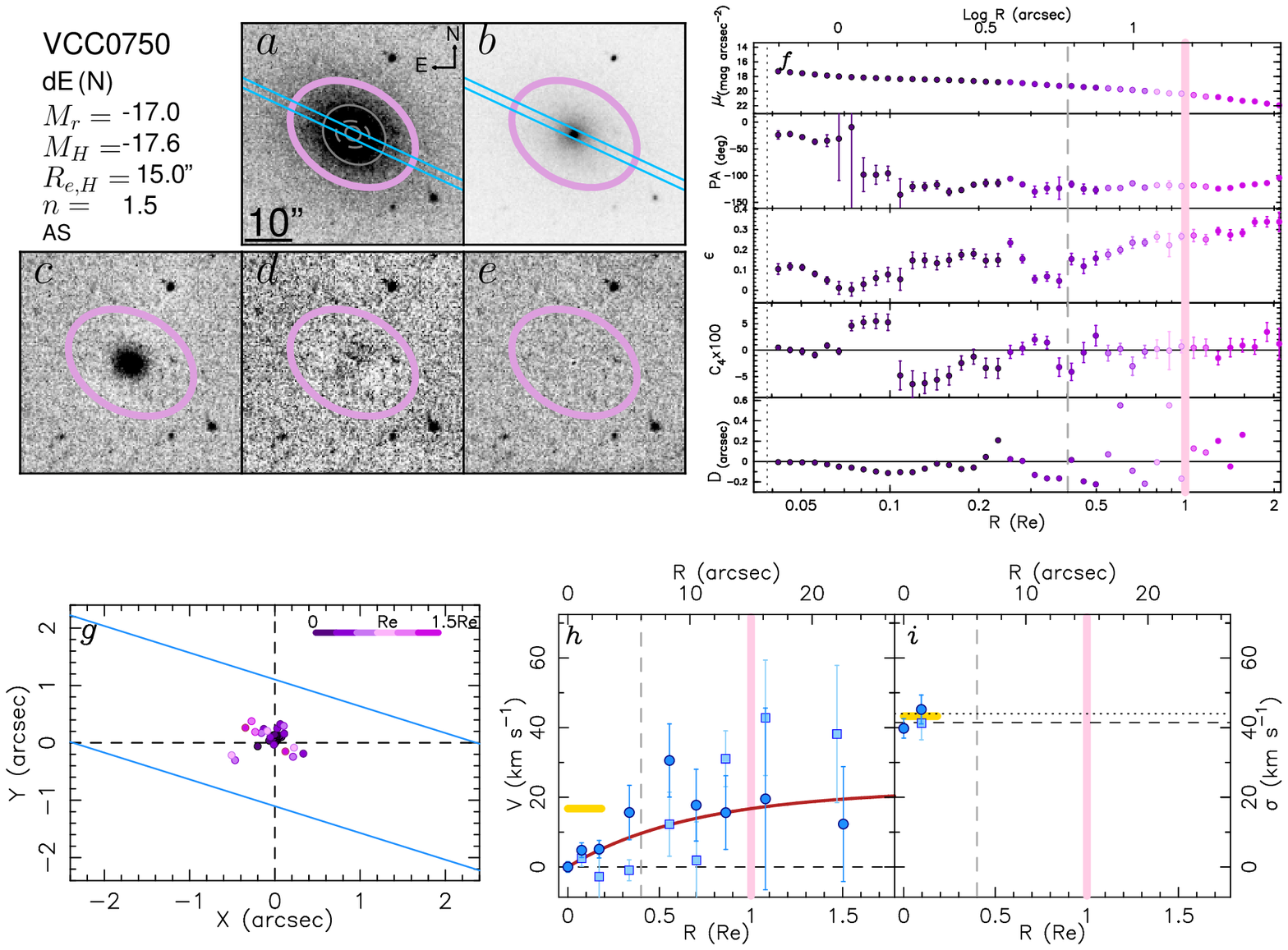}}
\caption{Same as Figure \ref{rotcurve_VCC0009} for VCC~750.}
\end{figure*}

\begin{figure*}
\centering
\resizebox{0.87\textwidth}{!}{\includegraphics[bb= 20 226 563 609,angle=0]{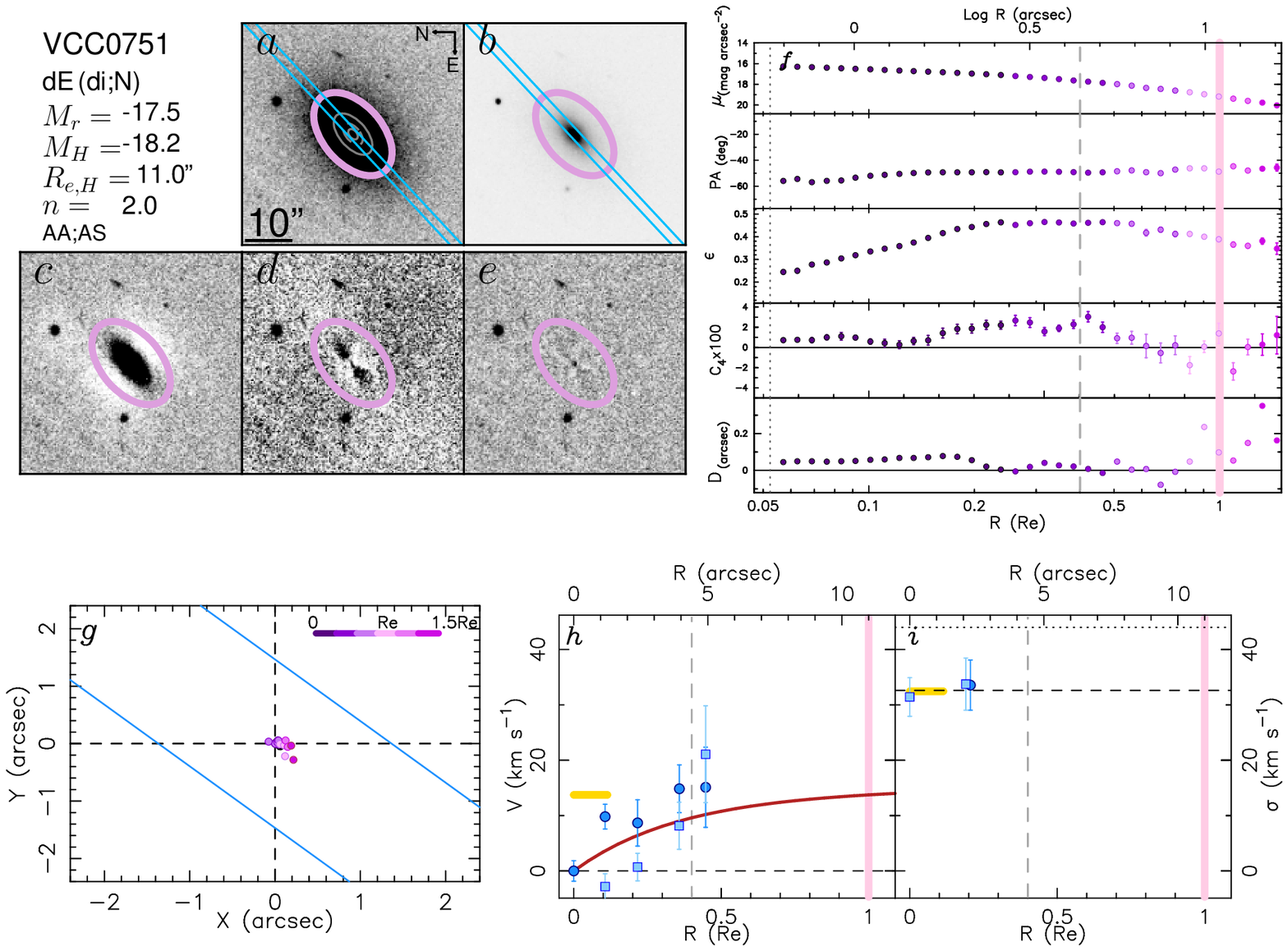}}
\caption{Same as Figure \ref{rotcurve_VCC0009} for VCC~751.}
\end{figure*}

\begin{figure*}
\centering
\resizebox{0.87\textwidth}{!}{\includegraphics[bb= 20 226 563 609,angle=0]{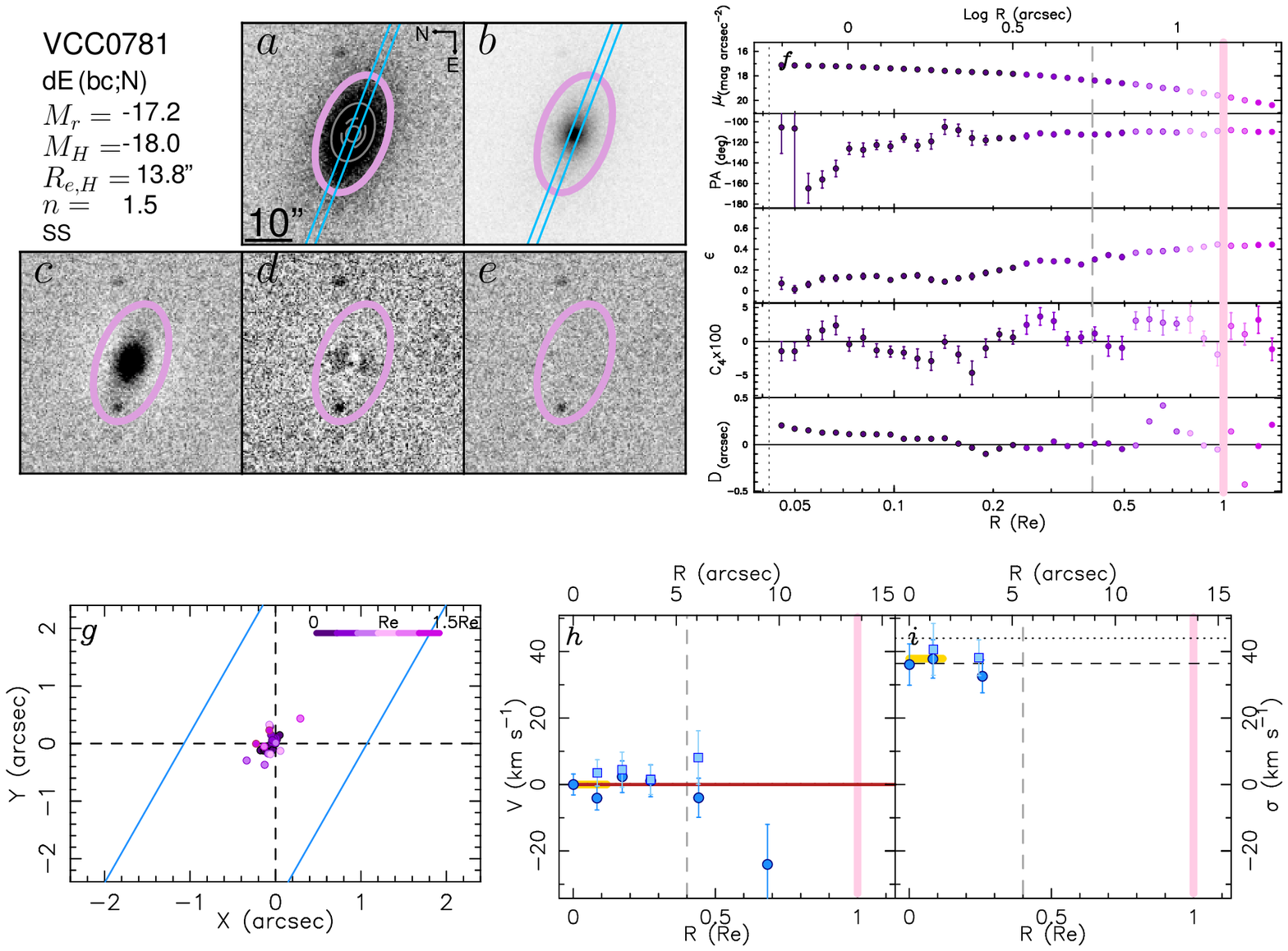}}
\caption{Same as Figure \ref{rotcurve_VCC0009} for VCC~781.}
\end{figure*}

\begin{figure*}
\centering
\resizebox{0.87\textwidth}{!}{\includegraphics[bb= 20 226 563 609,angle=0]{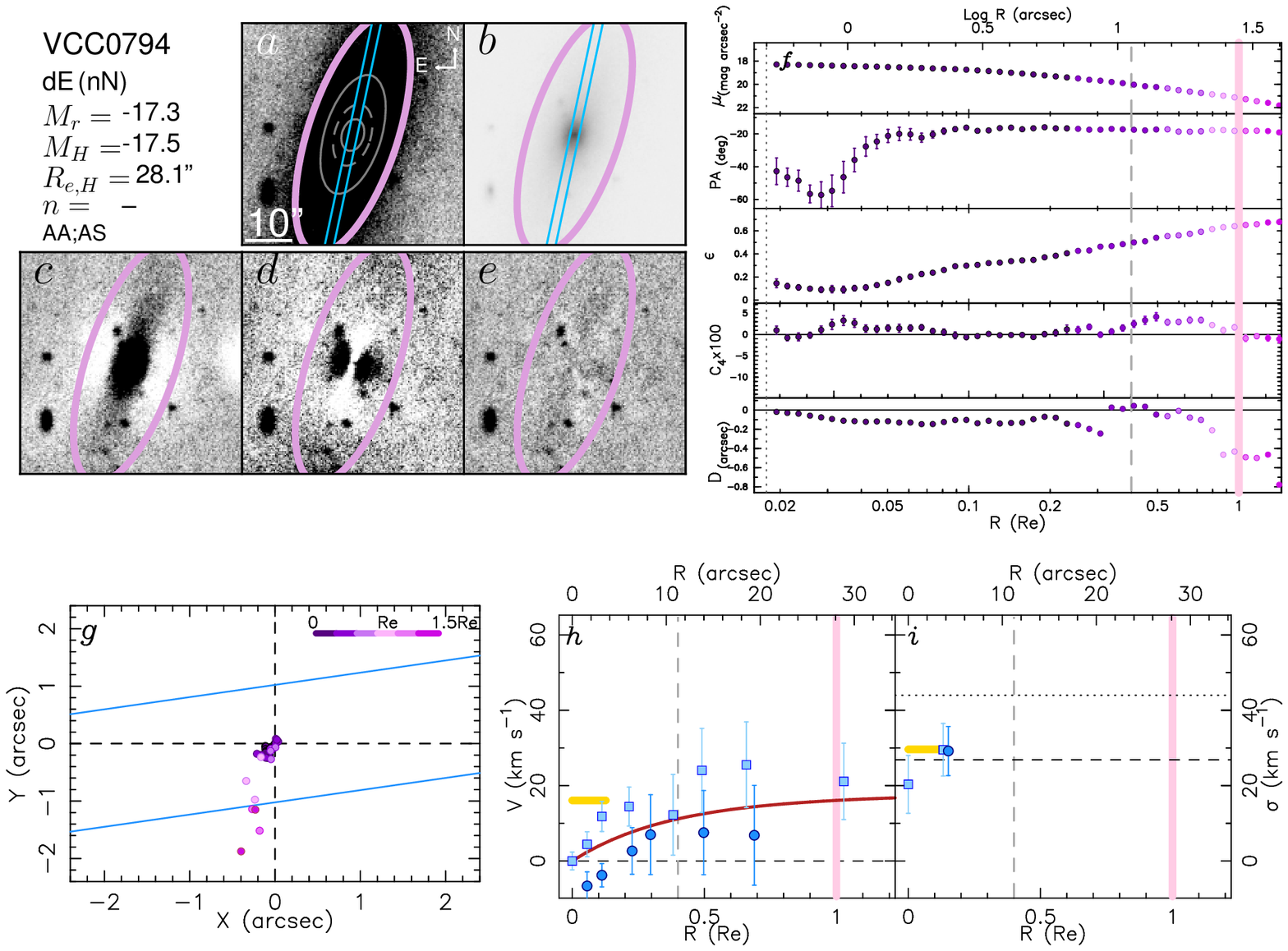}}
\caption{Same as Figure \ref{rotcurve_VCC0009} for VCC~794.}
\end{figure*}

\clearpage

\begin{figure*}
\centering
\resizebox{0.87\textwidth}{!}{\includegraphics[bb= 20 226 563 609,angle=0]{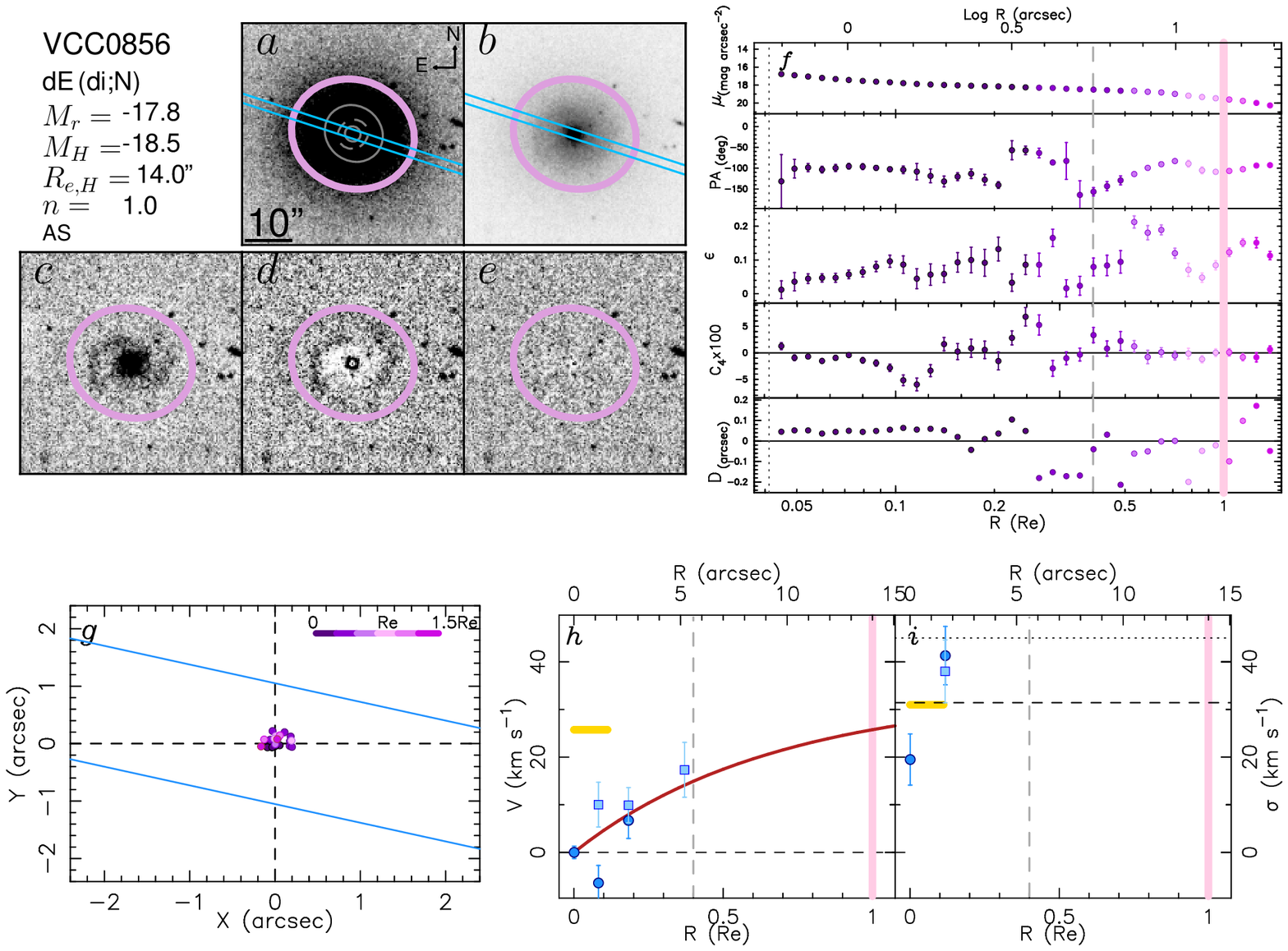}}
\caption{Same as Figure \ref{rotcurve_VCC0009} for VCC~856.}
\end{figure*}

\begin{figure*}
\centering
\resizebox{0.87\textwidth}{!}{\includegraphics[bb= 20 226 563 609,angle=0]{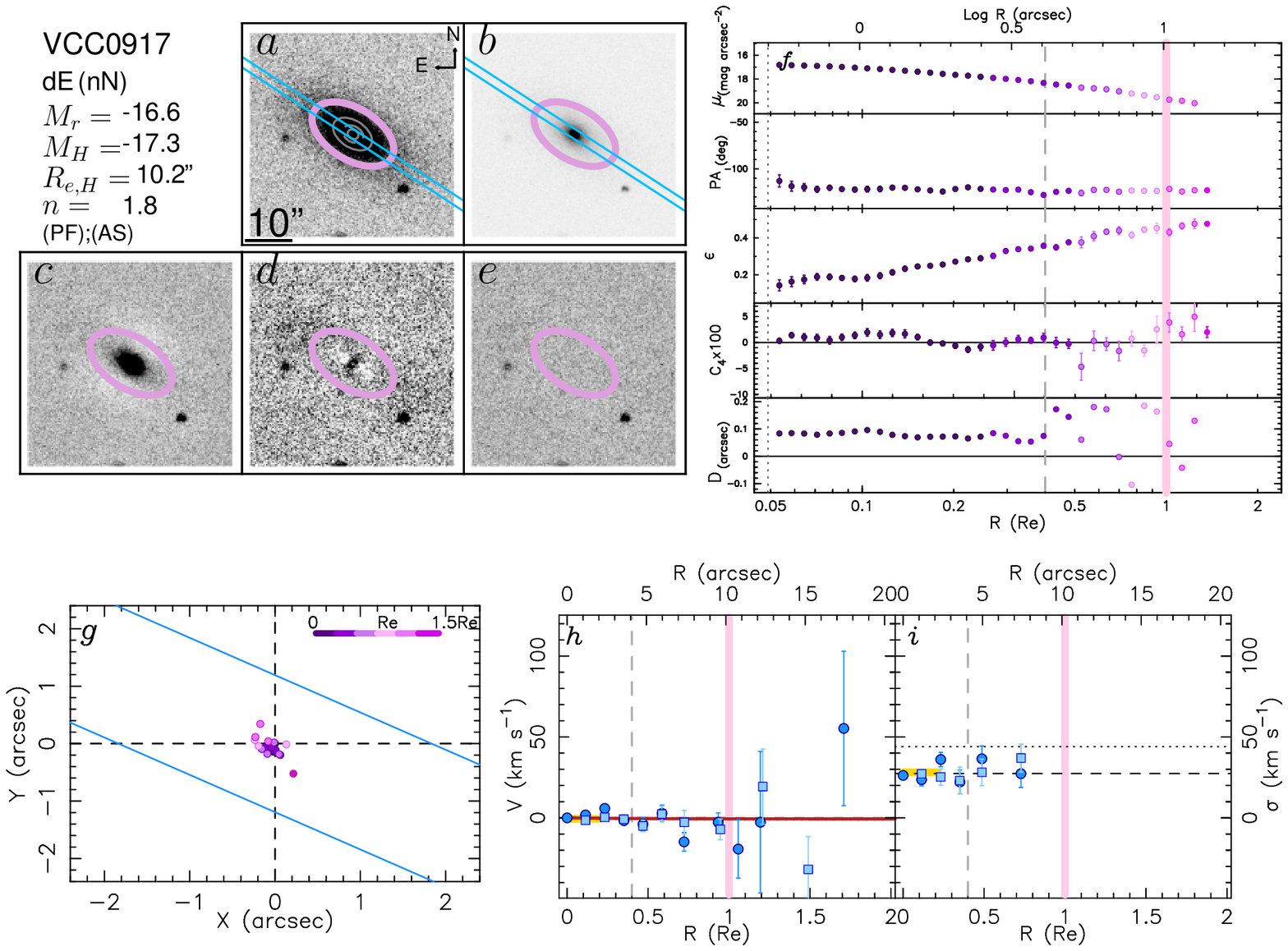}}
\caption{Same as Figure \ref{rotcurve_VCC0009} for VCC~917.}
\end{figure*}

\begin{figure*}
\centering
\resizebox{0.87\textwidth}{!}{\includegraphics[bb= 20 226 563 609,angle=0]{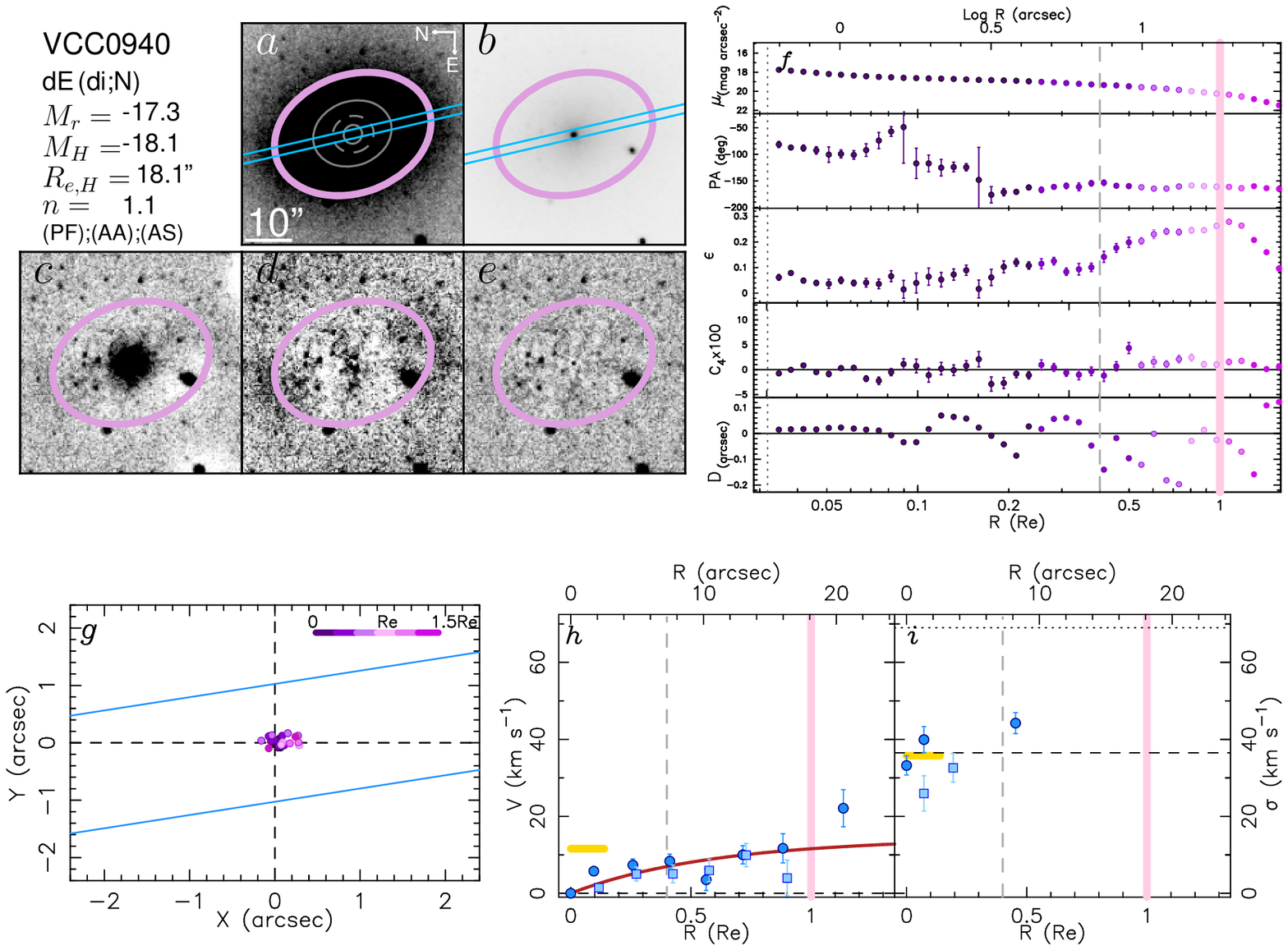}}
\caption{Same as Figure \ref{rotcurve_VCC0009} for VCC~940.}
\end{figure*}

\begin{figure*}
\centering
\resizebox{0.87\textwidth}{!}{\includegraphics[bb= 20 226 563 609,angle=0]{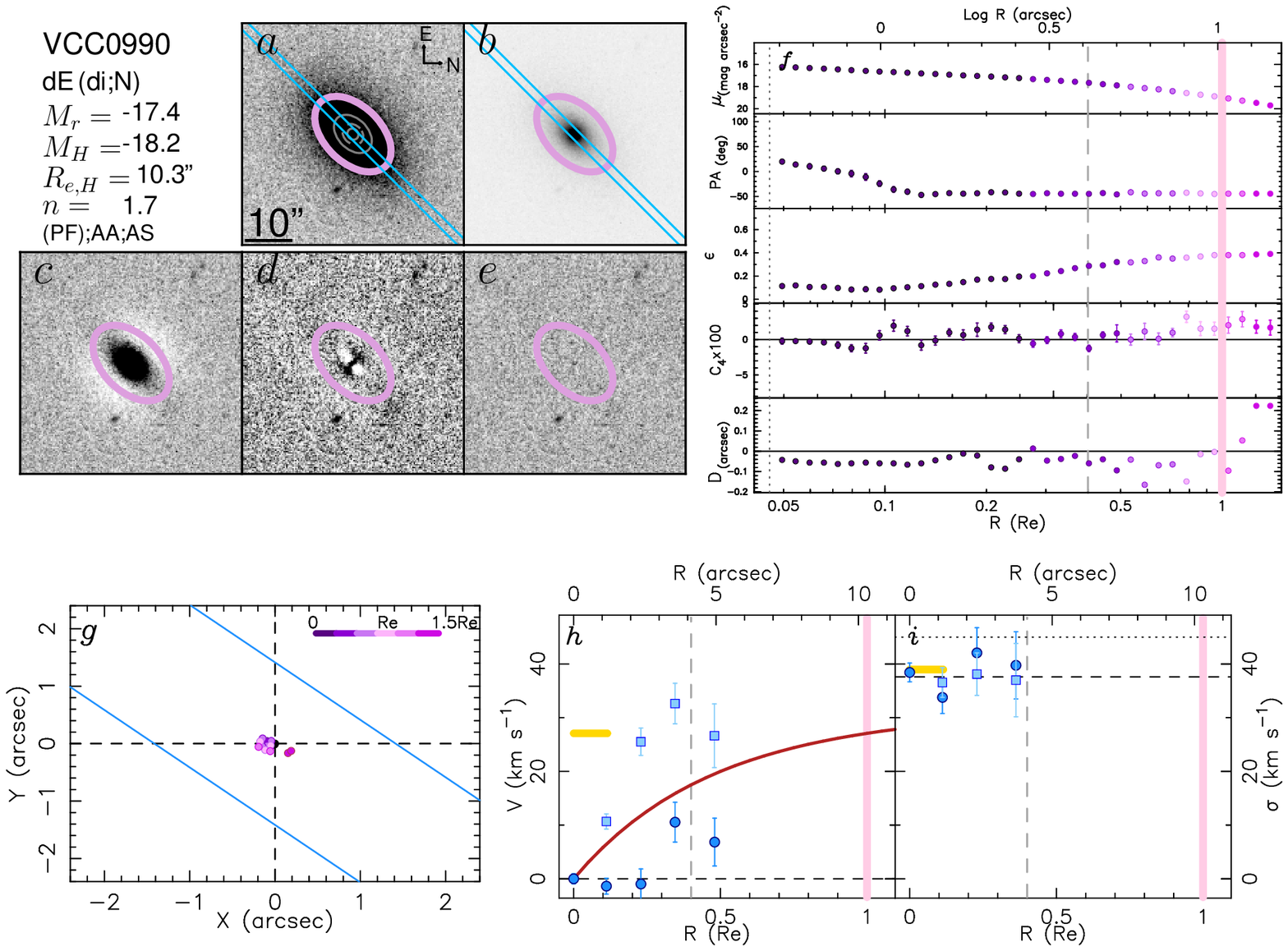}}
\caption{Same as Figure \ref{rotcurve_VCC0009} for VCC~990.}
\end{figure*}

\begin{figure*}
\centering
\resizebox{0.87\textwidth}{!}{\includegraphics[bb= 20 226 563 609,angle=0]{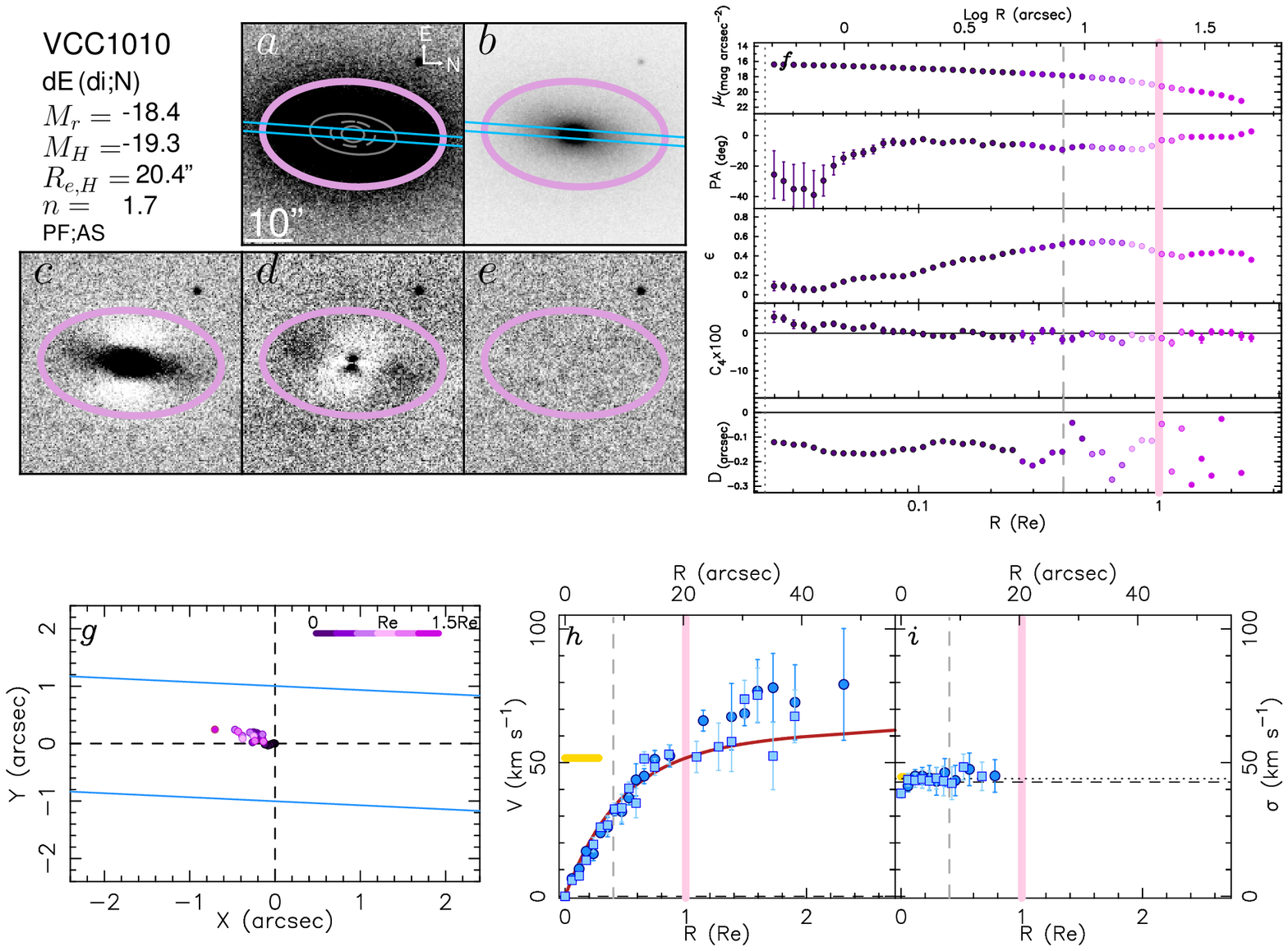}}
\caption{Same as Figure \ref{rotcurve_VCC0009} for VCC~1010.}
\end{figure*}

\begin{figure*}
\centering
\resizebox{0.87\textwidth}{!}{\includegraphics[bb= 20 226 563 609,angle=0]{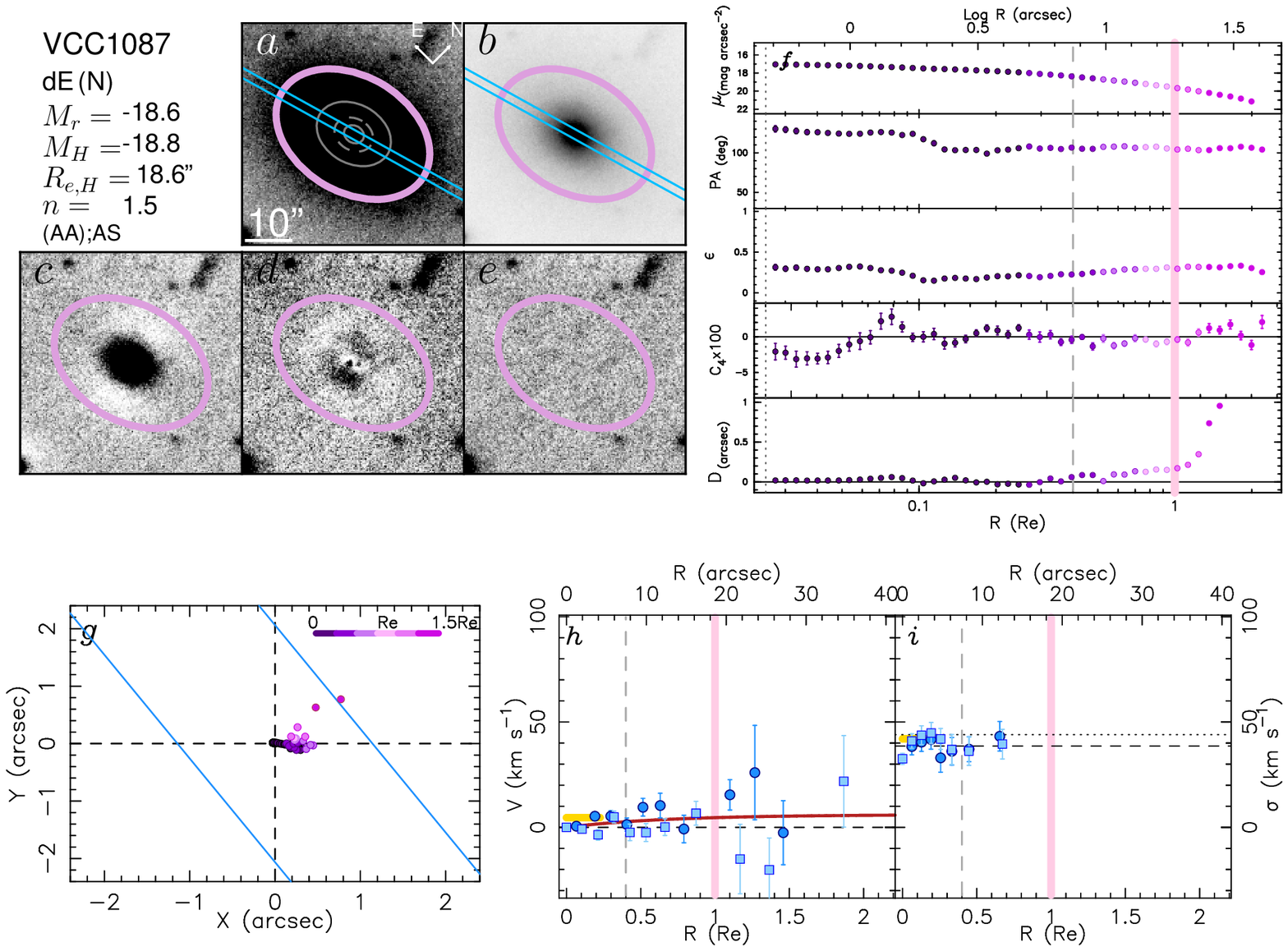}}
\caption{Same as Figure \ref{rotcurve_VCC0009} for VCC~1087.}
\end{figure*}

\begin{figure*}
\centering
\resizebox{0.87\textwidth}{!}{\includegraphics[bb= 20 226 563 609,angle=0]{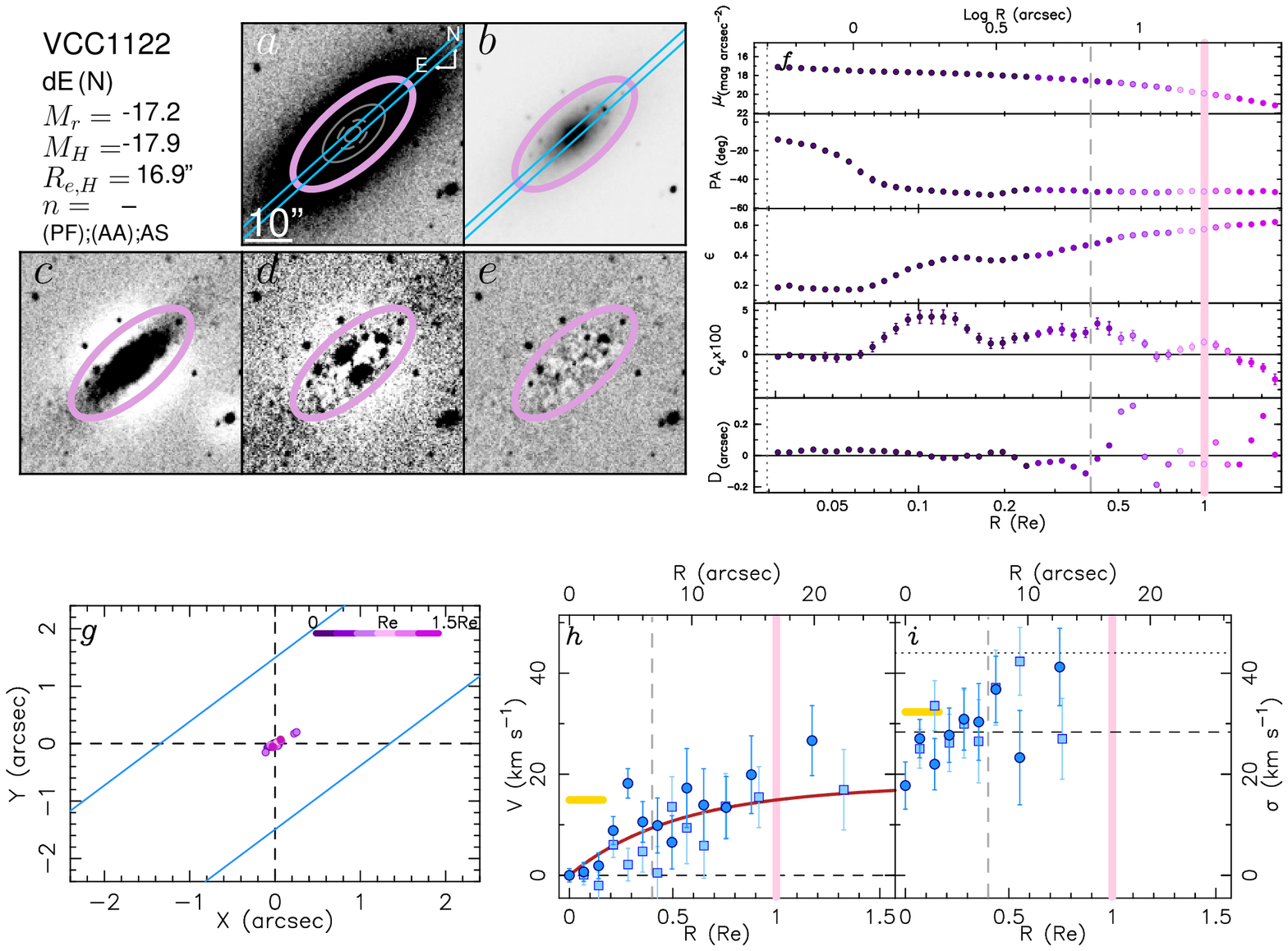}}
\caption{Same as Figure \ref{rotcurve_VCC0009} for VCC~1122.}
\end{figure*}

\begin{figure*}
\centering
\resizebox{0.87\textwidth}{!}{\includegraphics[bb= 20 226 563 609,angle=0]{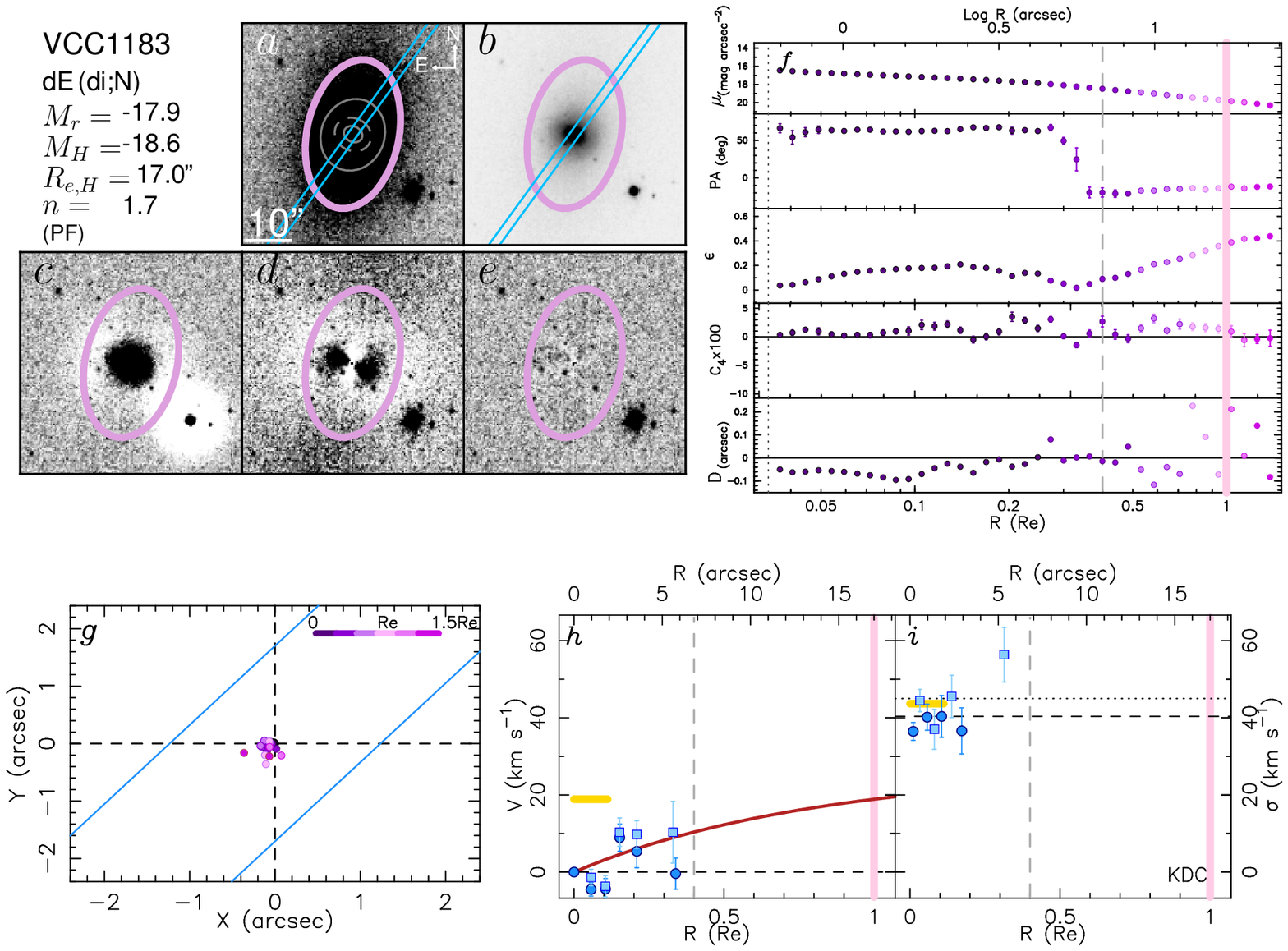}}
\caption{Same as Figure \ref{rotcurve_VCC0009} for VCC~1183. This galaxy has a kinematically decoupled core (see Paper~I).}
\end{figure*}

\begin{figure*}
\centering
\resizebox{0.87\textwidth}{!}{\includegraphics[bb= 20 226 563 609,angle=0]{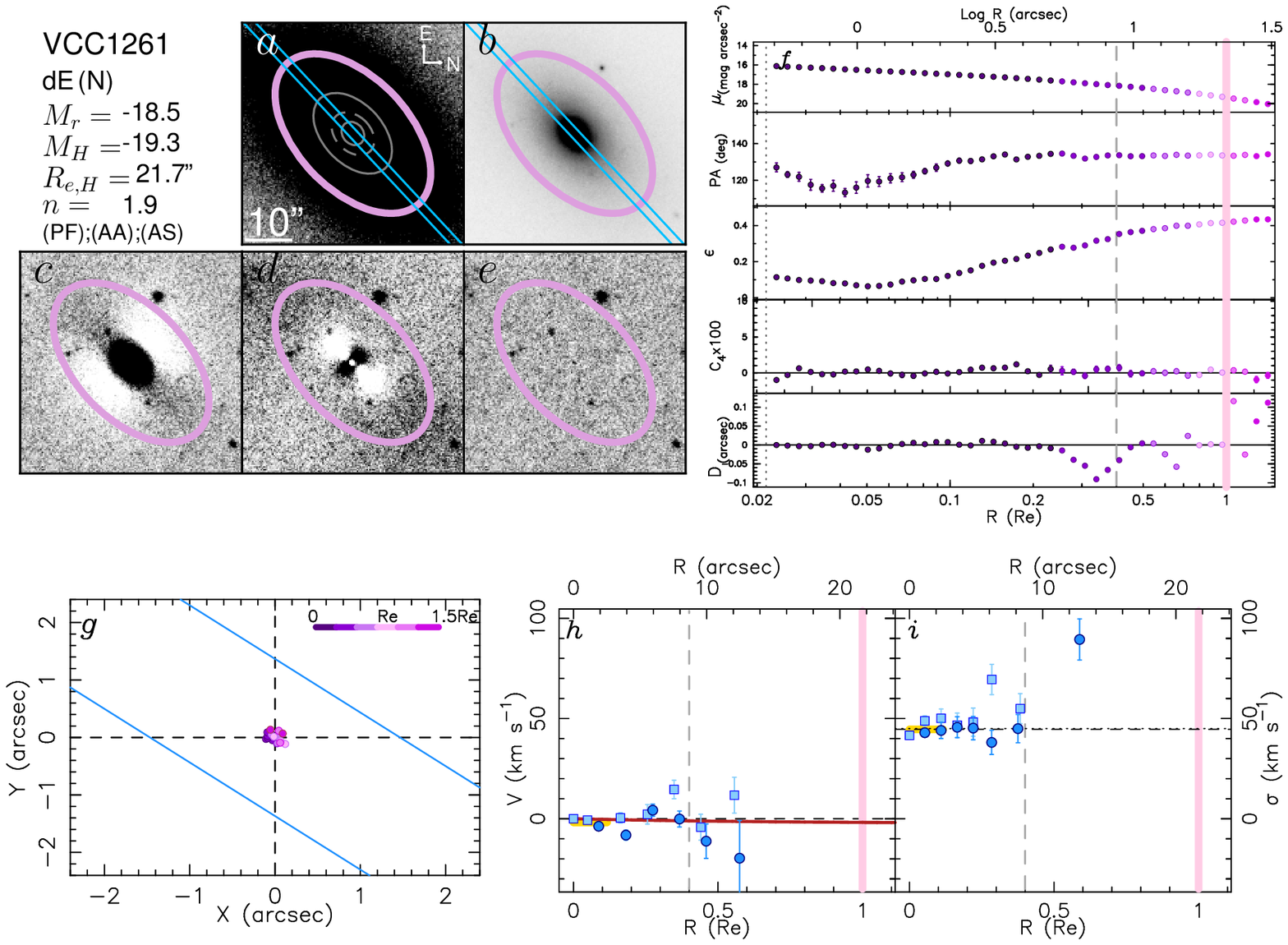}}
\caption{Same as Figure \ref{rotcurve_VCC0009} for VCC~1261.}
\end{figure*}

\begin{figure*}
\centering
\resizebox{0.87\textwidth}{!}{\includegraphics[bb= 20 226 563 609,angle=0]{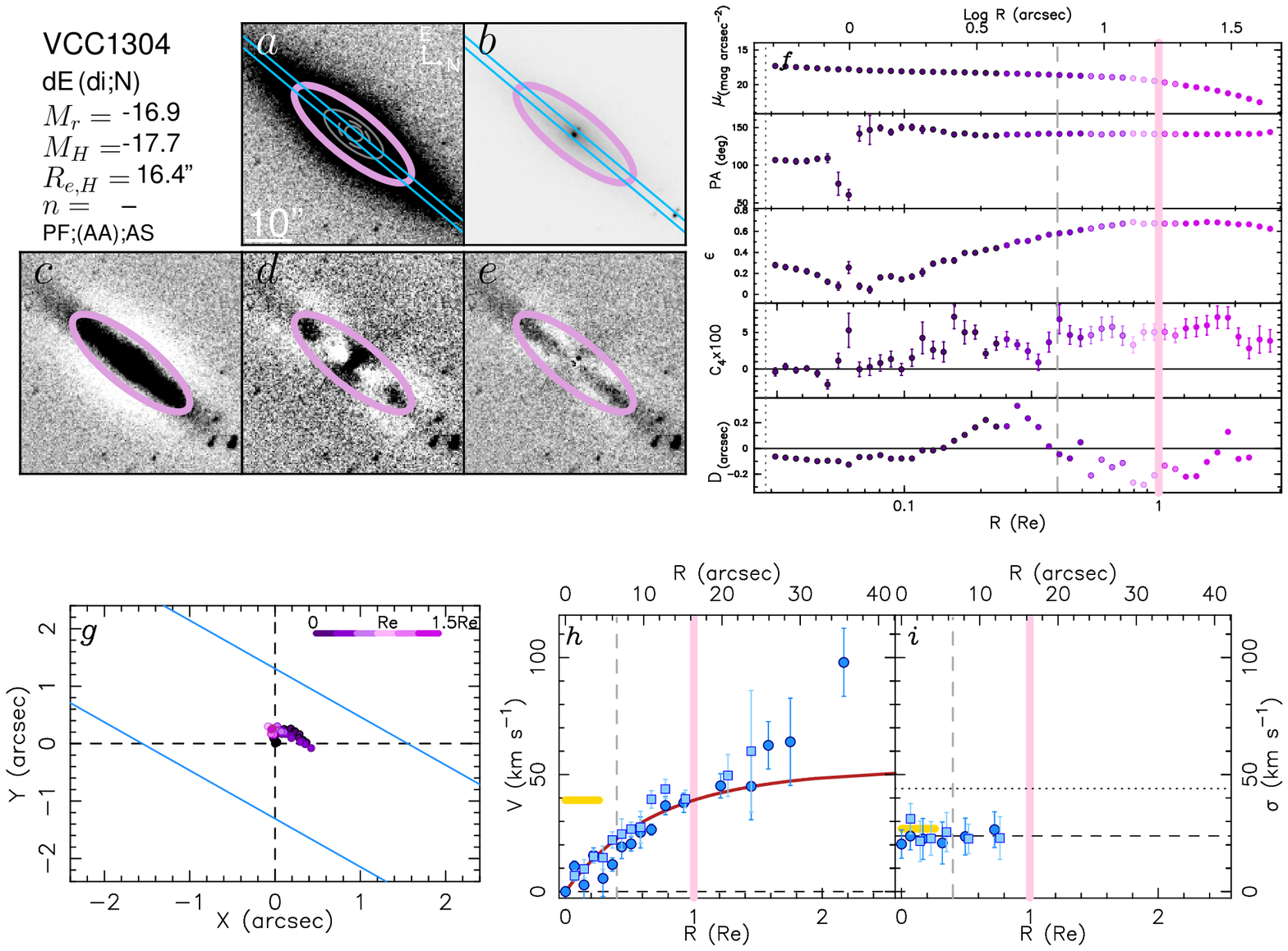}}
\caption{Same as Figure \ref{rotcurve_VCC0009} for VCC~1304.}
\end{figure*}

\begin{figure*}
\centering
\resizebox{0.87\textwidth}{!}{\includegraphics[bb= 20 226 563 609,angle=0]{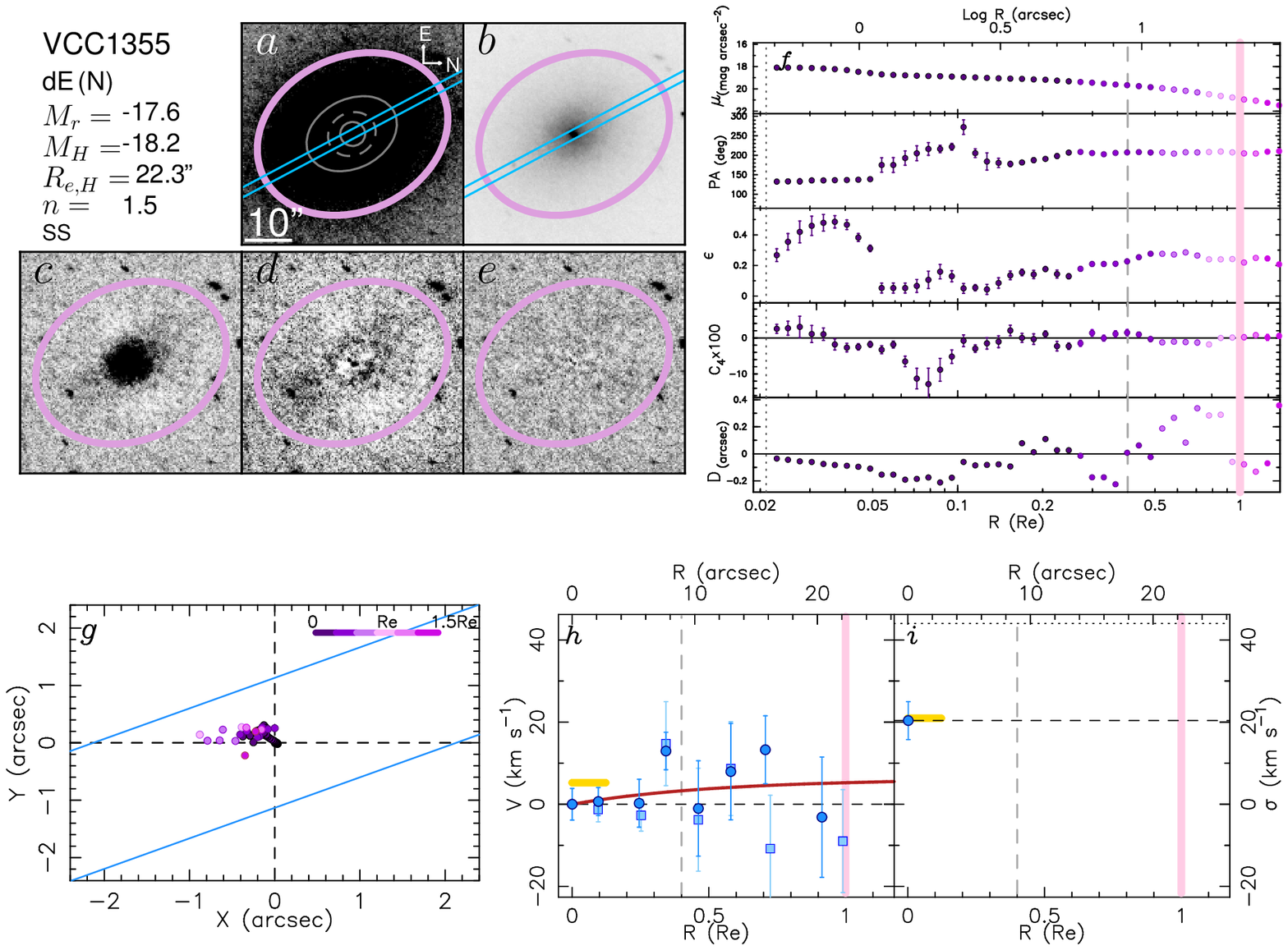}}
\caption{Same as Figure \ref{rotcurve_VCC0009} for VCC~1355.}
\end{figure*}

\begin{figure*}
\centering
\resizebox{0.87\textwidth}{!}{\includegraphics[bb= 20 226 563 609,angle=0]{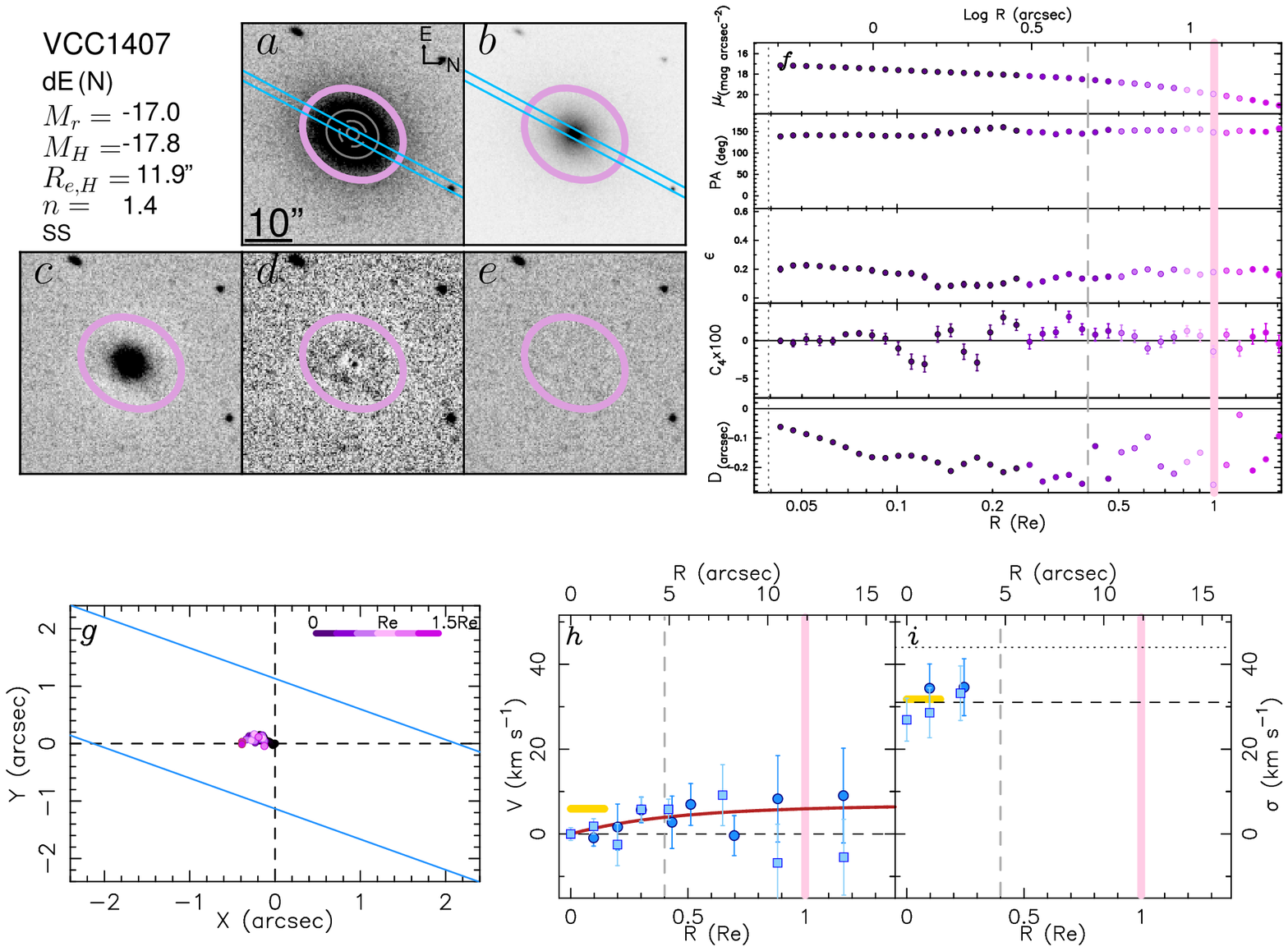}}
\caption{Same as Figure \ref{rotcurve_VCC0009} for VCC~1407.}
\end{figure*}

\newpage

\begin{figure*}
\centering
\resizebox{0.87\textwidth}{!}{\includegraphics[bb= 20 226 563 609,angle=0]{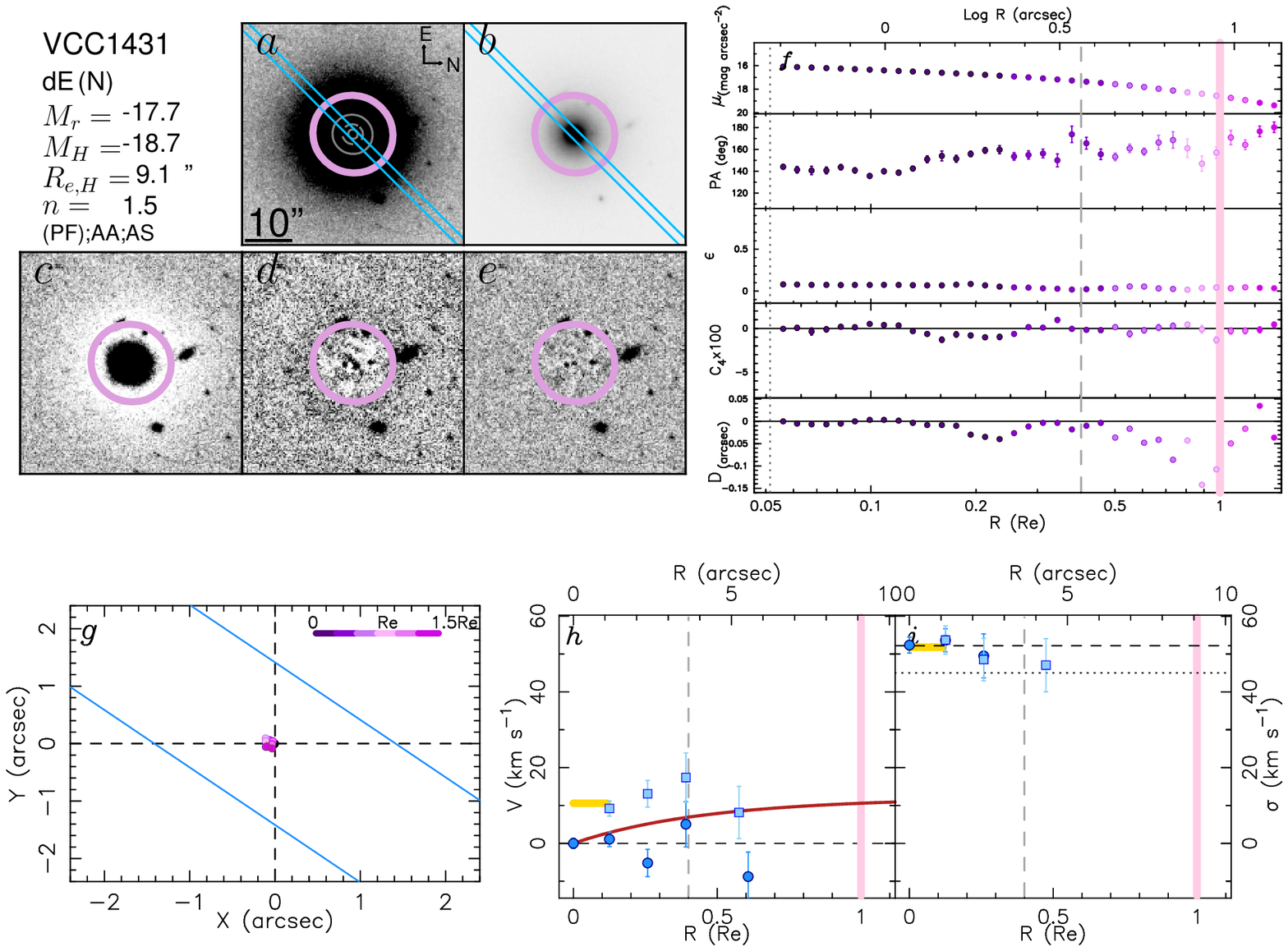}}
\caption{Same as Figure \ref{rotcurve_VCC0009} for VCC~1431.}
\end{figure*}

\begin{figure*}
\centering
\resizebox{0.87\textwidth}{!}{\includegraphics[bb= 20 226 563 609,angle=0]{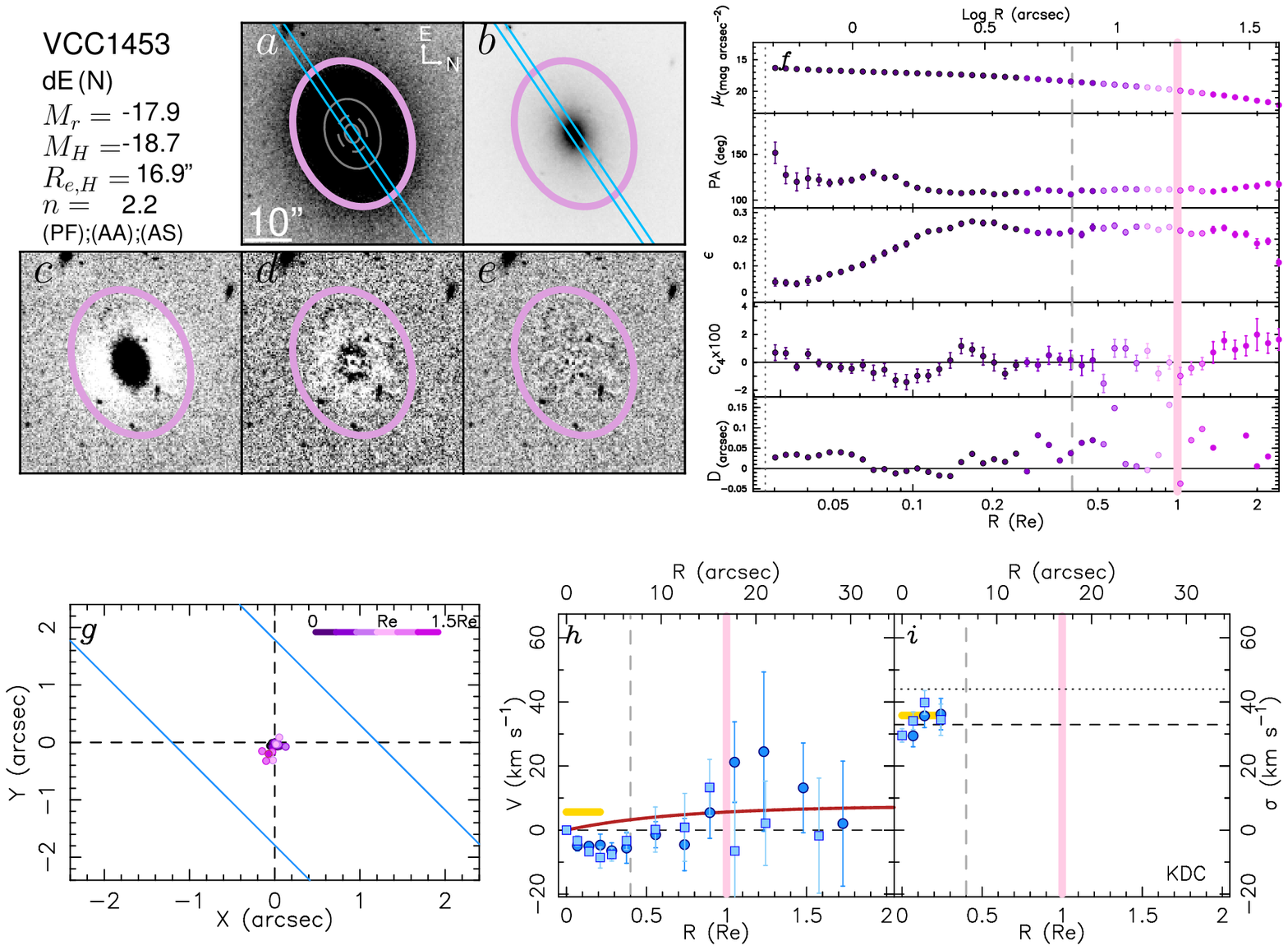}}
\caption{Same as Figure \ref{rotcurve_VCC0009} for VCC~1453. This galaxy has a kinematically decoupled core (see Paper~I).}
\end{figure*}

\begin{figure*}
\centering
\resizebox{0.87\textwidth}{!}{\includegraphics[bb= 20 226 563 609,angle=0]{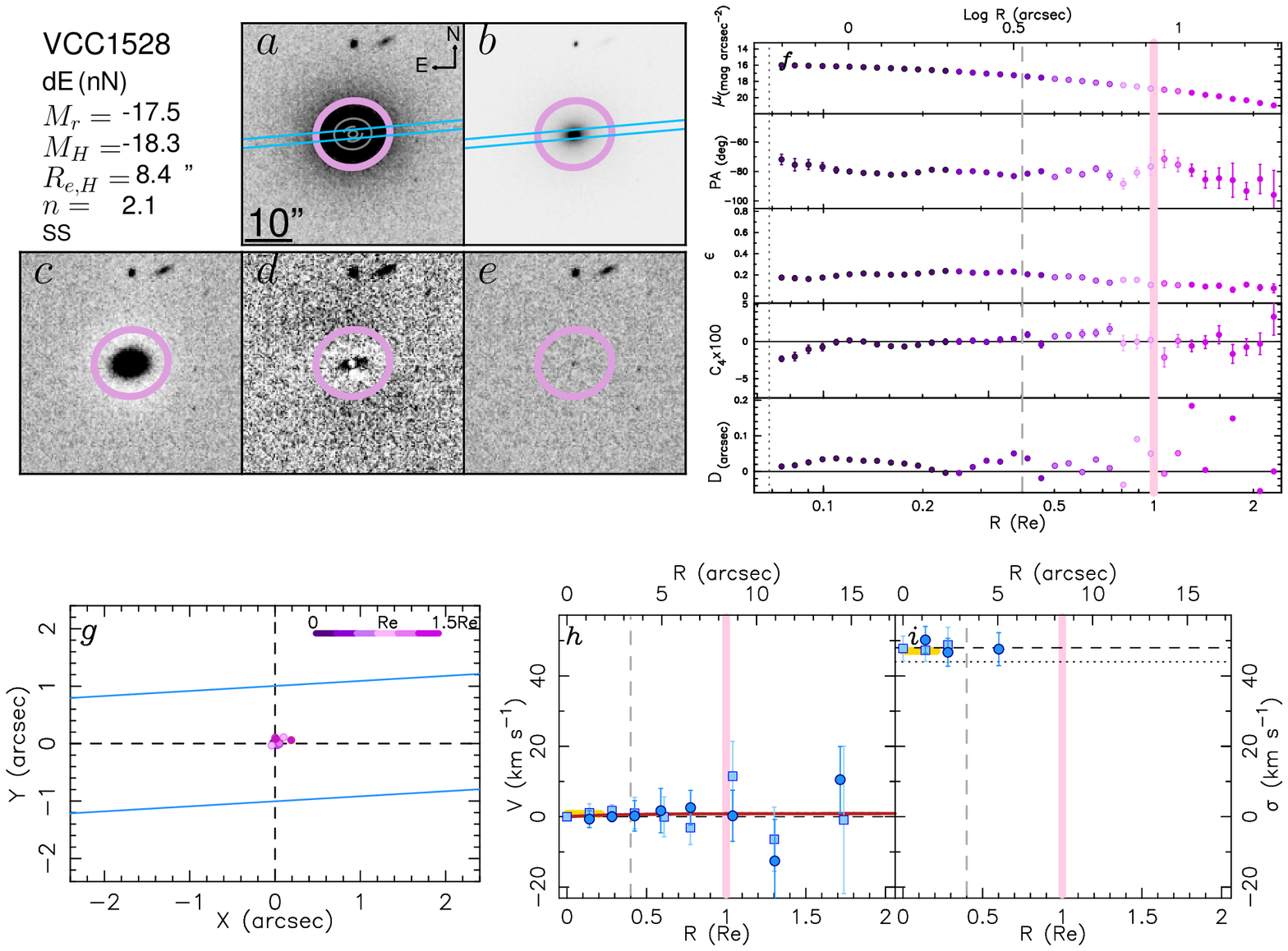}}
\caption{Same as Figure \ref{rotcurve_VCC0009} for VCC~1528.}
\end{figure*}

\begin{figure*}
\centering
\resizebox{0.87\textwidth}{!}{\includegraphics[bb= 20 226 563 609,angle=0]{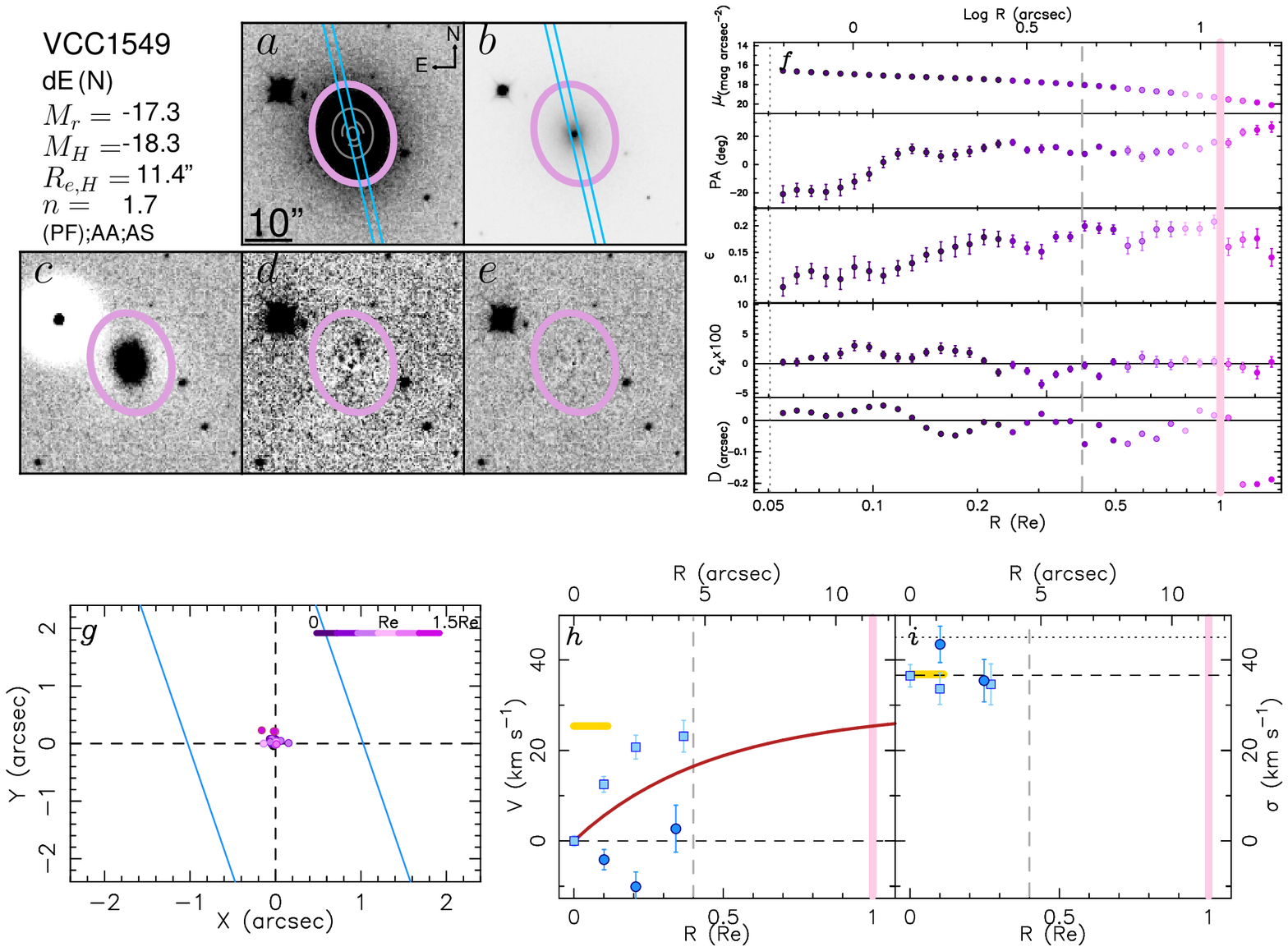}}
\caption{Same as Figure \ref{rotcurve_VCC0009} for VCC~1549.}
\end{figure*}

\begin{figure*}
\centering
\resizebox{0.87\textwidth}{!}{\includegraphics[bb= 20 226 563 609,angle=0]{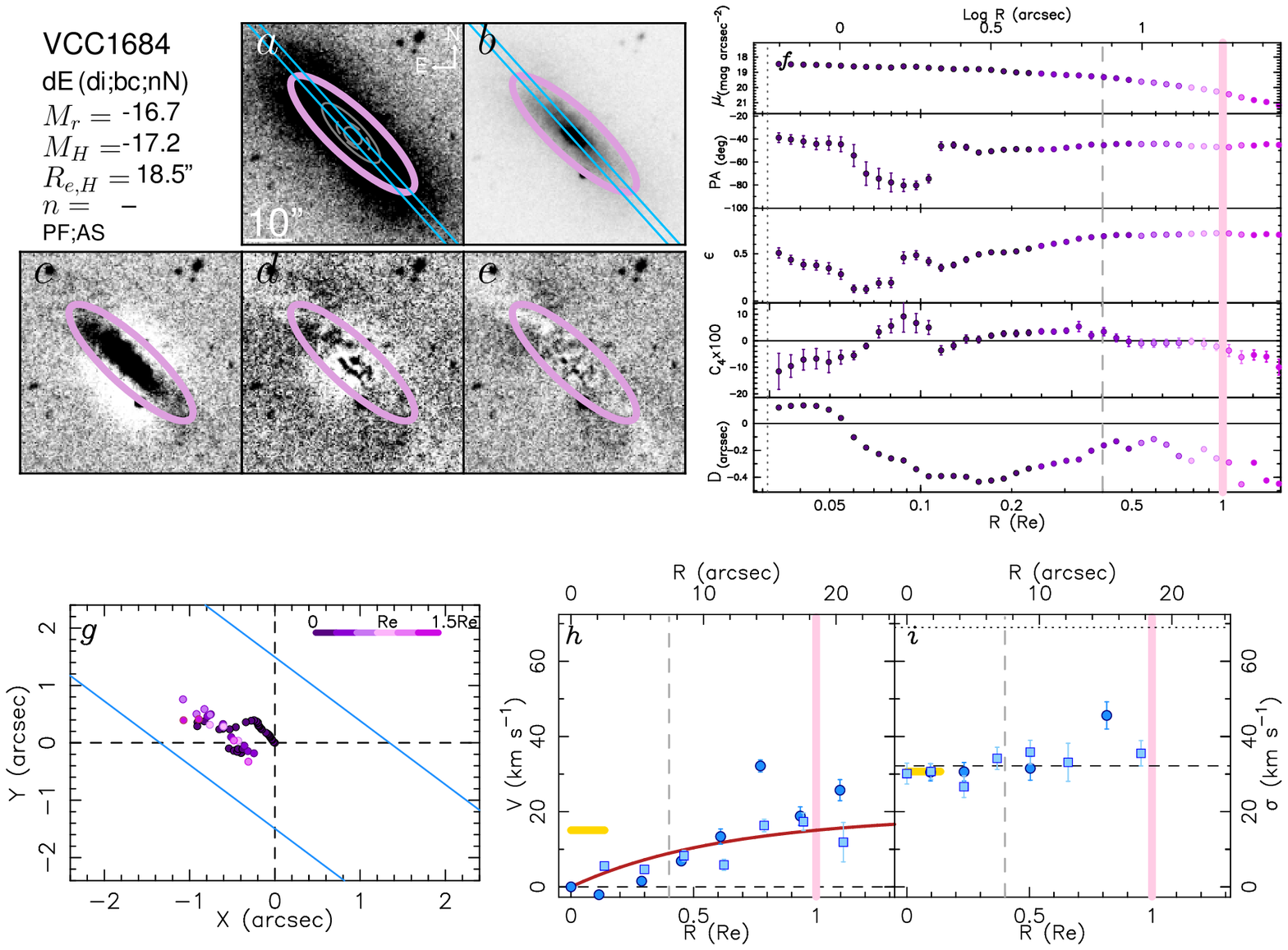}}
\caption{Same as Figure \ref{rotcurve_VCC0009} for VCC~1684.}
\end{figure*}

\begin{figure*}
\centering
\resizebox{0.87\textwidth}{!}{\includegraphics[bb= 20 226 563 609,angle=0]{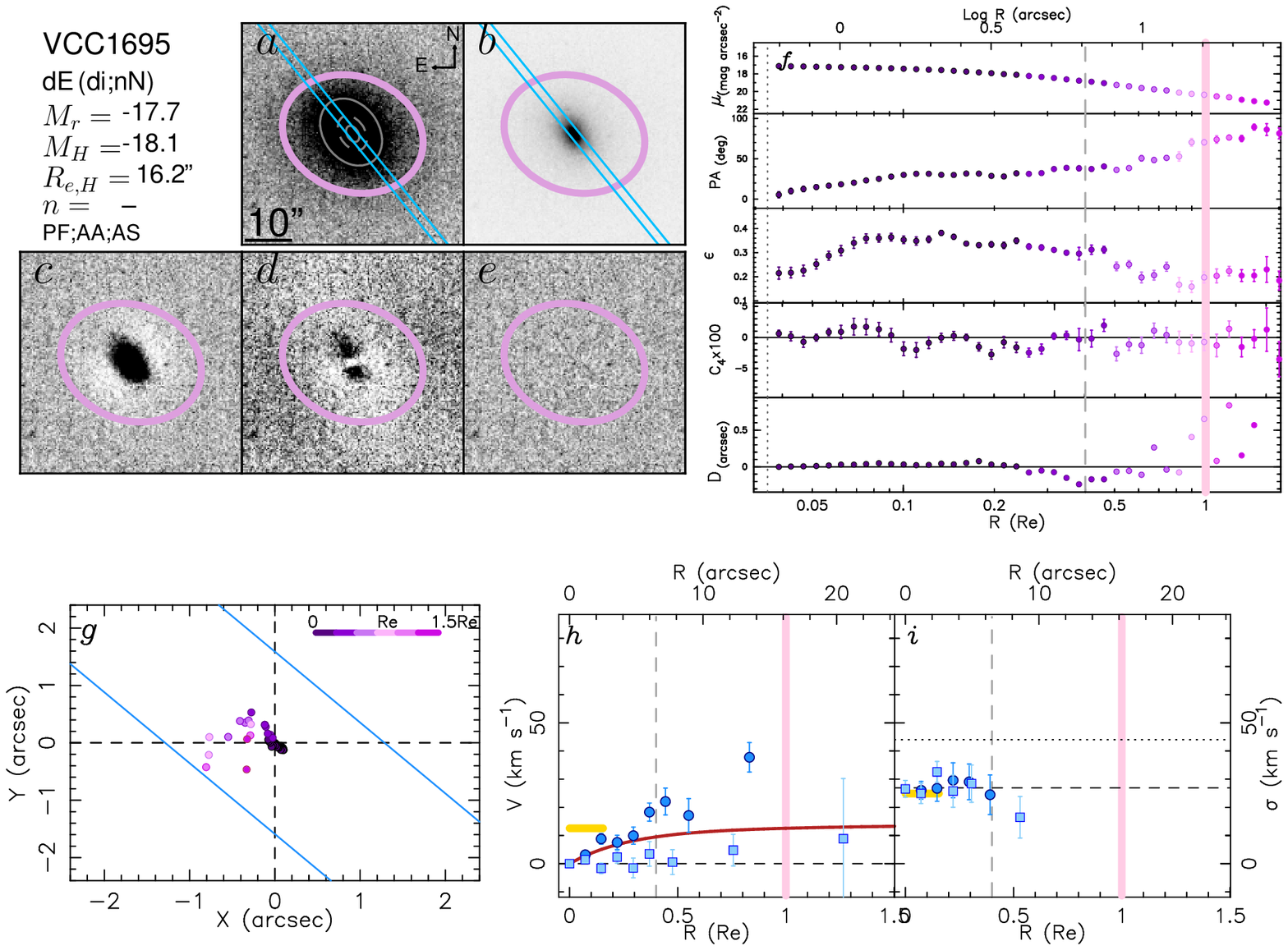}}
\caption{Same as Figure \ref{rotcurve_VCC0009} for VCC~1695.}
\end{figure*}

\clearpage

\begin{figure*}
\centering
\resizebox{0.87\textwidth}{!}{\includegraphics[bb= 20 226 563 609,angle=0]{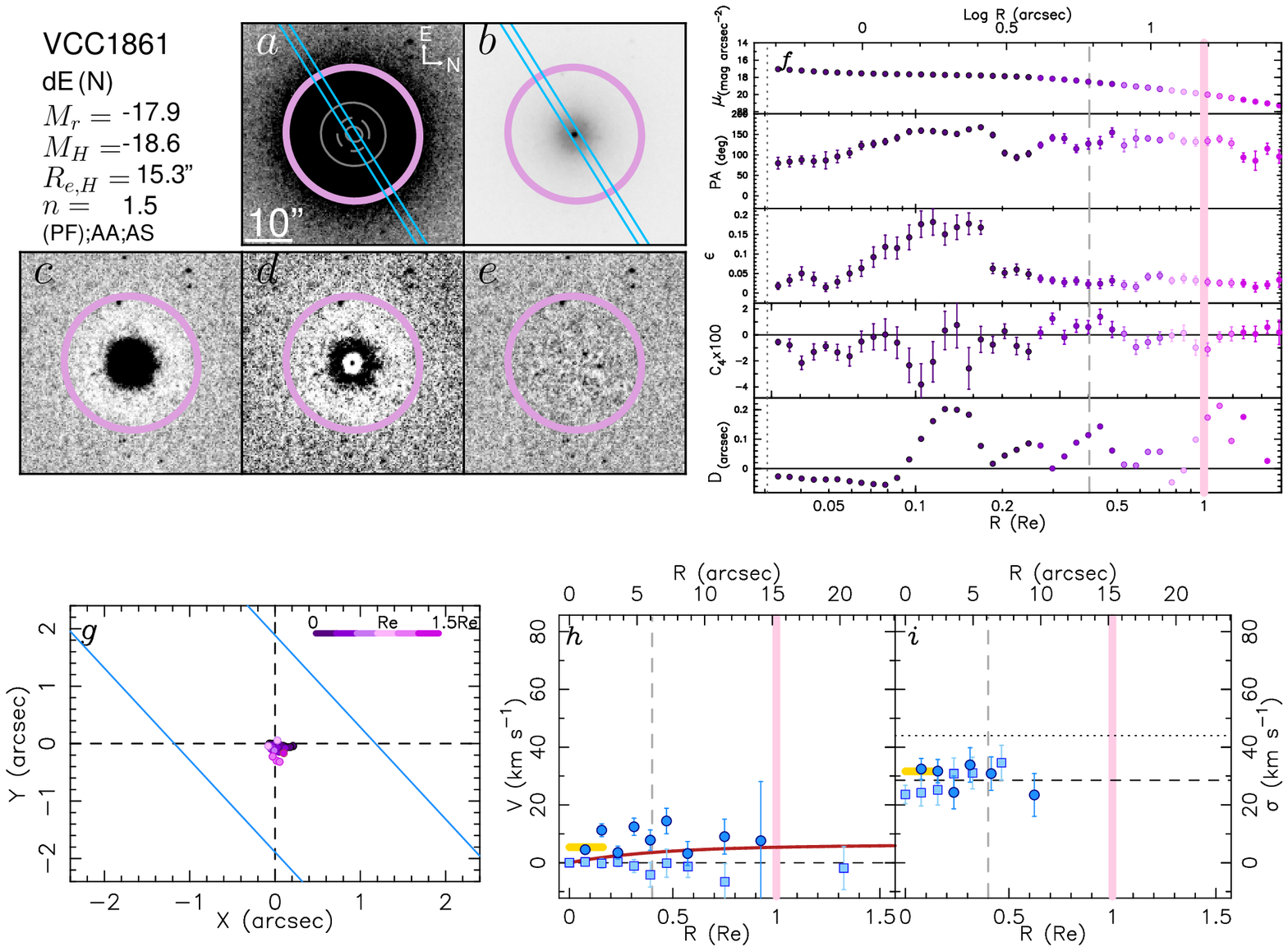}}
\caption{Same as Figure \ref{rotcurve_VCC0009} for VCC~1861.}
\end{figure*}

\begin{figure*}
\centering
\resizebox{0.87\textwidth}{!}{\includegraphics[bb= 20 226 563 609,angle=0]{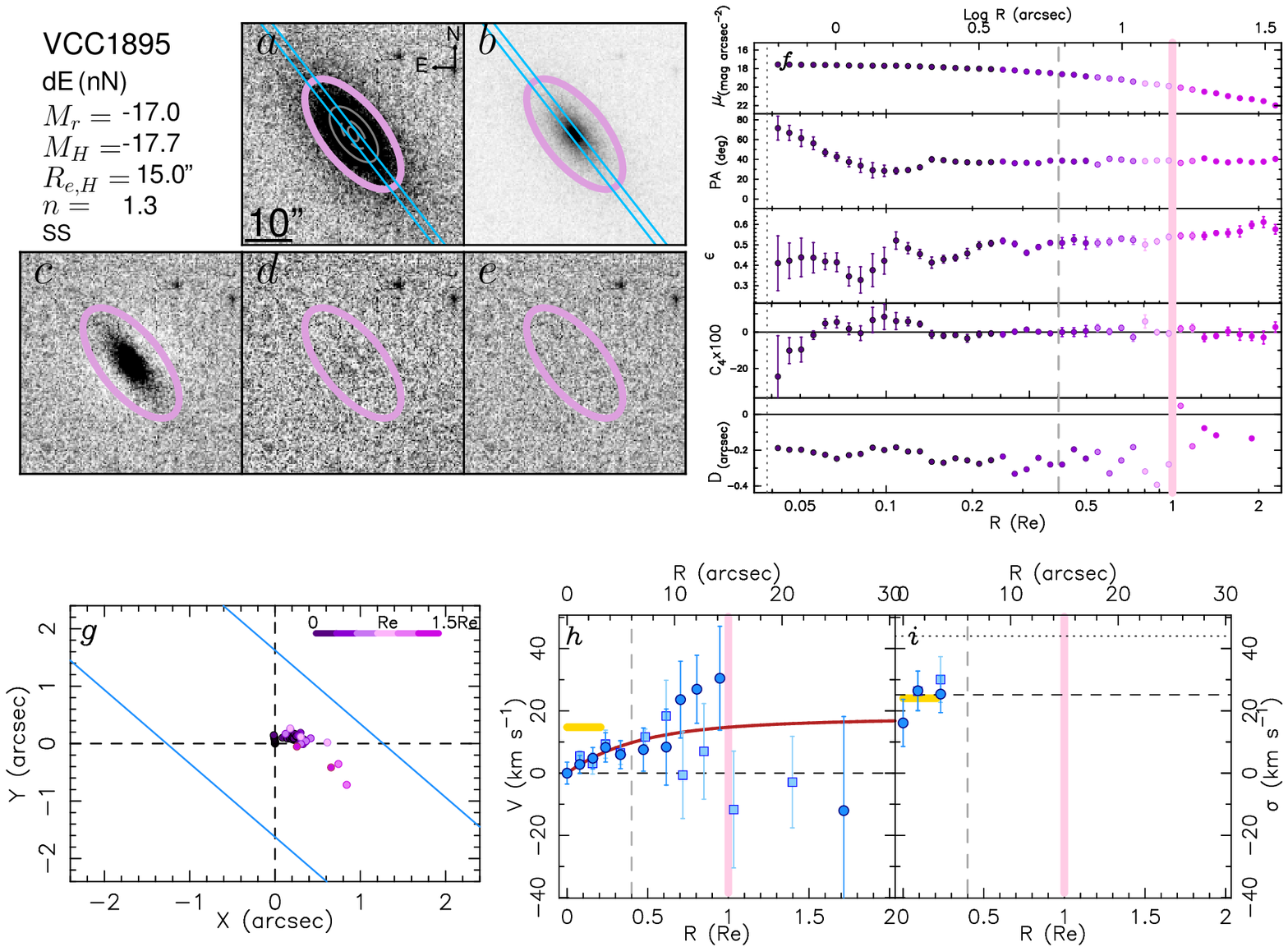}}
\caption{Same as Figure \ref{rotcurve_VCC0009} for VCC~1895.}
\end{figure*}

\begin{figure*}
\centering
\resizebox{0.87\textwidth}{!}{\includegraphics[bb= 20 226 563 609,angle=0]{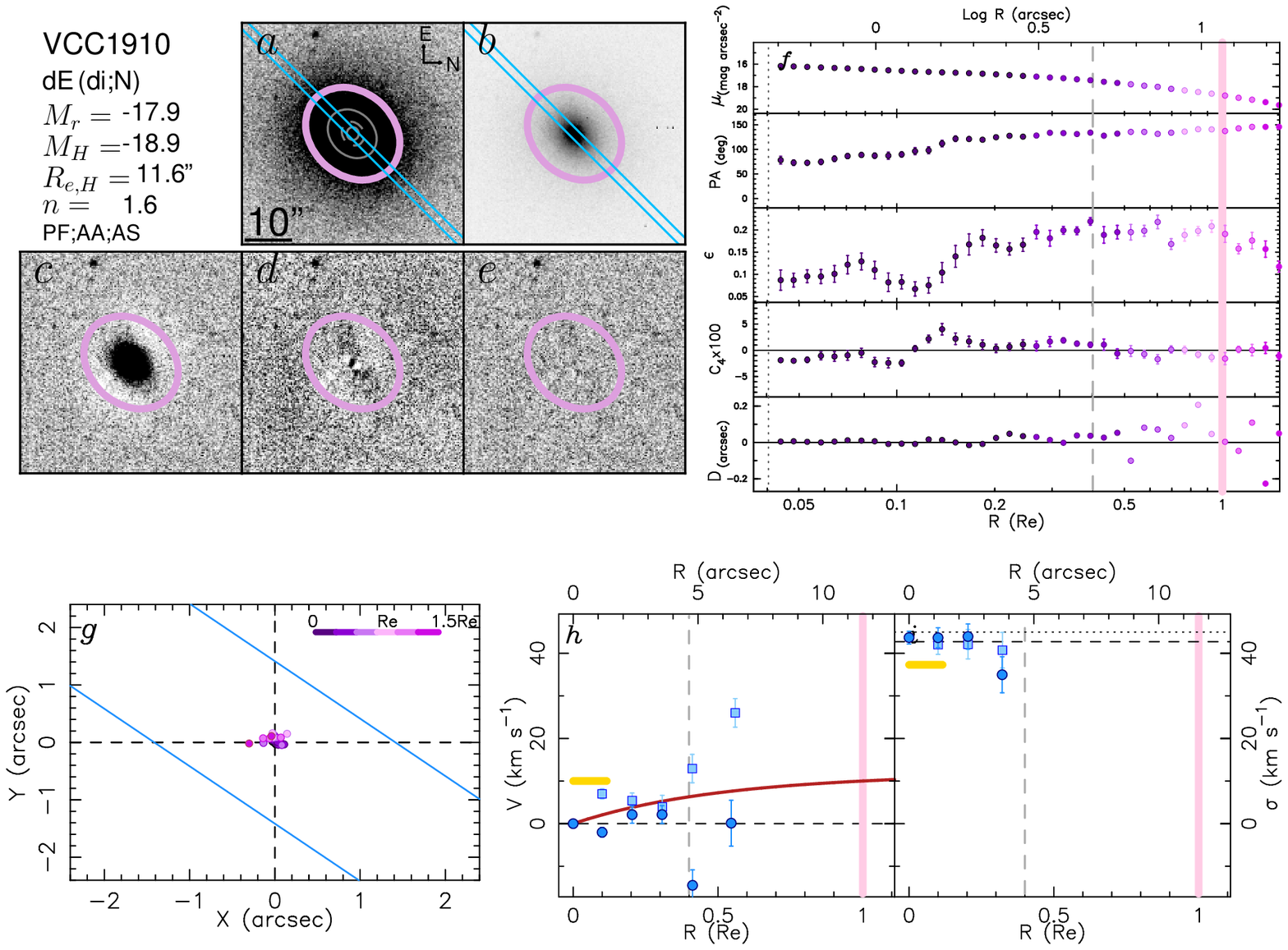}}
\caption{Same as Figure \ref{rotcurve_VCC0009} for VCC~1910.}
\end{figure*}

\begin{figure*}
\centering
\resizebox{0.87\textwidth}{!}{\includegraphics[bb= 20 226 563 609,angle=0]{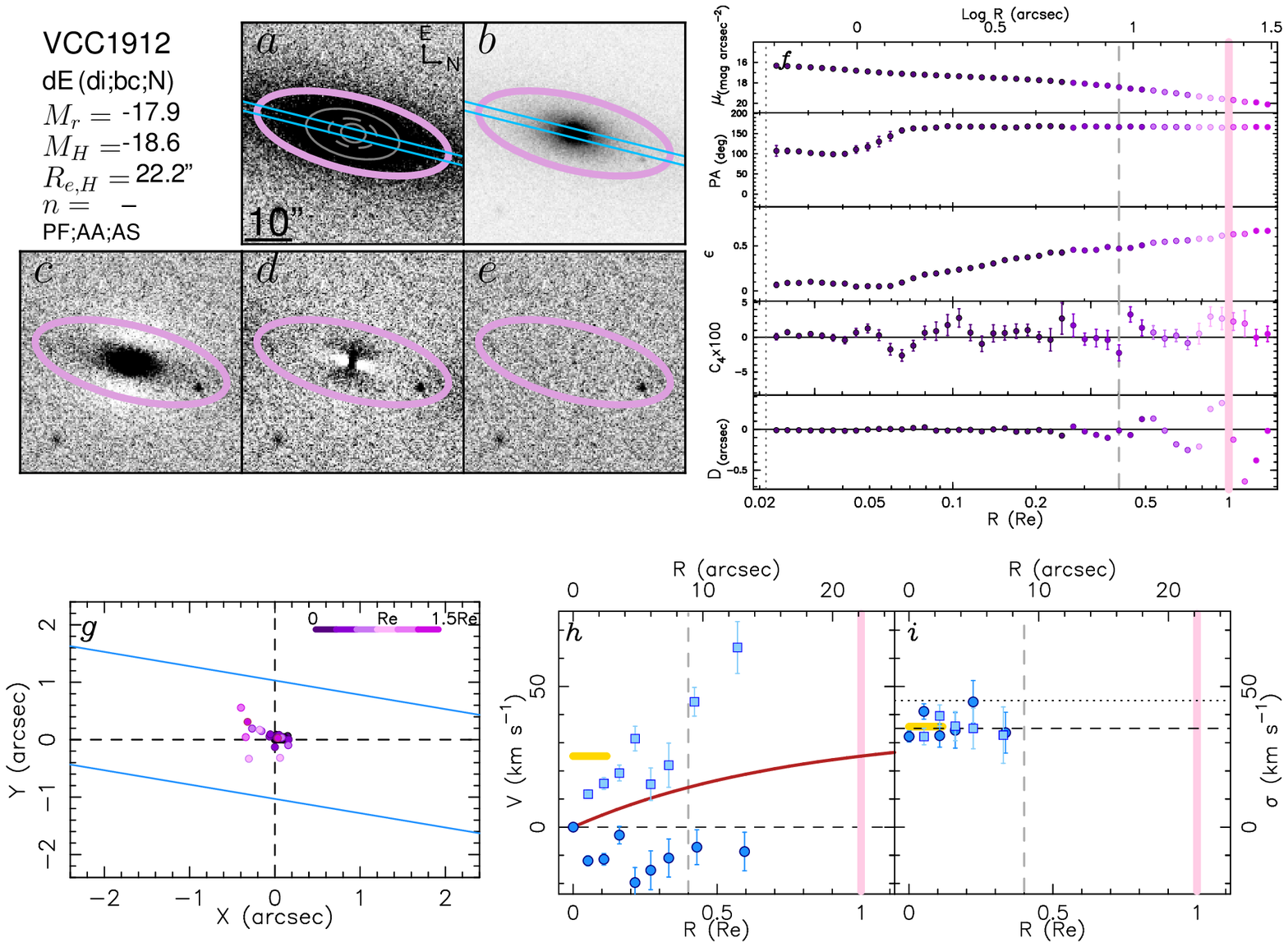}}
\caption{Same as Figure \ref{rotcurve_VCC0009} for VCC~1912.}
\end{figure*}

\begin{figure*}
\centering
\resizebox{0.87\textwidth}{!}{\includegraphics[bb= 20 226 563 609,angle=0]{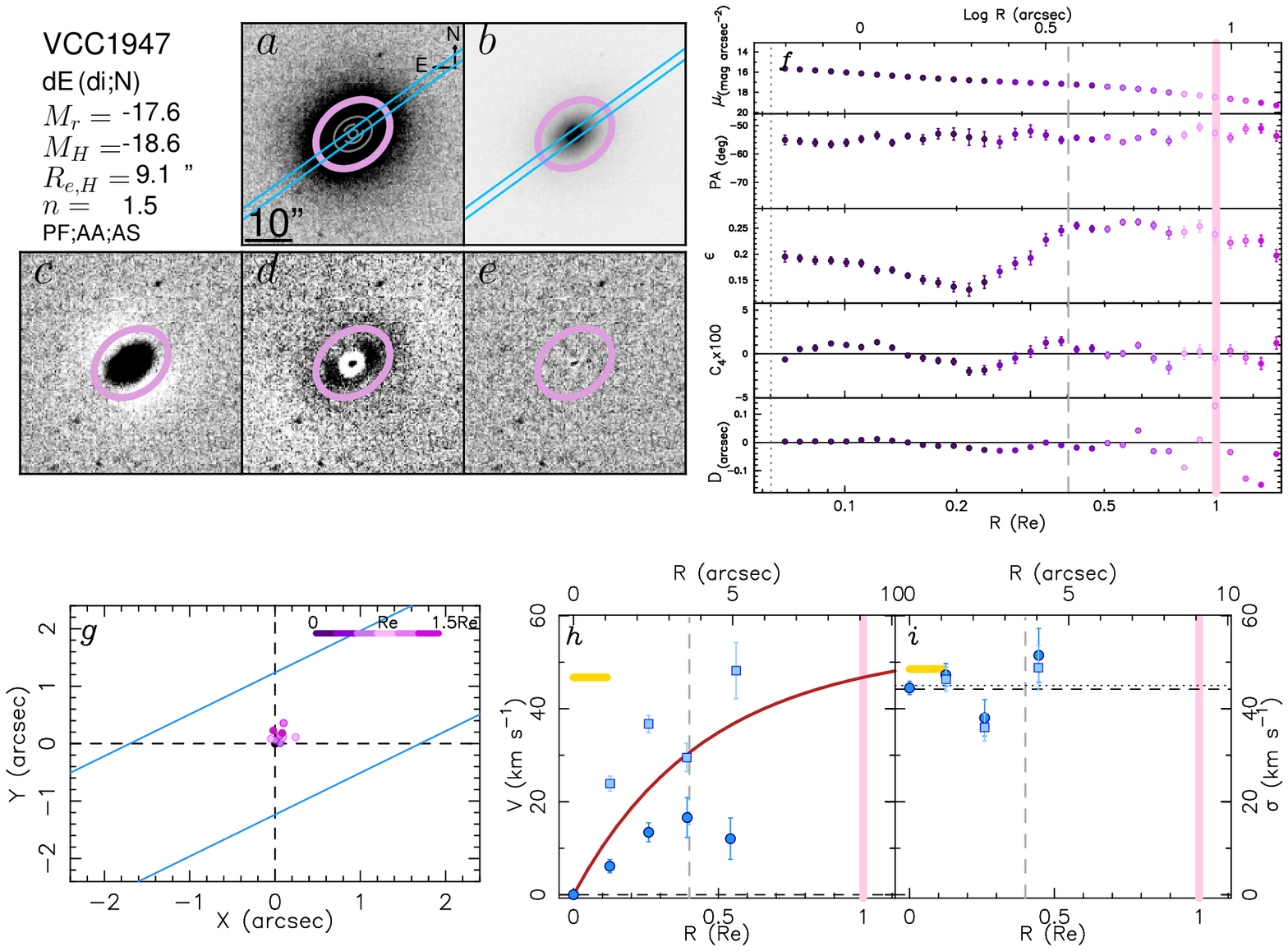}}
\caption{Same as Figure \ref{rotcurve_VCC0009} for VCC~1947.}
\end{figure*}

\begin{figure}
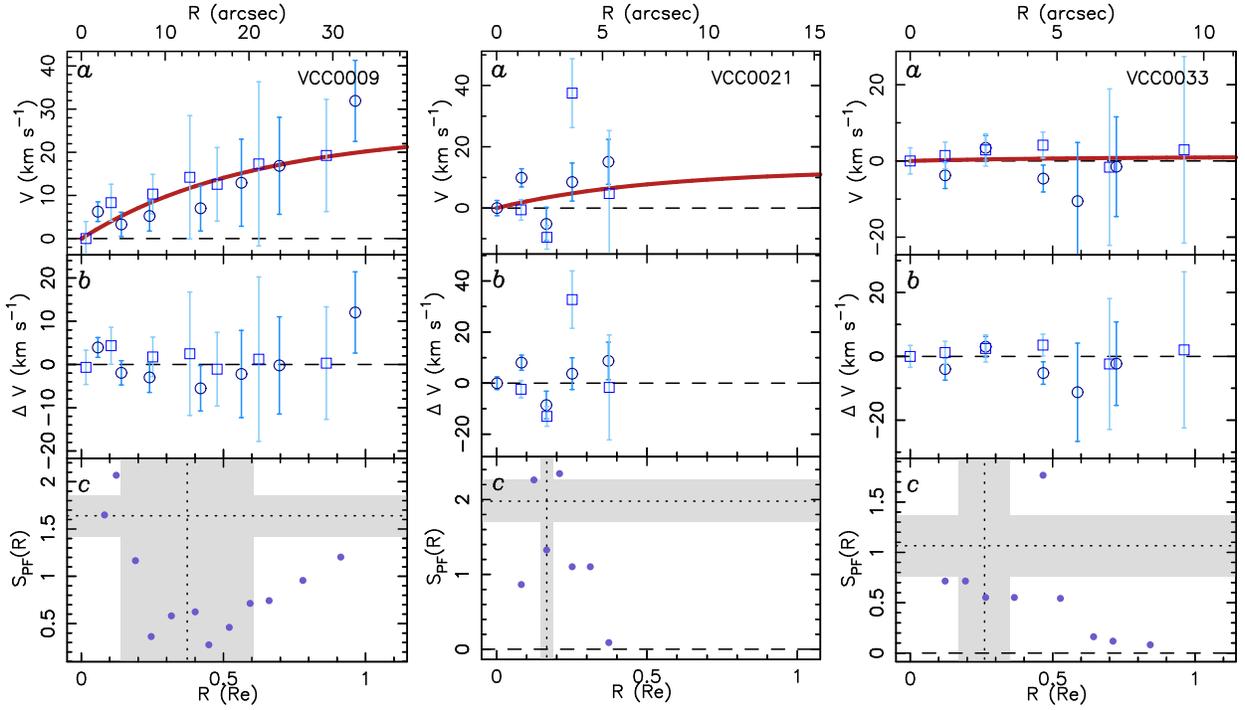

\centering
\resizebox{0.3\textwidth}{!}{\includegraphics[angle=-90,width=8cm]{fig54a.ps}}
\resizebox{0.3\textwidth}{!}{\includegraphics[angle=-90,width=8cm]{fig54b.ps}}
\resizebox{0.3\textwidth}{!}{\includegraphics[angle=-90,width=8cm]{fig54c.ps}}
\caption{Analysis of the shapes of the rotation curves of VCC~9,
  VCC~21, and VCC~33. {\bf Panel a:} Rotation curve and best fit {\it Polyex}
  function of panel {\it h} of Figures
  \ref{rotcurve_VCC0009}-\ref{rotcurve_VCC2083} for reference. {\bf
    Panel b:} Difference between the measured $V$ and the best fit {\it Polyex} function for the approaching (light blue squares) and receding (dark blue dots) sides of the rotation curve, respectively. {\bf Panel c:} Statistical parameter $S_{\rm PF}(R)$ used to quantify the significance of the anomalies in the rotation curves. The horizontal and vertical dotted lines and grey regions indicate the values for $\langle S_{{\rm max}}\rangle$, $\langle R_{S{\rm max}}\rangle$, and their uncertainties, respectively.}
             \label{Sstat_plots_VCC0009}
\end{figure}

\begin{figure}
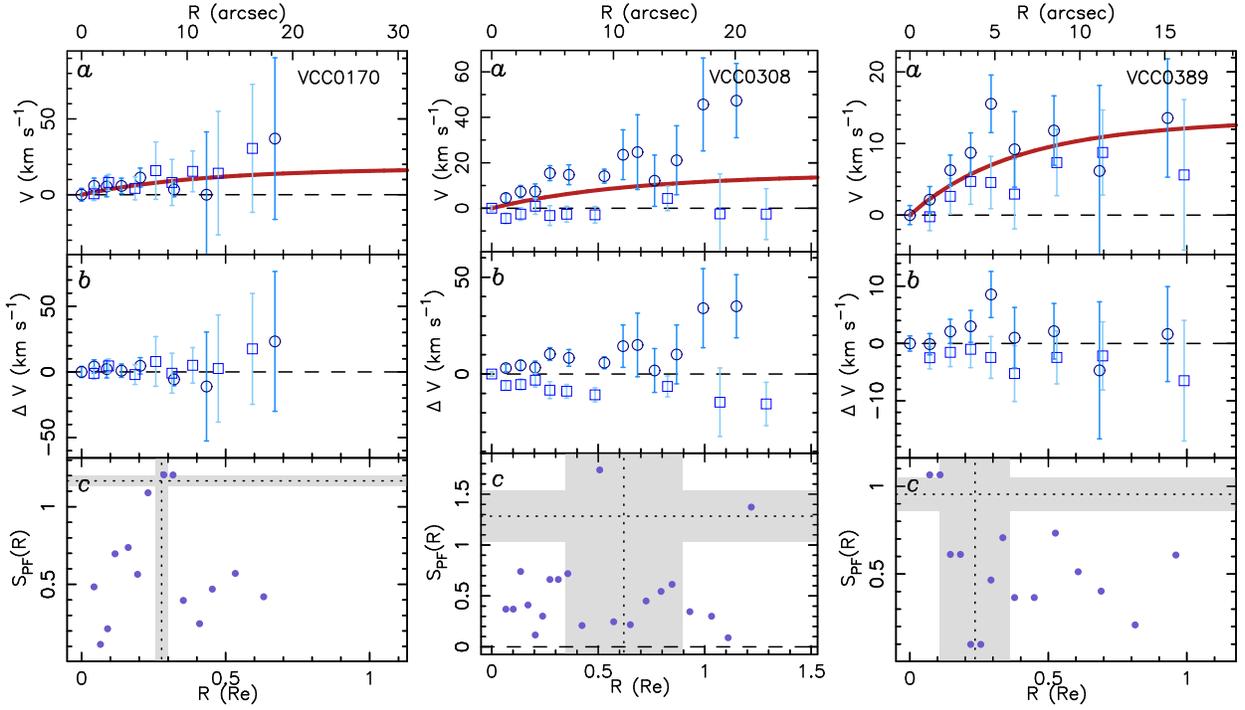

\centering
\resizebox{0.3\textwidth}{!}{\includegraphics[angle=-90,width=8cm]{fig55a.ps}}
\resizebox{0.3\textwidth}{!}{\includegraphics[angle=-90,width=8cm]{fig55b.ps}}
\resizebox{0.3\textwidth}{!}{\includegraphics[angle=-90,width=8cm]{fig55c.ps}}
\caption{Same as Figure \ref{Sstat_plots_VCC0009} for VCC~170, VCC~308, and VCC~389.}
\end{figure}

\begin{figure}
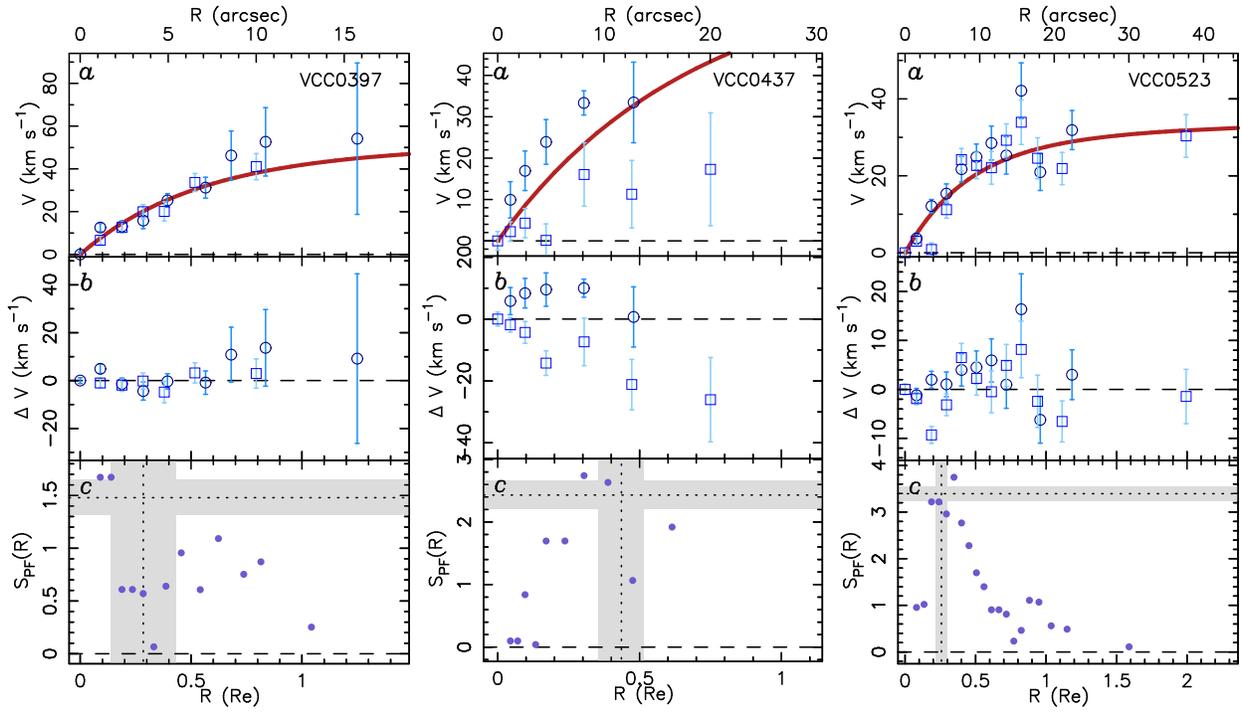

\centering
\resizebox{0.3\textwidth}{!}{\includegraphics[angle=-90,width=8cm]{fig56a.ps}}
\resizebox{0.3\textwidth}{!}{\includegraphics[angle=-90,width=8cm]{fig56b.ps}}
\resizebox{0.3\textwidth}{!}{\includegraphics[angle=-90,width=8cm]{fig56c.ps}}
\caption{Same as Figure \ref{Sstat_plots_VCC0009} for VCC~397, VCC~437, and VCC~523.}
\end{figure}

\begin{figure}
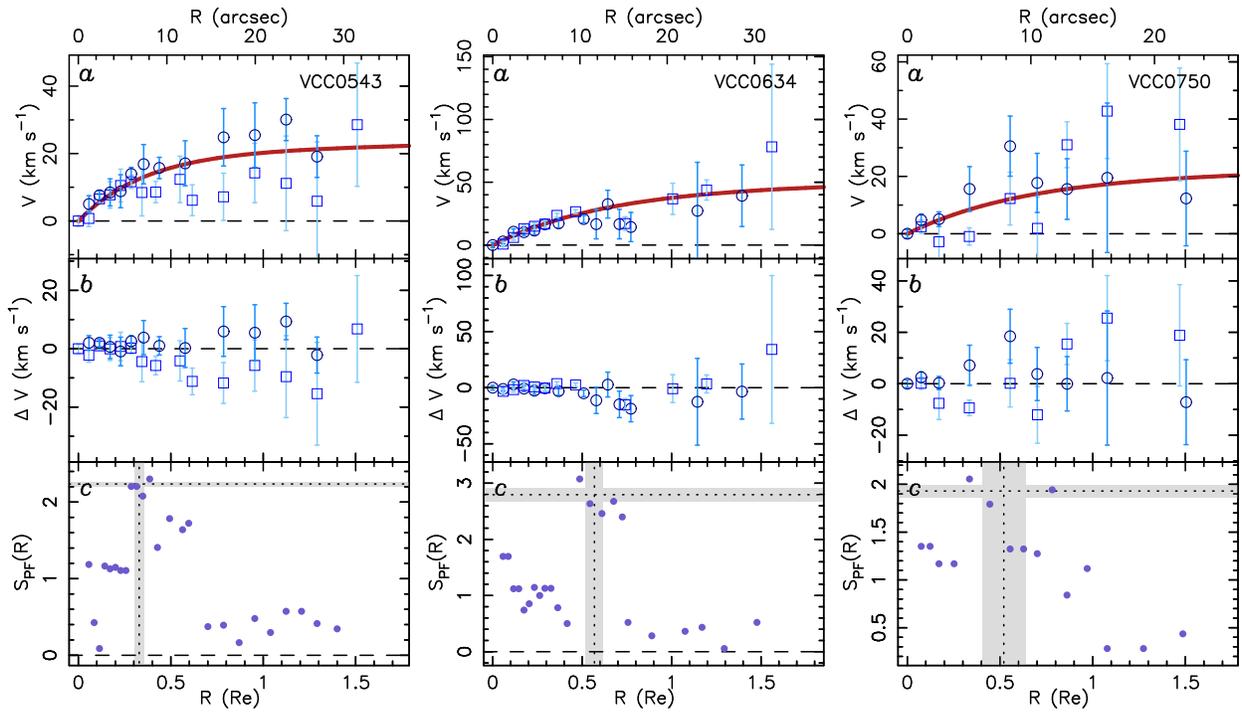

\centering
\resizebox{0.3\textwidth}{!}{\includegraphics[angle=-90,width=8cm]{fig57a.ps}}
\resizebox{0.3\textwidth}{!}{\includegraphics[angle=-90,width=8cm]{fig57b.ps}}
\resizebox{0.3\textwidth}{!}{\includegraphics[angle=-90,width=8cm]{fig57c.ps}}
\caption{Same as Figure \ref{Sstat_plots_VCC0009} for VCC~543, VCC~634, and VCC~750.}
\end{figure}

\begin{figure}
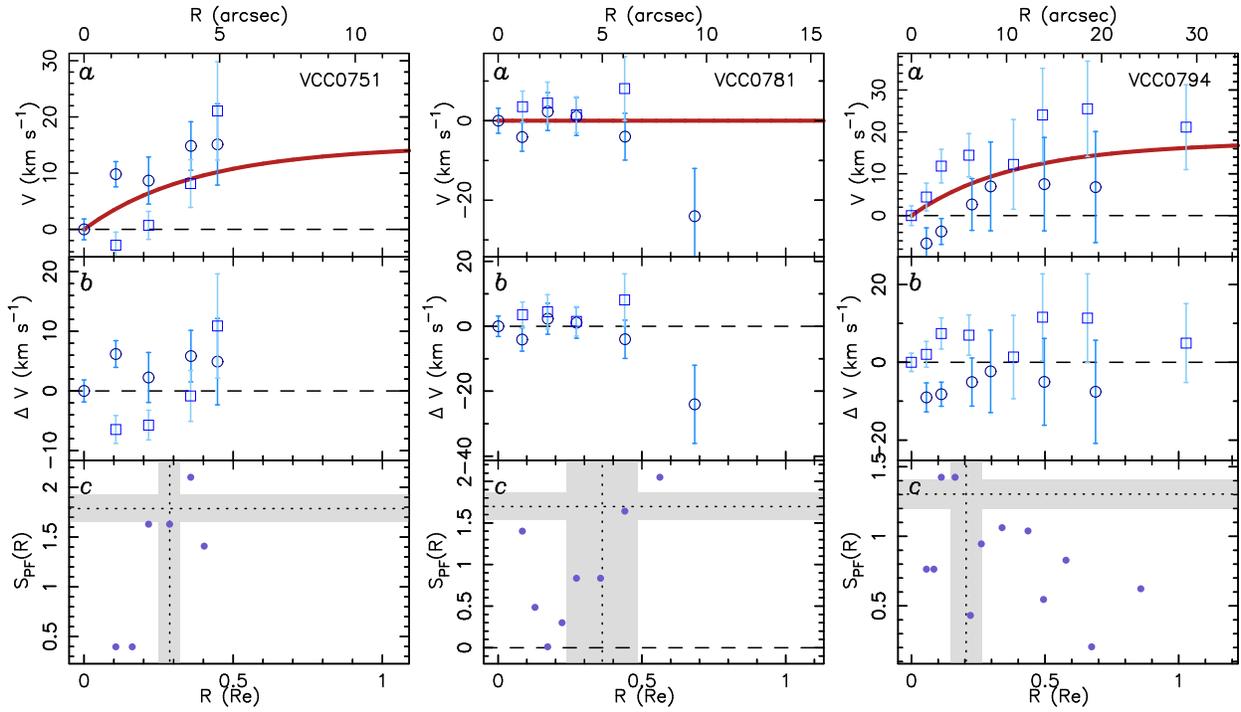

\centering
\resizebox{0.3\textwidth}{!}{\includegraphics[angle=-90,width=8cm]{fig58a.ps}}
\resizebox{0.3\textwidth}{!}{\includegraphics[angle=-90,width=8cm]{fig58b.ps}}
\resizebox{0.3\textwidth}{!}{\includegraphics[angle=-90,width=8cm]{fig58c.ps}}
\caption{Same as Figure \ref{Sstat_plots_VCC0009} for VCC~751, VCC~781, and VCC~794.}
\end{figure}

\begin{figure}
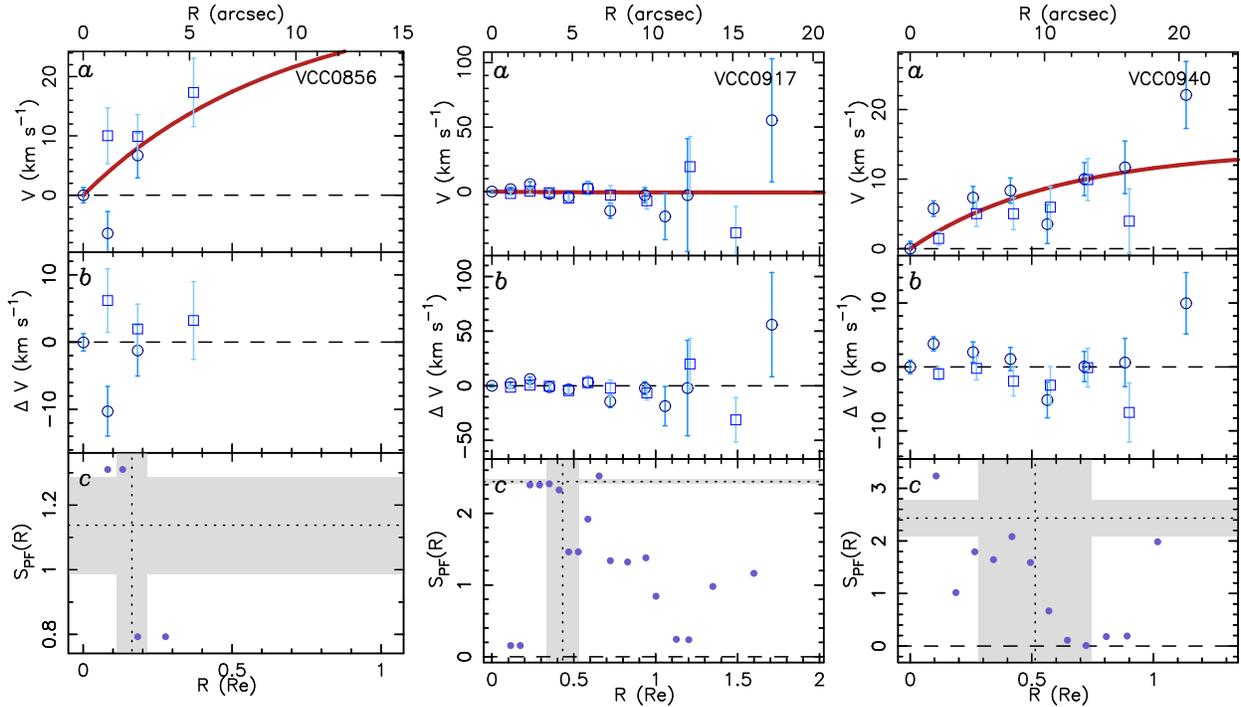

\centering
\resizebox{0.3\textwidth}{!}{\includegraphics[angle=-90,width=8cm]{fig59a.ps}}
\resizebox{0.3\textwidth}{!}{\includegraphics[angle=-90,width=8cm]{fig59b.ps}}
\resizebox{0.3\textwidth}{!}{\includegraphics[angle=-90,width=8cm]{fig59c.ps}}
\caption{Same as Figure \ref{Sstat_plots_VCC0009} for VCC~856, VCC~917, and VCC~940.}
\end{figure}

\begin{figure}
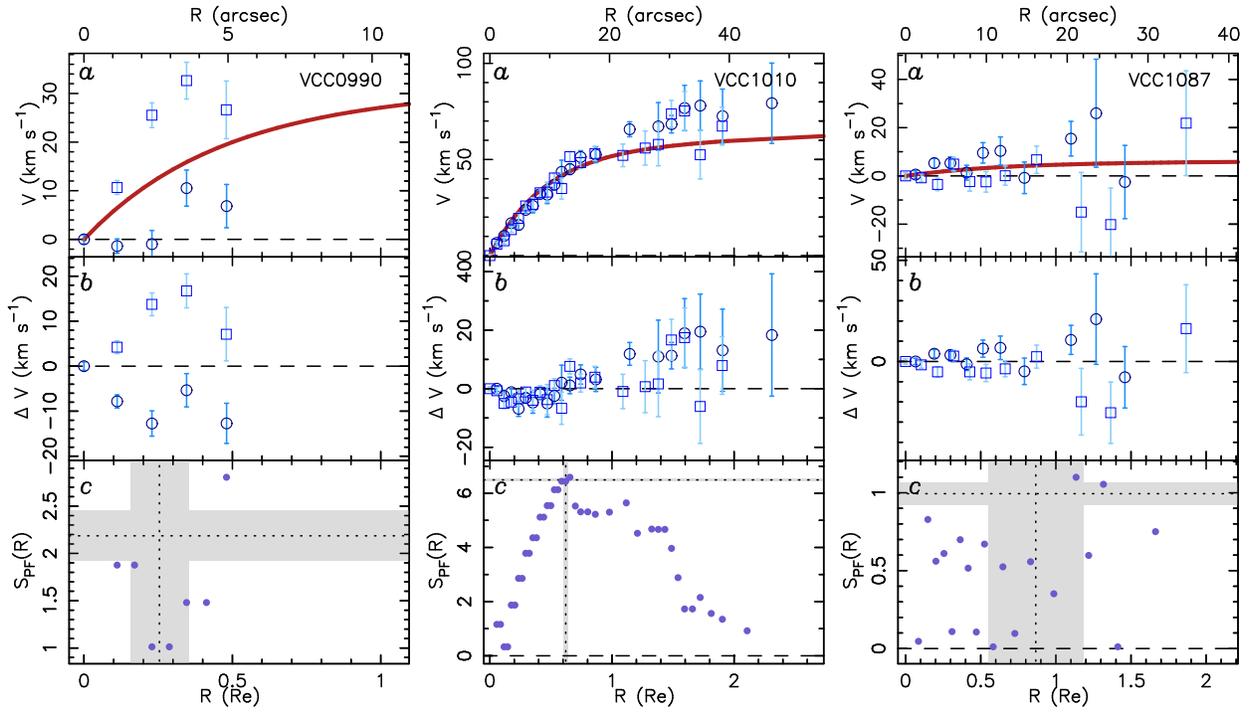

\centering
\resizebox{0.3\textwidth}{!}{\includegraphics[angle=-90,width=8cm]{fig60a.ps}}
\resizebox{0.3\textwidth}{!}{\includegraphics[angle=-90,width=8cm]{fig60b.ps}}
\resizebox{0.3\textwidth}{!}{\includegraphics[angle=-90,width=8cm]{fig60c.ps}}
\caption{Same as Figure \ref{Sstat_plots_VCC0009} for VCC~990, VCC~1010, and VCC~1087.}
\end{figure}

\begin{figure}
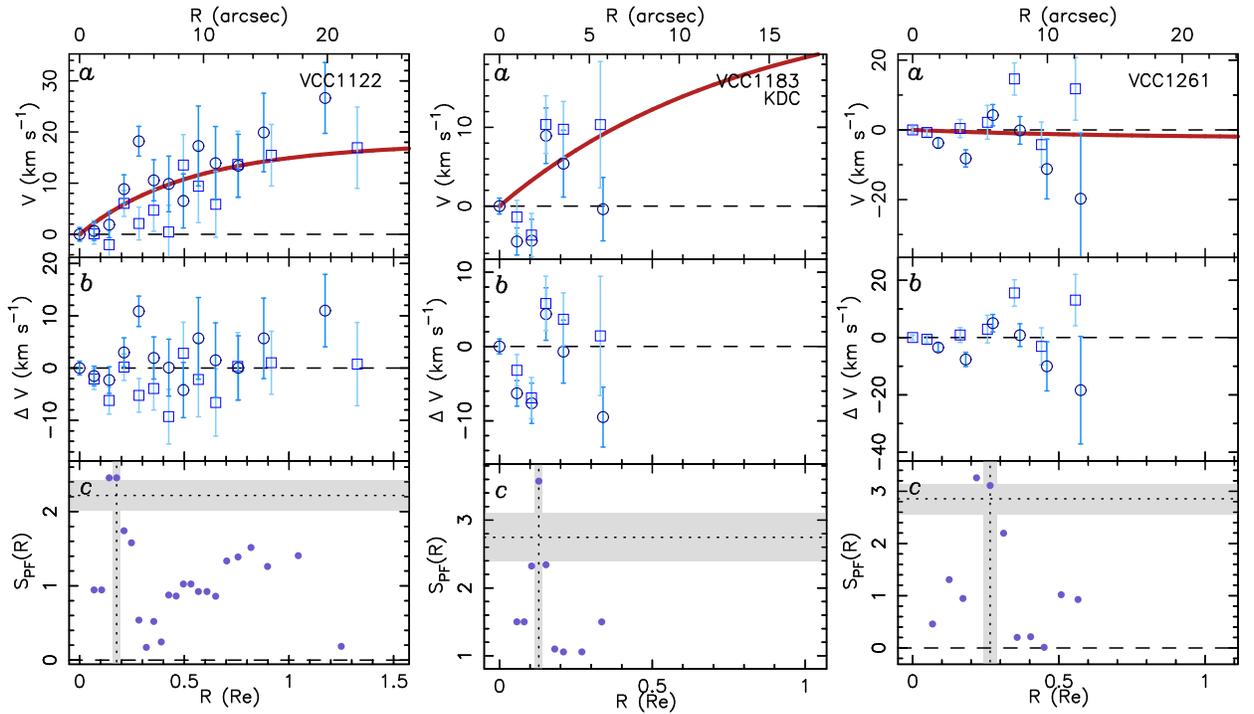

\centering
\resizebox{0.3\textwidth}{!}{\includegraphics[angle=-90,width=8cm]{fig61a.ps}}
\resizebox{0.3\textwidth}{!}{\includegraphics[angle=-90,width=8cm]{fig61b.ps}}
\resizebox{0.3\textwidth}{!}{\includegraphics[angle=-90,width=8cm]{fig61c.ps}}
\caption{Same as Figure \ref{Sstat_plots_VCC0009} for VCC~1122, VCC~1183, and VCC~1261.}
\end{figure}

\begin{figure}
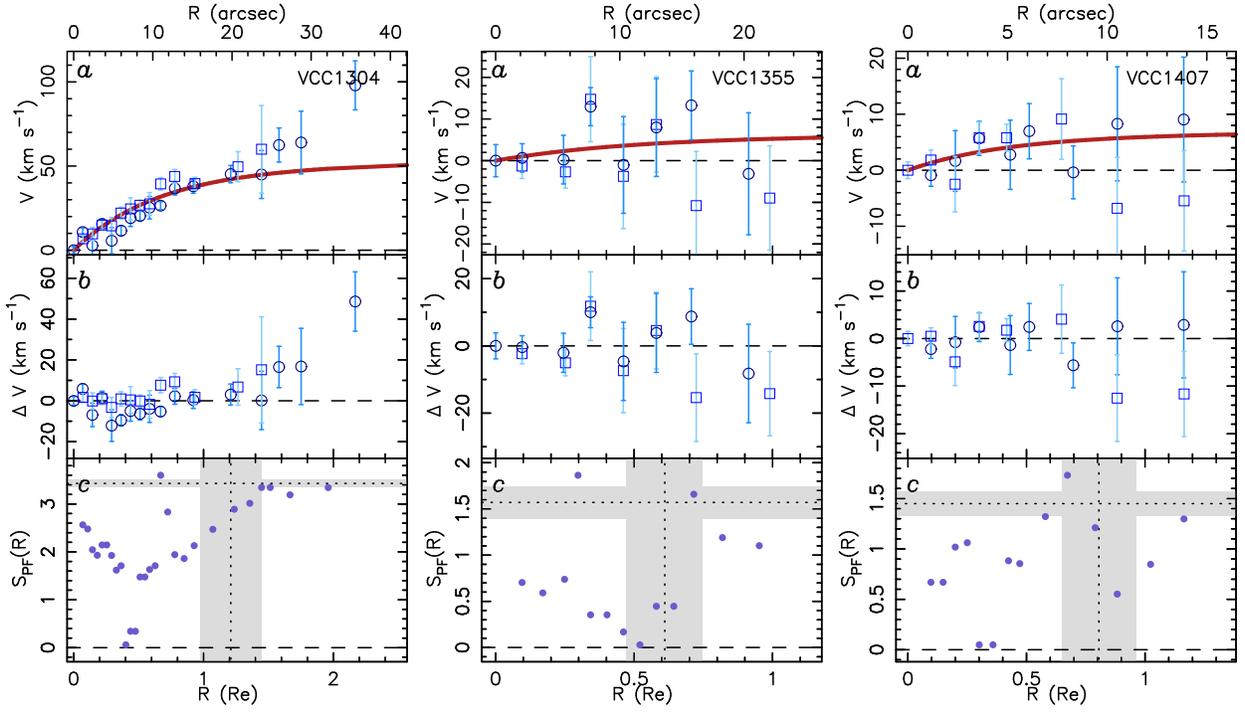

\centering
\resizebox{0.3\textwidth}{!}{\includegraphics[angle=-90,width=8cm]{fig62a.ps}}
\resizebox{0.3\textwidth}{!}{\includegraphics[angle=-90,width=8cm]{fig62b.ps}}
\resizebox{0.3\textwidth}{!}{\includegraphics[angle=-90,width=8cm]{fig62c.ps}}
\caption{Same as Figure \ref{Sstat_plots_VCC0009} for VCC~1304, VCC~1355, and VCC~1407.}
\end{figure}

\begin{figure}
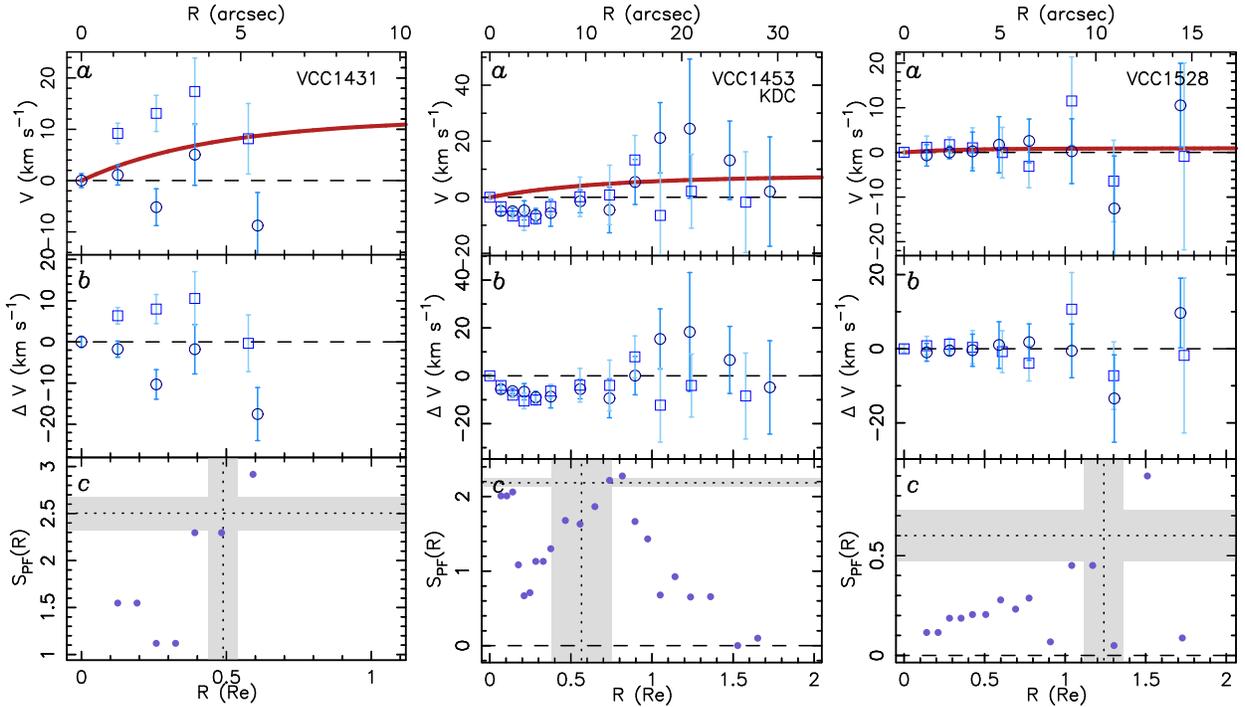

\centering
\resizebox{0.3\textwidth}{!}{\includegraphics[angle=-90,width=8cm]{fig63a.ps}}
\resizebox{0.3\textwidth}{!}{\includegraphics[angle=-90,width=8cm]{fig63b.ps}}
\resizebox{0.3\textwidth}{!}{\includegraphics[angle=-90,width=8cm]{fig63c.ps}}
\caption{Same as Figure \ref{Sstat_plots_VCC0009} for VCC~1431, VCC~1453, and VCC~1528.}
\end{figure}

\begin{figure}
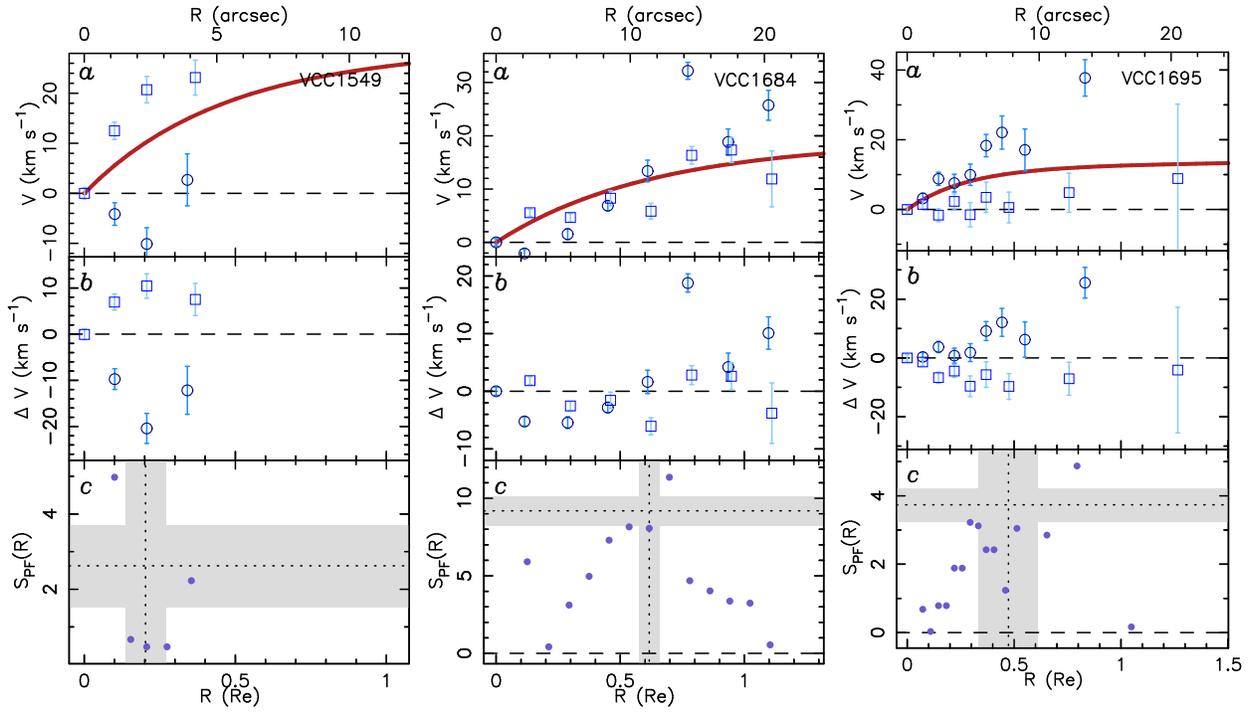

\centering
\resizebox{0.3\textwidth}{!}{\includegraphics[angle=-90,width=8cm]{fig64a.ps}}
\resizebox{0.3\textwidth}{!}{\includegraphics[angle=-90,width=8cm]{fig64b.ps}}
\resizebox{0.3\textwidth}{!}{\includegraphics[angle=-90,width=8cm]{fig64c.ps}}
\caption{Same as Figure \ref{Sstat_plots_VCC0009} for VCC~1549, VCC~1684, and VCC~1695.}
\end{figure}

\begin{figure}
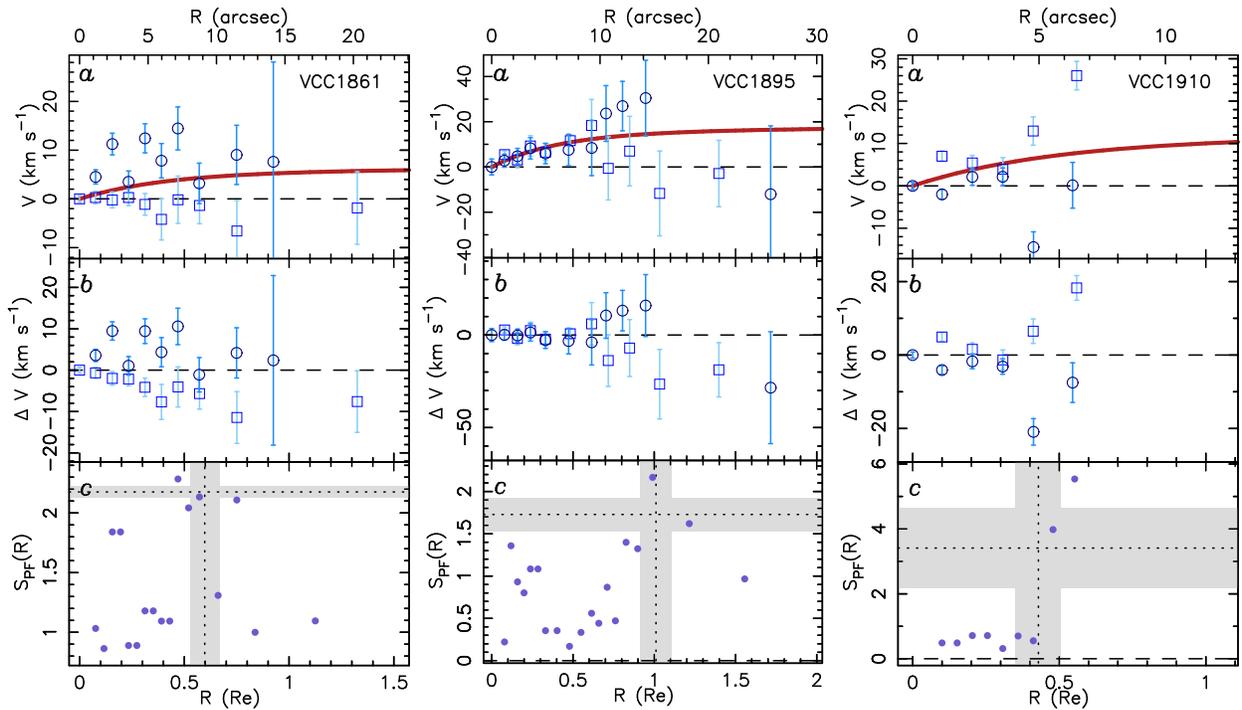

\centering
\resizebox{0.3\textwidth}{!}{\includegraphics[angle=-90,width=8cm]{fig65a.ps}}
\resizebox{0.3\textwidth}{!}{\includegraphics[angle=-90,width=8cm]{fig65b.ps}}
\resizebox{0.3\textwidth}{!}{\includegraphics[angle=-90,width=8cm]{fig65c.ps}}
\caption{Same as Figure \ref{Sstat_plots_VCC0009} for VCC~1861, VCC~1895, and VCC~1910.}
\end{figure}

\clearpage

\begin{figure}
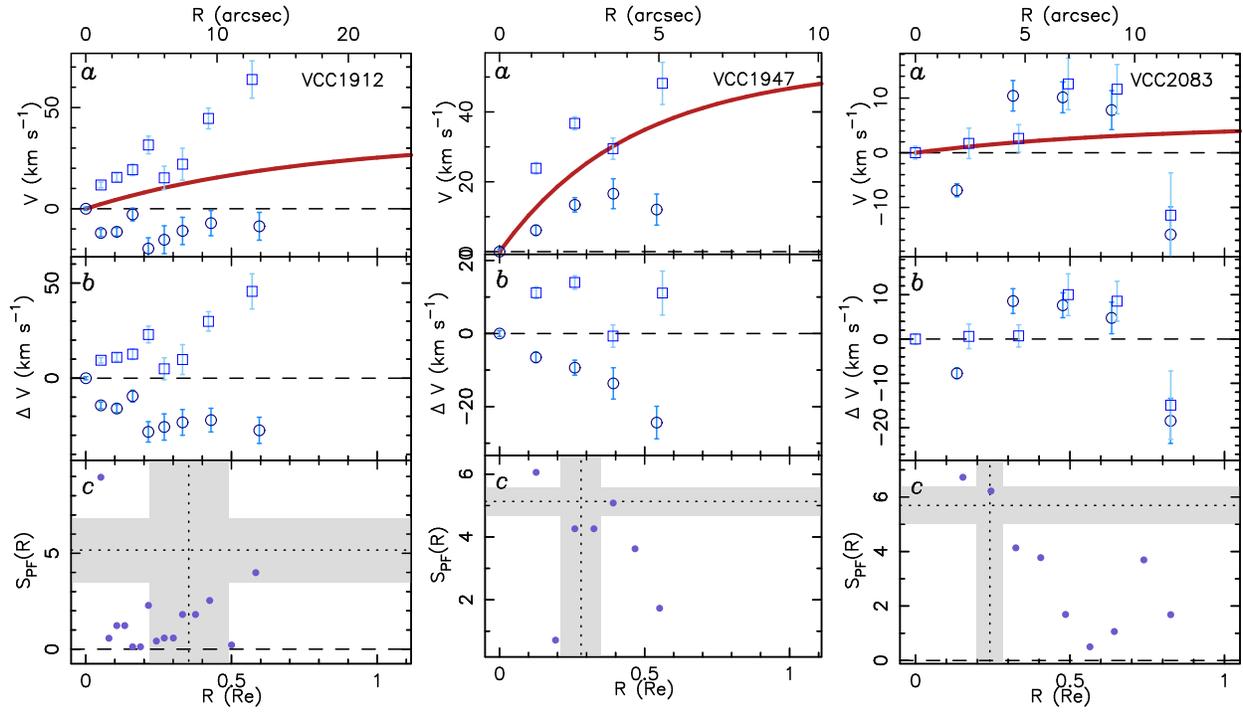

\centering
\resizebox{0.3\textwidth}{!}{\includegraphics[angle=-90,width=8cm]{fig66a.ps}}
\resizebox{0.3\textwidth}{!}{\includegraphics[angle=-90,width=8cm]{fig66b.ps}}
\resizebox{0.3\textwidth}{!}{\includegraphics[angle=-90,width=8cm]{fig66c.ps}}
\caption{Same as Figure \ref{Sstat_plots_VCC0009} for VCC~1912, VCC~1947, and VCC~2083.}
             \label{Sstat_plots_VCC2083}
\end{figure}

\clearpage
\begin{figure*}
\centering
\resizebox{0.35\textwidth}{!}{\includegraphics[angle=-90,width=8cm]{fig67a.ps}}
\resizebox{0.35\textwidth}{!}{\includegraphics[angle=-90,width=8cm]{fig67b.ps}}
\resizebox{0.35\textwidth}{!}{\includegraphics[angle=-90,width=8cm]{fig67c.ps}}
\resizebox{0.35\textwidth}{!}{\includegraphics[angle=-90,width=8cm]{fig67d.ps}}
\resizebox{0.35\textwidth}{!}{\includegraphics[angle=-90,width=8cm]{fig67e.ps}}
\resizebox{0.35\textwidth}{!}{\includegraphics[angle=-90,width=8cm]{fig67f.ps}}
\resizebox{0.35\textwidth}{!}{\includegraphics[angle=-90,width=8cm]{fig67g.ps}}
\resizebox{0.35\textwidth}{!}{\includegraphics[angle=-90,width=8cm]{fig67h.ps}}
\resizebox{0.35\textwidth}{!}{\includegraphics[angle=-90,width=8cm]{fig67i.ps}}
\resizebox{0.35\textwidth}{!}{\includegraphics[angle=-90,width=8cm]{fig67j.ps}}
\resizebox{0.35\textwidth}{!}{\includegraphics[angle=-90,width=8cm]{fig67k.ps}}
\caption{Comparison of the rotation curves for the 11 dEs in common with the literature (see Section \ref{lit_comp}). The grey region shows the rotation curves measured in this work. The blue dots indicate the rotation curve measured by \citet{Geha02,Geha03}. The green squares indicate the rotation curve measured by \citet{Chil09}. The position angle of the long-slit used by each work is indicated in the left upper corner in the corresponding color. After the name of the galaxy, the telescope used for that galaxy in this work is indicated within brackets. The rotation curves agree within the $1\sigma_G$ uncertainties between the different works independently from the telescope used in the observations. The difference found for VCC~1947 is likely to be due to the large difference in the position angles used to place the long-slits.}
             \label{Rotation_comparison}
\end{figure*}

\clearpage

\begin{figure*}
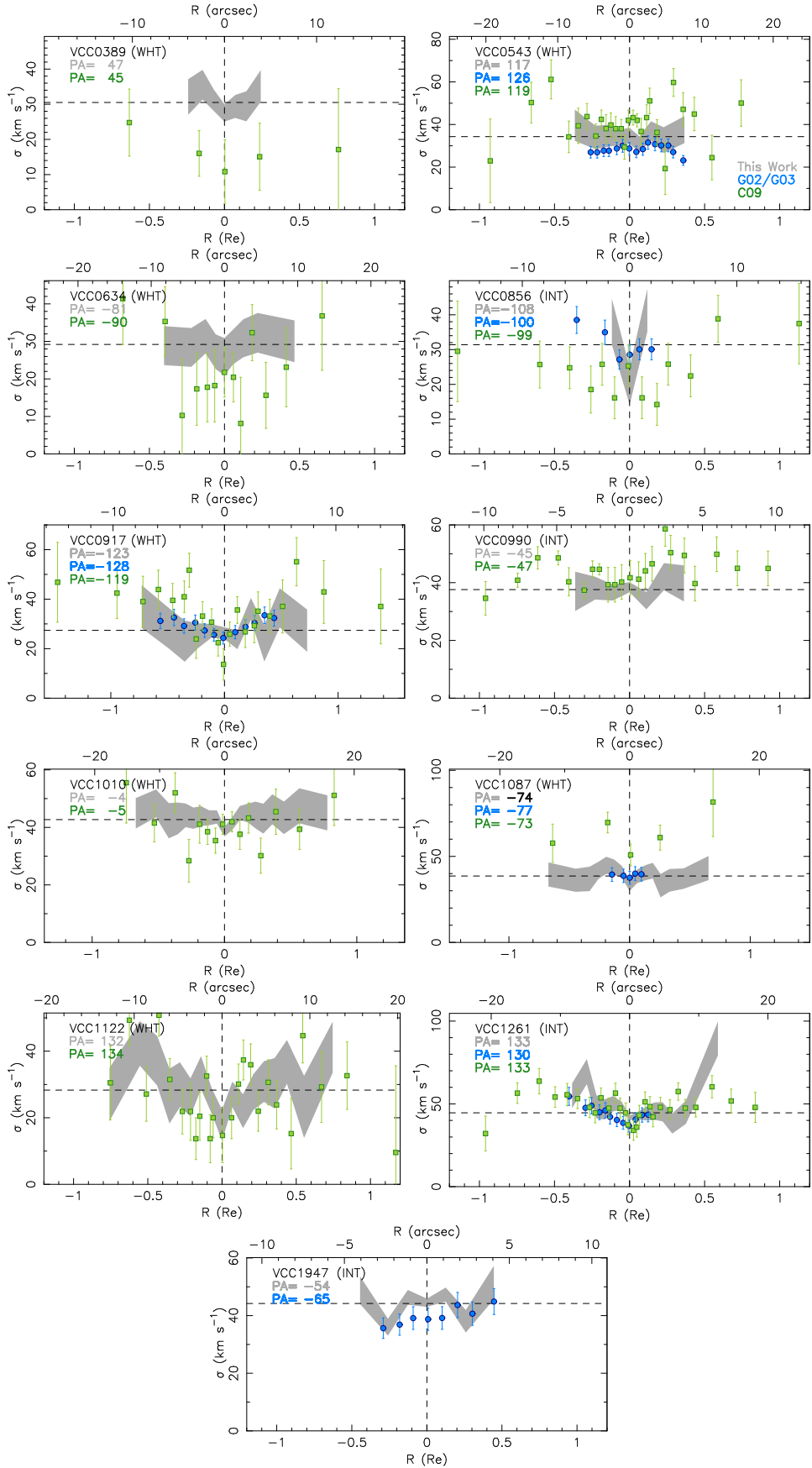

\centering
\resizebox{0.35\textwidth}{!}{\includegraphics[angle=-90,width=8cm]{fig68a.ps}}
\resizebox{0.35\textwidth}{!}{\includegraphics[angle=-90,width=8cm]{fig68b.ps}}
\resizebox{0.35\textwidth}{!}{\includegraphics[angle=-90,width=8cm]{fig68c.ps}}
\resizebox{0.35\textwidth}{!}{\includegraphics[angle=-90,width=8cm]{fig68d.ps}}
\resizebox{0.35\textwidth}{!}{\includegraphics[angle=-90,width=8cm]{fig68e.ps}}
\resizebox{0.35\textwidth}{!}{\includegraphics[angle=-90,width=8cm]{fig68f.ps}}
\resizebox{0.35\textwidth}{!}{\includegraphics[angle=-90,width=8cm]{fig68g.ps}}
\resizebox{0.35\textwidth}{!}{\includegraphics[angle=-90,width=8cm]{fig68h.ps}}
\resizebox{0.35\textwidth}{!}{\includegraphics[angle=-90,width=8cm]{fig68i.ps}}
\resizebox{0.35\textwidth}{!}{\includegraphics[angle=-90,width=8cm]{fig68j.ps}}
\resizebox{0.35\textwidth}{!}{\includegraphics[angle=-90,width=8cm]{fig68k.ps}}
\caption{Comparison of the velocity dispersion profiles for the 11 dEs in common with the literature (see Section \ref{lit_comp}). The colors and symbols are the same as in Figure \ref{Rotation_comparison}. This work and G02/G03 agree well within the $1\sigma_G$ uncertainties. However, there are systematic offsets with respect to C09. These offsets, shown in Figure \ref{comp_sig}, place the data points along the line with slope $-1$ which also results in a Gaussian function whose width is broader than 1 in the right panel of Figure \ref{hists_comp_lit}.}
             \label{Dispersion_comparison}
\end{figure*}

\end{document}